\documentclass[prb, twocolumn, showpacs]{revtex4}         

\usepackage{bm,amsmath,graphicx,subfigure,amssymb,latexsym,mathrsfs,enumerate,color}
\usepackage[mathcal]{euscript}
\begin{document}
\title{Topological solitons in highly anisotropic two dimensional ferromagnets}

\author{B. A. Ivanov}
\email{bivanov@i.com.ua} \affiliation{Institute of Magnetism,
04071 Kiev, Ukraine} \affiliation{National Taras Shevchenko
University of Kiev, 03127 Kiev, Ukraine}

\author{A. Yu. Merkulov}
\affiliation{FOM-Institute for Plasma Physics Rijnhuizen, Postbus
1207, 3430 BE Nieuwegein, The Netherlands}

\author{V. A. Stephanovich}
\homepage{http://cs.uni.opole.pl/~stef}
\email{stef@math.uni.opole.pl} \affiliation{Institute of
Mathematics and Informatics, Opole University, 45-052, Opole,
Poland}

\author{C. E. Zaspel}
\affiliation{University of Montana-Western, Dillon, MT 59725, USA}

\begin{abstract}

We study the solitons, stabilized by spin precession in a classical
two--dimensional lattice model of Heisenberg ferromagnets with
non-small easy--axis anisotropy. The properties of such solitons are
treated both analytically using the continuous model including
higher then second powers of magnetization gradients, and
numerically for a discrete set of the spins on a square lattice. The
dependence of the soliton energy $E$ on the number of spin
deviations (bound magnons) $N$ is calculated.  We have shown that
the topological solitons are stable if the number $N$ exceeds some
critical value $N_{\rm{cr}}$. For $N < N_{\rm{cr}}$ and the
intermediate values of anisotropy constant $K_{\mathrm{eff}} <0.35J$
($J$ is an exchange constant), the soliton properties are similar to
those for continuous model; for example, soliton energy is
increasing and the precession frequency $ \omega (N)$ is decreasing
monotonously with $N$ growth. For high enough anisotropy
$K_{\mathrm{eff}} > 0.6 J$  we found some fundamentally new soliton
features absent for continuous models incorporating even the higher
powers of magnetization gradients. For high anisotropy, the
dependence of soliton energy $E(N)$ on the number of bound magnons
become non-monotonic, with the minima at some "magic" numbers of
bound magnons. Soliton frequency $\omega (N)$ have quite irregular
behavior with step-like jumps and negative values of $\omega $ for
some regions of $N$. Near these regions, stable static soliton
states, stabilized by the lattice effects, exist.
\end{abstract}

\pacs{75.10.Hk, 75.30.Ds, 05.45.-a}

\maketitle

\section{Introduction}
\label{s:introduction}

An analysis of two-dimensional (2D) magnetic solitons has been an
active area of research for more then 30 years, see
Refs.~\onlinecite{Kosevich90, Bar'yakhtar93, Bar'yakhtar94,
ZaspelDrumh, Mertens00} for review. Such solitons are known to play
an important role in the physics of 2D magnetic systems. In the
easy--plane magnets with continuously degenerate ground state there
appear magnetic vortices, responsible for the
Berezinskii--Kosterlitz--Thouless (BKT) phase transition
\cite{Berezinsky72,Kosterlitz73}. The presence of vortices leads to
the emergence of a central peak in dynamical response functions of a
magnet,\cite{Mertens89} which can be observed experimentally
\cite{Wiesler89}. There are no vortices in the easy--axis magnets
with discretely degenerate ground state, but various types of
localized topological solitons appear there. Belavin and Polyakov
were the first to construct the exact analytical solutions for 2D
topological solitons in a continuous model of the isotropic
magnet\cite{Belavin75}. The energy of such Belavin - Polyakov (BP)
solitons $E_{BP}$ in a magnet with exchange constant $J$ is finite
and described by the universal relation
\begin{equation}
E_{\rm {BP}}=4\pi JS^2, \label{ebp}
\end{equation}
$S$ is the atomic spin. The structure of such solitons is described
by a topologically nontrivial distribution of the magnetization
field $\vec{m}(x,y)$,\cite{Kosevich90} which is determined by the
$\pi _2$ topological invariant, see below Eq.~(\ref{eq:Pontryagin})
and Refs.\onlinecite{Topology} for more details.  Belavin and
Polyakov have also proved that such solitons are responsible for the
destruction of the long--range magnetic order in purely continuous
isotropic models at any finite temperature. \cite{Belavin75}  The
studies of topological solitons has become interesting now due to
their possible application in high--energy physics,\cite{Walliser00}
and the quantum Hall effect \cite{Prange90}. Also, such soliton
solutions after the change $y \to c\tau$ ($\tau $ is an imaginary
time, $c$ is the speed of magnons) determine so-called instantons,
describing non-small quantum fluctuations in 1D isotropic
antiferromagnets\cite{Affleck}.

Note that the properties of the solitons in this isotropic model is
rather academic problem since all real magnets have a discrete
lattice structure and non-zero anisotropy. The role of uniaxial
anisotropy has been investigated in a number of articles, see.
Refs.~\onlinecite{Kosevich90,Bar'yakhtar93} for review. Topological
solitons of the same $\pi _2$ topological structure are also
inherent for standard continuum (with accounting of terms quadratic
on the magnetization field gradients, like the term $W_2$ in the
first line of  Eq.~\eqref{en4ok} below) models of anisotropic
magnets. The basic problem of soliton physics in 2D magnets is
related to soliton stability. According to the famous
Hobard--Derrick theorem,\cite{Hobard63,Derrick64} the stable static
non--one--dimensional soliton with finite energy and finite radius
does not exist in the standard nonlinear field models; the soliton
is unstable against collapse. This is, in particular, true for the
uniaxial 2D ferromagnet with the anisotropy energy density ${\cal
W}_a\propto m_x^2+m_y^2$.

The possibility to construct non--one--dimensional solitons stable
against collapse is due to the presence of additional integrals of
motion. For example, such solitons can be realized in the uniaxial
ferromagnet due to the conservation of $z$--projection of the total
spin \cite{Kosevich90,Bar'yakhtar93}. This leads to the appearance
of so--called \emph{precessional} solitons characterizing by
time-independent projection of magnetization onto the easy axis
($z$-axis hereafter), with the precession of the magnetization
vector $\vec m$ at constant frequency around the $z$ axis. The
analogies of such precessional solitons are known to occur in
different models of field theory and condensed matter physics, see
Ref.~\onlinecite{Makhankov78} for review. Here the stability of 2D
solitons is not related directly to their topological properties;
the non--topological dynamical solitons may also exist in magnets.
Such solitons are characterized by the spatially localized function
$\vec{m}(x,y)$, which does not have nontrivial topological
properties. All stable precessional solitons, topological and
non-topological, realize the minimum of energy for a given number of
spin $S_z$ deviations. In the semiclassical approximation, this
value can take integer values $N$ only, and can be interpreted as a
number of magnons excited in the magnet. Thus, we naturally arrive
at the concept of a soliton as a bound state of a large number of
magnons \cite{Kosevich90}.

Two - parameter (parameters are precession frequency and velocity of
translational motion of a soliton) small-amplitude
\emph{non-topological} magnetic solitons moving with arbitrary
velocity in a two-dimensional easy - axis ferromagnet have been
constructed in Ref.\onlinecite{IvZaspYastr}. Their minimal energy
$E_{\rm{NT}} = 11.7JS^2$ depends only on combination $JS^2$ similar
to Eq. ({\ref{ebp}), which is a bit smaller then the energy of above
discussed Belavin-Polyakov soliton, $E_{\rm{NT}} = 0.93E_{\rm
{BP}}$.  For such solitons, the relation between their energy (for
given $N$) and momentum  can be thought of as their dispersion law.
Near the minimal energy $E_{\rm{NT}}$, this dispersion law has a
form $E(P, N)\simeq \varepsilon(P)\cdot N$, where $\varepsilon(P)$
is a dispersion law for linear magnons. For $P=0$, this gives the
critical number of bound magnons $N_c=E_{\rm{NT}}/\varepsilon(0)$.
Thus, these solitons are nothing but weakly coupled magnon clouds,
see Fig. \ref{fig:v1} below. The expression for the soliton
dispersion law is used to calculate the soliton density and the
soliton contributions to thermodynamic quantities (response
functions) like specific heat. The signature of soliton contribution
to the response functions of a magnet is an Arrhenius temperature
dependence like $\exp(-E_0/T)$ with the characteristic value $E_0$
as a soliton energy.  Such behavior with $E_{\rm{NT}}$ $ \leq E_0
\leq$ $E_{\rm{BP}}$ has been observed experimentally in
Refs.~\onlinecite{Waldner+,Zaspel+}, see
Ref.~\onlinecite{ZaspelDrumh} for review.  Comparison of
contributions from solitons and free magnons shows that there is a
wide temperature range where the solitons give more important
contribution to thermodynamic functions such as  heat capacity or
density of spin deviations.

We note here, that the structure of these non-topological solitons
for $E \geq E_{\rm{NT}}$ is essentially different from that of
topological solitons in uniaxial magnets. Namely,  as $E\to
E_{\rm{BP}}$, the radius of topological soliton in continuous model
of anisotropic magnets diminishes, making them "more localized",
contrary to non--topological solitons, which become delocalized  as
$E\to E_{\rm{NT}}$. Hence, although the energy of topological
solitons is a little larger then that of non--topological solitons
(0.89 and 1 in the units of $E_{\rm {BP}}$, see above), it is
possible that only BP-type topological solitons would contribute to
response functions measured by neutron scattering in the region of
non-small momentum transfer. On the other hand, recent Monte-Carlo
simulations for 2D discrete models of easy-axis magnets did not show
any signatures of small-radius topological solitons \cite{Wysin+}.

For real magnets, which are discrete spin systems on a lattice,
there is an additional problem of application of the topological
arguments, which, strictly speaking, can be applied only to
continuous functions ${\vec m}({\vec r},t)$. It is widely accepted
that the continuous description is valid for discrete systems if the
characteristic scale $l_0$ of magnetization ${\vec m}({\vec r},t)$
variation, $|\nabla {\vec m}|\sim {\vec m}/l_0$, is much larger than
lattice constant $a$. The analysis of the magnetic vortices have
shown that $\pi _1-$ topological charge of a vortex is determined by
the behavior at infinities only so that the topological structure of
such a vortex survives even in "very discrete" models with $l_0<a$.
As for our case of $\pi _2$ topological charge, the situation is not
so simple and obvious. From one side, the continuous approach
describes quantitatively the magnetization distribution in the
vortex core already at $l_0\approx 1.5a$\cite{Ivanov96}. On the
other side, so-called cone state vortices, with different energies
for two possible spin directions in the vortex core are much more
sensitive to the anisotropy, in fact, to the parameter $a/l_0$. Even
for $l_0 > 10a$ their  $\pi _2-$ topological charge (polarization
$p\equiv m_z(0)=\pm 1$), characterizing the core structure of
vortices, for heavy vortices with higher energy can change so that
they convert into more preferable light vortices with  opposite
polarization $p$,\cite{IvWysin02} that never happened for continuum
model \cite{IvSheka}. Thus, the role of the discreteness effects is
quite ambiguous.

The above situation resembles the one-dimensional case - there are
also kinks and breathers (non--topological solitons) with smaller
energy. But it is well-known, that in the 1D case for low anisotropy
only kinks contribute to the response functions since small energy
breathers transit continuously into weakly coupled magnon
conglomerates. We note that for such systems in one space dimension,
the difference between topological and non-topological solitons for
high anisotropy is not that large. For instance, the spin complexes
with several $N\sim 10$ magnons have been observed in a chain
material CoCl$_2 \cdot$ 2H$_2$O with high Ising - type anisotropy
\cite{{ExperCsCl2}}. These complexes can be interpreted as
non-topological one-dimensional solitons.

The present work is devoted to the analysis of 2D solitons, both
topological and non-topological, in the strongly anisotropic magnets
accounting  for discreteness effects. In other words, here we
investigate the influence of finiteness of $a/l_0$ on the soliton
structure. For intermediate values of anisotropy we found the
presence of the critical number of bound magnons $N_{\mathrm{cr}}$:
the topological soliton is stable at $N > N_{\mathrm{cr}}$ only. For
very large anisotropy we found the specific effects of non-monotonic
dependence of soliton properties on the number of bound magnons,
caused by discreteness leading to the presence of "magic" magnon
numbers.

\section{The discrete model and its continuous description}
\label{s:model}
     We consider the model of a classical 2D ferromagnet
with uniaxial anisotropy, described by the following Hamiltonian
\begin{eqnarray} \label{eq:H-discrete}
&&{\mathcal H} = -\sum_{{\vec n},{\vec a}}\! \left(J\vec{S}_{\vec
n}\!\cdot\! \vec{S}_{{\vec n}+{\vec a}}
+\kappa S^z_{\vec n} S^z_{{\vec n}+{\vec a}} \right)+ \nonumber \\
&&+K\sum_{\vec n} [(S^x_{\vec n})^2+(S^y_{\vec n})^2].
\end{eqnarray}
Here  $\vec{S}\equiv\left(S^x, S^y, S^z\right)$ is a classical spin
vector with fixed length $S$ on the site $\vec{n}$ of a 2D square
lattice. The first summation runs over all nearest--neighbors ${\vec
a}$, $J > 0$ is the exchange integral, and the constant $\kappa $
describes the anisotropy of spin interaction. In subsequent
discussion, we will refer to this type of anisotropy as exchange
anisotropy (ExA). Additionally, we took into account single-ion
anisotropy (SIA) with constant $K$. We consider the $z$--axis as the
easy magnetization direction so that $K>0$ or $\kappa >0$.

In quantum case, the Hamiltonian (\ref{eq:H-discrete}) commutes with
$z-$projection of total spin. It is more convenient to use
semiclassical terminology, and to present it as a number of bound
magnons in a soliton $N$, defined by the equation
\begin{equation} \label{eq:N-discrete}
N = \sum_{\vec{n}} \left(S-S^z_{\vec{n}} \right).
\end{equation}

The spin dynamics is described by Landau--Lifshitz equations
\begin{equation} \label{eq:LL-discrete}
\frac{d \vec{S}_{\vec{n}} }{dt} =  - \frac1\hbar \left[
\vec{S}_{\vec{n}}\times \frac{\partial \mathcal{H} }{\partial
\vec{S}_{\vec{n}}}\right].
\end{equation}
In the case of weak anisotropy, $K,\ \kappa \ll  J$, the
characteristic size of excitations $l_0 \gg a$, see Eq.
\eqref{lzero} below, so that the magnetization varies slowly in a
space. In this case we can introduce the smooth function $\vec
S(x,y,t)$ instead of variable $\vec S_n(t)$ and use a continuous
approximation for the Hamiltonian \eqref{eq:H-discrete}. It is based
on the expansion of a classical magnetic energy $E$ in power series
of magnetization ${\vec S}$ gradients,
\begin{equation}\label{enw1}
   E=W_2+W_4+...,
\end{equation}
where $W_2$ contains zeroth and second order contributions to
magnetic energy and $W_4$ contains the fourth powers. These are
given by

\begin{subequations}
\begin{eqnarray} \label{enw}
W_2&=&\int d^2x \Biggl\{\frac {K_{\rm{eff}}}{a^2}(S^2-S_z^2)+\frac
{J}{2}\left(\nabla {\vec S}\right)^2 + \nonumber \\
&+&\frac{\kappa}{2} \left(\nabla S_z\right)^2\Biggl\}, \label{w2} \\
  W_4&=&-\frac{a^4}{24} \int d^2x\Biggl\{J \left[ \left( \frac{\partial ^2 {\vec
S}}{\partial x^2}\right)^2 + \left(\frac{\partial ^2 {\vec
S}}{\partial y^2}\right)^2 \right]+  \nonumber \\
&+& \kappa \left[ \left( \frac{\partial ^2 S_z }{\partial
x^2}\right)^2 + \left( \frac{\partial ^2 S_z }{\partial
y^2}\right)^2 \right] \Biggr\},\label{w4}
\end{eqnarray}
\end{subequations}
where $\nabla $ is a 2D gradient of the function ${\vec S}({\vec
r},t)$. Here, we used integrations by parts with respect to the fact
that our soliton texture is spatially localized. We omit unimportant
constants and limit ourselves to the terms of fourth order only as
they are playing a decisive role in stabilization of solitons, see
for details \cite{Ivanov86,IvWysin02}. Also, we introduce the
effective anisotropy constant
\begin{equation}\label{keff}
K_{\rm{eff}}=K+2\kappa.
\end{equation}
We note here, that single-ion anisotropy enters only $W_2$, but not
$W_4$ and higher terms, while exchange anisotropy enters every term
of the above expansion over powers of magnetization gradients.
Usually, this difference is not important for small anisotropy, $K,\
\kappa \ll J$, but, as we will see below, this fact gives
qualitatively different behavior near the soliton stability
threshold.

Introducing the angular variables for normalized magnetization
\begin{equation}\label{ang}
\vec{m} = \frac{\vec{S}}{S} =
\left(\sin\theta\cos\phi;\sin\theta\sin\phi;\cos\theta\right),
\end{equation}
we obtain the following form of the classical magnetic energy
\begin{widetext}
\begin{eqnarray}
E[\theta,\phi]&=&W_2+W_4,\ W_2=JS^2\int
d^2x\left\{\frac{K_{\rm{eff}} }{a^2}\sin^2\theta+ \frac 12\left[
\left( \nabla \theta \right) ^{2}\left( 1+\kappa \sin ^{2}\theta
\right) +\left( \nabla \varphi
\right) ^{2}\sin ^{2}\theta \right] \right\},\nonumber \\
W_4&=&-\frac{1}{24}Ja^2S^2\int d^2x\Biggl\{\left( \nabla^2 \theta
\right) ^{2}\left[ 1+\kappa \sin ^{2}\theta \right] +\left( \nabla
\theta \right)
^{4}\left[ 1+\kappa \cos ^{2}\theta \right] +  \nonumber \\
&&+\sin ^{2}\theta \left( \nabla \varphi \right) ^{2}\left[ \left(
\nabla \varphi \right) ^{2}+2\left( \nabla \theta \right)
^{2}\right] +2\sin \theta \cos \theta (\nabla^2 \theta ) \left[
\kappa \left( \nabla \theta \right) ^{2}-\left( \nabla \varphi
\right) ^{2}\right] \Biggr\}. \label{en4ok}
\end{eqnarray}
\end{widetext}
In this long expression we omitted the terms with scalar product
of gradients like $(\nabla \theta, \, \nabla\phi)$, because they
do not contribute to the trial function we will use for analysis,
see below.

In terms of fields $\theta$ and $\phi$, the continuous analog of
Landau--Lifshitz equations \eqref{eq:LL-discrete} read
\begin{equation} \label{eq:LL-continuum}
\sin\theta\frac{\partial\phi}{\partial t} = \frac{a^2}{\hbar S}
\frac{\delta E}{\delta \theta}, \quad
\sin\theta\frac{\partial\theta}{\partial t} = -\frac{a^2}{\hbar
S}\frac{\delta E}{\delta \phi}.
\end{equation}

These equations can be derived from the Lagrangian
\begin{equation} \label{eq:Lagrangian}
\mathcal{L}[\theta,\phi] = -\frac{\hbar S}{a^2}\int d^2 x
(1-\cos\theta) \frac{\partial\phi}{\partial t} - E[\theta,\phi].
\end{equation}

The simplest nonlinear excitation of the model \eqref{eq:Lagrangian}
with $E\equiv W_2$ is the 2D localized soliton, characterized by the
homogeneous distribution of magnetization far from its core.
Topological properties of the soliton are determined by the mapping
of physical $XY$--plane to the $\mathbb{S}^2$--sphere given by the
equation $\mathbf{m}^2=1$ of the order parameter space. This mapping
is described by the homotopy group $\pi_2(\mathbb{S}^2) =
\mathbb{Z}$, see, \cite{Topology} which is characterized by the
topological invariant (Pontryagin index)
\begin{multline}\label{eq:Pontryagin}
Q=\frac 1{8\pi }\int\limits_{S^2} d^2 x\;\varepsilon _{\alpha \beta
} \left[\vec m\cdot\bigl(\nabla _\alpha \vec m\times \nabla _\beta
\vec m \bigr)\right]= \\ =\frac 1{4\pi }\int\limits_{S^2}\sin \theta
\;d\theta \;d\phi ,
\end{multline}
taking integer values, $Q\in\mathbb{Z}$. Here $\varepsilon _{\alpha
\beta }$ is Levi-Civita tensor.

To visualize the structure of a topological soliton, we consider the
case of a purely isotropic magnet with $K_{\rm{eff}}=0 $ and
$W_4=0$. For this isotropic continuous model the soliton solution is
aforementioned BP soliton of the form \cite{Belavin75}
\begin{equation} \label{eq:BP-soliton}
\tan\frac{\theta}{2} = \left(\frac{R}{r}\right)^{|Q|}, \quad \phi
= \varphi_0 + Q\chi,
\end{equation}
where $r$ and $\chi$ are polar coordinates in  the $XY$--plane,
$\varphi_0$ is an arbitrary constant. The energy (\ref{ebp}) of this
soliton does not depend of its radius $R$, which is arbitrary
parameter for the isotropic magnet, see Ref.~\onlinecite{Belavin75}.
However, even small anisotropy breaks the scale invariance of the
above model since now the scale
\begin{equation}\label{lzero}
l_0^2=\frac{a^2J}{2K_{\rm{eff}}}
\end{equation}
enters the problem.

In the latter case, the soliton energy with respect to $W_2$ term
only has the form $E=E_{\rm {BP}}+{\rm {const}}\cdot(R/l_0)^2$ and
hence has a minimum at $R=0$ only, which signifies the instability
of the static soliton against collapse in this model. Obviously, the
scale invariance is also broken in the initial discrete model. The
"trace of discreteness" in our continuous model
(\ref{eq:Lagrangian}) is the presence of a contribution $W_4$. This
term gives the contribution to the soliton energy proportional to
$JS^2a^2/R^2$. For some  rather exotic models containing higher
powers of a magnetization field gradients like $ \left(\vec{\nabla}
\vec{m}\right)^4$ (with \emph{positive} sign),
\cite{Skirm61,Hobard65,Enz77a,Enz78} such term might be able to
stabilize even static soliton against collapse, see \cite{Ivanov86}
for details. However, for Heisenberg magnets this kind of
stabilization is very problematic. For example, discrete magnetic
models with the Heisenberg interaction of nearest neighbors only
(i.e. those without biquadratic exchange and/or
next--nearest--neighbors interaction) have \emph{negative} $W_4 <
0$, and the higher powers of magnetization gradients do not
stabilize a static soliton in this case. Moreover, as we will show
below, the presence of discreteness ruins the stability of the
precessional soliton with $N<N_{\mathrm{cr}}$ even in the case, when
it is stable in the simplest model with $W=W_2$ only.

We now discuss the stability of precessional solitons. For a purely
uniaxial ferromagnet the energy functional $E[\theta,\phi]$ does not
depend explicitly on the variable $\phi$ so that there exists an
additional integral of motion,\cite{Kosevich90} which is the
continuous analog of Eq.(\ref{eq:N-discrete})
\begin{equation}\label{eq:N}
N = \frac{S}{a^2}\int d^2 x \left(1-\cos\theta\right).
\end{equation}
The conservation law \eqref{eq:N} can provide a conditional (for
constant $N$) minimum of the energy functional $E$, stabilizing the
possible soliton solution. Namely, we may look for an extremum of
the expression
\begin{equation}\label{cond}
L=E-\hbar \omega N,
\end{equation}
where $\omega $ is an internal soliton precession frequency, that
can be regarded as a Lagrange multiplier. Note that this functional
is nothing but the Lagrangian \eqref{eq:Lagrangian} calculated with
respect to specific time dependence
\[
\phi=\omega t+Q\chi+\varphi _0,
\]
which holds instead of (\ref{eq:BP-soliton}) in this case. This
condition leads to the relation,\cite{Kosevich90}
\begin{equation}\label{dEdN}
\hbar \omega= \frac{dE}{dN},
\end{equation}
which makes clear the microscopic origin of the precessional
frequency $\omega$. Namely, an addition of one extra spin deviation
(bound magnon) to the soliton changes its energy by $\hbar \omega$.
Thus, the dependence $\omega (N)$ is extremely important for the
problem of a soliton stability. For the general continuous model of
a ferromagnet, even containing the terms like $W_4$, the sufficient
and necessary condition of soliton stability reads $d\omega/ dN<
0$,\cite{Zhmudsky97} but for the discrete model the validity of this
condition is not clear yet. The point is that the known analytical
methods of a soliton stability analysis rely essentially on the
presence of a zeroth (translational) mode, which is obviously
present for any continuous model, but is absent for discrete models,
where lattice pinning effects are present. To solve the problem of
soliton stability we will investigate explicitly the character of
conditional extremum of the energy with $N$ fixed.

\section{The methods of a soliton structure investigations}

To get the explicit soliton solution and investigate its stability
we have to solve the Landau-Lifshitz equations
(\ref{eq:LL-continuum}) with respect to the energy (\ref{en4ok}).
For the simplest model accounting for $W_2$ only, an exact {\em
ansatz}
\begin{equation}\label{ansatz}
\theta = \theta(r), \ \phi=Q\chi + \omega t\;,
\end{equation}
can be used, leading to the ordinary differential equation for the
function $\theta(r)$. This equation can be easily solved numerically
by shooting procedure,  using the value of $d \theta/d r $ at $r=0$
as a shooting parameter, see Ref.~\onlinecite{Kosevich90}. The shape
of the soliton essentially depends on the number $N$ of bound
magnons. In the case of the soliton with large $N\gg N_2\equiv 2\pi
S(l_0/a)^2$), the approximate ''domain wall'' solution works pretty
well. This solution has the shape of a circular domain wall of
radius $R$
\begin{equation}\label{eq:domain-wall}
\cos\theta_0(r) = \tanh\frac{r-R}{l_0}.
\end{equation}
Using this simple structure one can obtain the number of bound
magnons, which is proportional to the size of the soliton, $N\approx
2 \pi S (R/a)^2$ and energy,  $E=4\pi S^2 \sqrt{2JK_{\rm{eff}}}R/a$.
Note that such a solution is the same for topological and
non-topological solitons except for the behavior near the soliton
core. This means, that the characteristics of such solitons are
pretty similar. In the case of the small radius soliton ($R \ll
l_0$), the following asymptotically exact solution works well
\cite{Voronov83}
\begin{equation}\label{eq:smallR-structure}
\tan\frac{\theta_0(r)}{2} = \frac{R}{r_0}\ K_1
\left(\frac{r}{r_0}\right), \qquad r_0 = \frac{l_0}{\sqrt{1-\omega
/\omega_0}},
\end{equation}
where $K_1(x)$ is the McDonald function, and $\omega_0$ is a gap
frequency for linear magnons. It provides correct behavior at
$r<R\ll l_0$, where it converts to the Belavin--Polyakov solution
\eqref{eq:BP-soliton}. For large distances ($r\gg R$), this
expression gives an exponential decay (instead of power decay for a
Belavin-Polyakov soliton) with characteristic scale $r_0$. For
solution (\ref{eq:smallR-structure}) $\omega \to \omega_0$ as $N\to
0$ so that small radius solitons in anisotropic magnets have two
different scales, the core size  $R \ll l_0$ and the scale of the
exponential ``tail'' $r_0=l_0/ \sqrt{1-\omega /\omega_0} \gg l_0$
\cite{Voronov83,Ivanov86}.

For even minimal accounting for the discreteness on the basis of a
generalized model with fourth spatial derivatives, the problem
becomes much more complicated. The complexity of the problem is not
only due to the fact that for the energy \eqref{en4ok} it is
necessary to solve the fourth order differential equation, and use a
much more complicated three-parameter shooting method. But the basic
complication here is the fact that in general fourth-derivative
terms contain anisotropic contributions like $(\partial ^2 \theta/
\partial x^2)^2 + (\partial ^2 \theta/ \partial y^2)^2$, which cannot be
reduced to the powers of radially-symmetric Laplace operator (these
terms are omitted in Eq. \eqref{en4ok} for simplicity). In this
case, the radially-symmetric ansatz with $\theta = \theta(r), \,
\phi=\phi(\chi )$ is not valid, and we have to solve a set of
partial nonlinear differential equation for the functions $\theta =
\theta(r,\chi), \ \phi=\phi(r,\chi)$. To the best of our knowledge,
an exact method for construction of soliton (separatrix) solutions
for such type of equations do not exist so that some other
approximate methods should be used for this problem.

\subsection{Variational approach for general continuum model}
  One of the approaches, which we use for the approximate analysis of
the solitons in the model \eqref{en4ok}, is the direct variational
method.  For the minimization of the energy $E=W_2+W_4$ we use a
trial function,
\begin{equation}\label{macd}
\tan \frac{\theta }{2}=\Lambda RK_1\left( \Lambda r\right).
\end{equation}
Here, we consider  the case $Q=1$, and the case of higher
topological invariants is qualitatively similar. Note, that trial
function (\ref{macd}) is based on the asymptotically exact soliton
solution (\ref{eq:smallR-structure}). The trial function
\eqref{macd} gives correct asymptotics both for $x\to 0$
(corresponding to BP soliton) and for $x\to\infty$ (exponential
decay with characteristic scale $1/\Lambda $), see above. Our
analysis shows that the same results can be obtained using a
simplified trial function, which also captures the asymptotic
behavior of the soliton. This function has the form
\begin{equation}\label{simpl}
\tan \frac{\theta}{2}=\frac{R}{r}\exp(-\Lambda r).
\end{equation}
    In the spirit of  the minimization method above discussed we
consider the parameter $\Lambda $ as variational, keeping the
parameter $R$ constant as it is related to $N$, $N\propto R^2$, see,
e.g.. \cite{Kosevich90} In other words, we minimize the energy
(\ref{en4ok}) with the trial function (\ref{macd}) or (\ref{simpl})
over $\Lambda $ for constant $R$. This approach has the advantage
that simultaneously with equation solving, it permits investigation
of the stability of obtained solution on the base of simple and
obvious criterion. Namely, a soliton is stable if it corresponds to
the conditional minimum of the energy at fixed $N$, and it is
unstable otherwise.

\subsection{Numerical analysis  of the lattice model} \label{s:numerics}

Since the continuous description fails for the case of a high
anisotropy, one needs to elaborate the discrete energy
(\ref{eq:H-discrete}) on a lattice. As we are not able to solve the
problem analytically, we discuss the numerical approach to study the
soliton-like spin configurations on a discrete square lattice.
Obviously, the direct molecular dynamics simulations on a lattice is
a powerful tool for the soliton investigation for any values of
anisotropy. Direct spin dynamics simulations of 2D solitons have
been recently performed in
Refs.~\onlinecite{Kamppeter01,Sheka+ivanov+EuJP06}. After the
relaxation scheme, where an initial trial solution was fitted to the
lattice, the spin--dynamical simulations of the discrete
Landau-Lifshitz equations were performed. This approach, however, is
quite computer intensive, requiring powerful computers. To minimize
the calculation time, the parallel algorithms have been used
\cite{Sheka+ivanov+EuJP06}.

For this reason, we obtained  desired spin configuration by direct
minimization of the energy $E$, keeping $N$ constant. For our
modeling we choose the square lattices with "circular" boundary
conditions. We fix the values of spins on the boundary to the ground
state ($\theta =0$), those values have been kept intact during
minimization. The method of minimization is the simplex type method
with non-linear constraints. This method is based on the steepest
descent routine applied to the functions of a large number of
variables. The above method is able to find the conditional minimum
of a given function with several (usually small number) constraints,
consisting of relations between variables. In our problem, such
variables are the directions of each spin, parameterized by the
angular variables $\theta _n$ and $\varphi_n$, and the conditional
minimum of the energy, $E$ has been obtained for a fixed value of
$z-$ projection of the total spin. As this additional constraint
slows down the calculations substantially, we use other method,
valid for initial configurations where $N$ differs from the
necessary quantity by one spin. The idea of the method is as
follows: the angle $\theta$ of an arbitrarily chosen "damper" spin
was excluded from the minimization procedure, and its value has been
kept constant, to achieve the necessary $N$ value throughout whole
minimization with respect to all other variables.

The method in fact is dealing formally with a static problem, but it
gives the possibility to find the precessional frequency directly.
To find $\omega$, we used the discrete Landau-Lifshitz equation
\eqref{eq:LL-discrete} rewritten for the angular variable $\phi_n$,
in the way used in equation \eqref{eq:LL-continuum}. Using relation
$\partial \phi_n/\partial t = \omega$, we obtain
\begin{equation} \label{eq:omega-discr}
\omega=\frac{1}{\sin\theta_n}\cdot\frac{1}{\hbar S} \frac{\partial
\mathcal{H}}{\partial \theta_n} \ ,
\end{equation}
which does not depend on index $n$ throughout all the system. Here
$\mathcal{H}$ is the discrete Hamiltonian \eqref{eq:H-discrete}.

To find the above local minimum, we start from  the BP initial
configuration. The size of lattice clusters varied from $20\times
20$ (for large anisotropies, where discreteness of a lattice is
revealed most vividly) to $32\times32$ for small anisotropy when
system is well described by a continuous model. The criterion of the
presence of a truly local soliton configuration was its independence
of the system size. Another important criterion was the constancy of
the frequency $\omega $, calculated from the equation
\eqref{eq:omega-discr} throughout the soliton configuration.
Sometimes we found the minimum over the variables $\theta _n$ only,
considering $\phi _n$ to obey Eq. (\ref{ansatz}) and choosing the
reference frame origin in the symmetric points between the lattice
sites. Our analysis has shown that for moderate anisotropies $K_{\rm
{eff}}\lesssim 0.5J$ the above partial minimization gives the same
energy and frequency as well as the instability point position, as
complete minimization over $\theta _n$ and $\phi _n$. The use of
partial minimization over $\theta _n$ permits not only to accelerate
the numerical calculations, but turns out to be useful for
construction of "quasitopological" \ textures for extremely high
anisotropies, see below, Section \ref{s:extreme}.
\begin{figure}
\vspace*{-5mm} \hspace*{-5mm} \centering{\
\includegraphics[width=1.1\columnwidth]{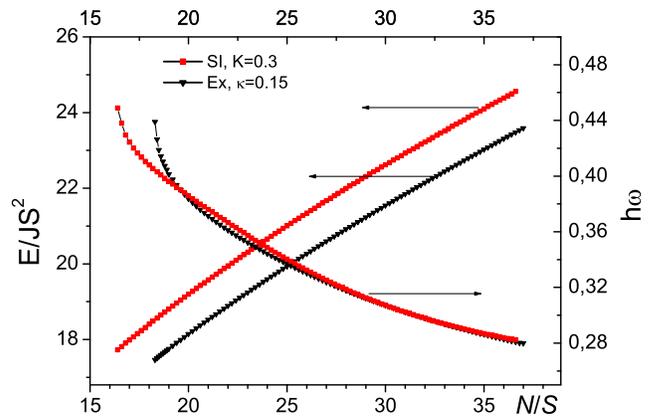}}
\caption{Dependence $E(N)$ and $\omega(N)$ for a soliton with
$K_{\rm{eff}}=0.3$, obtained from numerical simulations.}
\label{fig:v6}
\end{figure}
   In Fig.\ref{fig:v6}, the dependence of a soliton energy and
precession frequency, is shown as a function of the bound magnon
number for $K_{\rm{eff}}=0.3$ for single-ion and exchange
anisotropies. The frequency has been calculated directly from the
equation \eqref{eq:omega-discr} and also found by differentiation of
the energy with respect to $N$, see Eq.~\eqref{dEdN}. Here, far from
the instability point $N_{{\rm{cr}}}$, the behavior of $\omega (N)$
is almost similar, while near this point it is different. For both
types of anisotropy (single-ion and exchange) the "cusps" in the
dependencies $\omega (N)$ as $N\to N_{{\rm{cr}}}$ are well seen.

For $N<N_{{\rm{cr}}}$ the minimal configuration cannot be found. In
fact, any attempt to find the topological soliton with
$N<N_{{\rm{cr}}}$ leads to appearance of non-topological solitons
with $\partial\phi /\partial \chi \simeq 0$. This occurs even for
the above \ "partial" \ (i.e. over $\theta _n $ only) minimization.
In this case, some spins rotate by non-small angles to organize the
structure with zero $Q$. This demonstrates the instability of
topological solitons for $N<N_{{\rm{cr}}}$ quite vividly. These
features, as well as the values of $N_{{\rm{cr}}}$ or
$E_{{\rm{cr}}}$, can be described also on the basis of a variational
approach with simple trial functions, see the next section.

\section{Solitons for moderate values of anisotropy.}
To perform specific calculations for the generalized continuous
model \eqref{enw}, it is convenient to introduce the following
dimensionless variables
\begin{equation}\label{dimvar}
x=\Lambda r, \  \lambda=a\Lambda,\ z= \Lambda R,
\end{equation}
as before, $a$ is a lattice constant. Further we may express both
trial functions (\ref{macd}) and (\ref{simpl}) in the following
universal form
\begin{equation}
\tan \frac{\theta }{2} =zf(x),\ f_1(x)=K_1(x), \
f_2(x)=\frac{\exp(-x)}{x}.\label{trf}
\end{equation}
Then using the equations (\ref{eq:N}) and (\ref{en4ok}), we can
calculate the number of bound magnons in the soliton
\begin{equation}\label{n1}
\frac{N}{S}=4\pi \frac{z^2}{\lambda ^2}\psi (z), \ \psi
(z)=\int_0^{\infty}xdx\frac{f^2(x)}{1+z^2f^2(x)}\;,
\end{equation}
and the soliton energy
\begin{equation}\label{dimen}
  \frac{E}{2\pi JS^2}= \frac{K_{\rm{eff}}}{\lambda
^2}\gamma _0(z)+\gamma _2(z)-\frac{1}{24}\lambda ^2\gamma _4(z).
\end{equation}

\begin{figure}
\vspace*{-5mm} \hspace*{-5mm} \centering{\
\includegraphics[width=1.1\columnwidth]{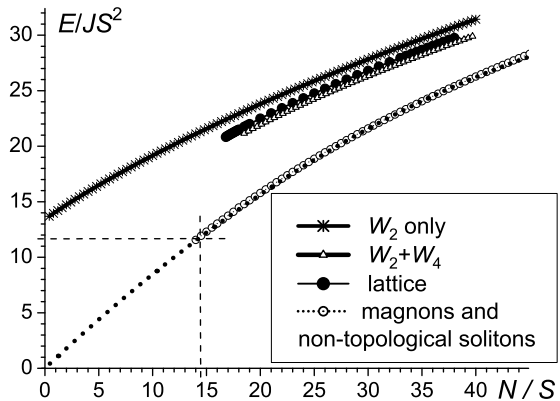}}
\caption{Dependence $E_{\rm{min}}(N)$ for topological and
non-topological solitons for the case of exchange anisotropy only,
$\kappa=0.2$. Dashed lines, parallel to the axes, correspond to
characteristic number of bound magnons $N=N_c$ for non-topological
soliton, see the text above. Curves, corresponding to contribution
of $W_2$ only are shown for comparison.} \label{fig:v1}
\end{figure}

Here we introduced the following notations
\begin{widetext}
\begin{eqnarray}
&&\gamma _0(z)=\int_{0}^{\infty }\sin ^{2}\theta xdx, \quad \gamma
_2(z)=\frac{1}{2}\int_0^{\infty }xdx\left[ \theta^{\prime 2}
\left( 1+\kappa \sin ^2\theta \right) +\frac{%
\sin ^2\theta }{x^2}\right] ,  \nonumber \\
&&\gamma _4(z)=\int_0^{\infty }xdx\Biggl\{\left( \Delta _x\theta
\right)^2\left( 1+\kappa \sin ^2\theta \right) +\theta^{\prime 4}
\left( 1+\kappa \cos ^2\theta \right) +\frac{\sin ^2\theta }{x^2}%
\left( \frac{1}{x^{2}}+2\theta ^{\prime 2}\right) +  \nonumber \\
&&+\Delta _{x}\theta \sin 2\theta \left( \kappa \theta ^{\prime 2}-\frac{%
1}{x^2}\right) \Biggr\},\ \ \theta '=\frac{d\theta}{dx}, \  \Delta
_x\theta = \frac{d ^2 \theta}{dx^2} +\frac 1x\frac{d \theta}{dx}\;.
\label{asm7}
\end{eqnarray}
\end{widetext}

Thus, we express the energy and the number of magnons via two
parameters, $\lambda$ and $z$. It turns out, that initial
dimensional variables $\Lambda $ and $R$ enter the problem only in
the form of their product $z$. The dependence of $N$ and $E$ on $z$
enters the problem via a few complicated functions $\psi, \gamma_0 ,
\gamma_2 $ and $\gamma_4$, which can be written only implicitly in
the form of integrals. However, in terms of these functions, the
dependence on $\lambda $ (\ref{dimen}) turns out to be quite simple.
This permits reformulation of the initial variational problem in
terms of variables $z$ and $N$ only. Namely, we express
\begin{equation}\label{tyi}
  \lambda ^2=4\pi \frac{z^2}{(N/S)}\psi (z)
\end{equation}
and substitute this expression in the dimensionless energy
(\ref{dimen}). This gives us the expression for the energy of a
soliton with given $N$, as a function of variational parameter $z$.
Then we can find a minimum of $E$ with respect to $z$, keeping $N$
constant. The result of such minimization in the form of the
dependence $E_{\rm {min}}(N)$ is shown in Fig.\ref{fig:v1} for a
magnet with purely exchange anisotropy (i.e. for that with $K=0$).
We show this dependence for $\kappa=0.2$. It is seen that there is a
good correspondence between the dependencies $E(N)$ found by
variational and numerical (full symbol) minimizations in the region
of parameters where the soliton is stable. This justifies the
applicability of the variational approach with the trial functions
of the form (\ref{trf}) to the problem under consideration. The
approach based on simpler trial functions $f_2$ reproduce well the
particular feature found for numerical analysis of the discrete
model; namely, the value of the threshold number $N_{\rm {cr}}$.

\begin{figure}
\vspace*{-5mm} \hspace*{-5mm} \centering{\
\includegraphics[width=80mm]{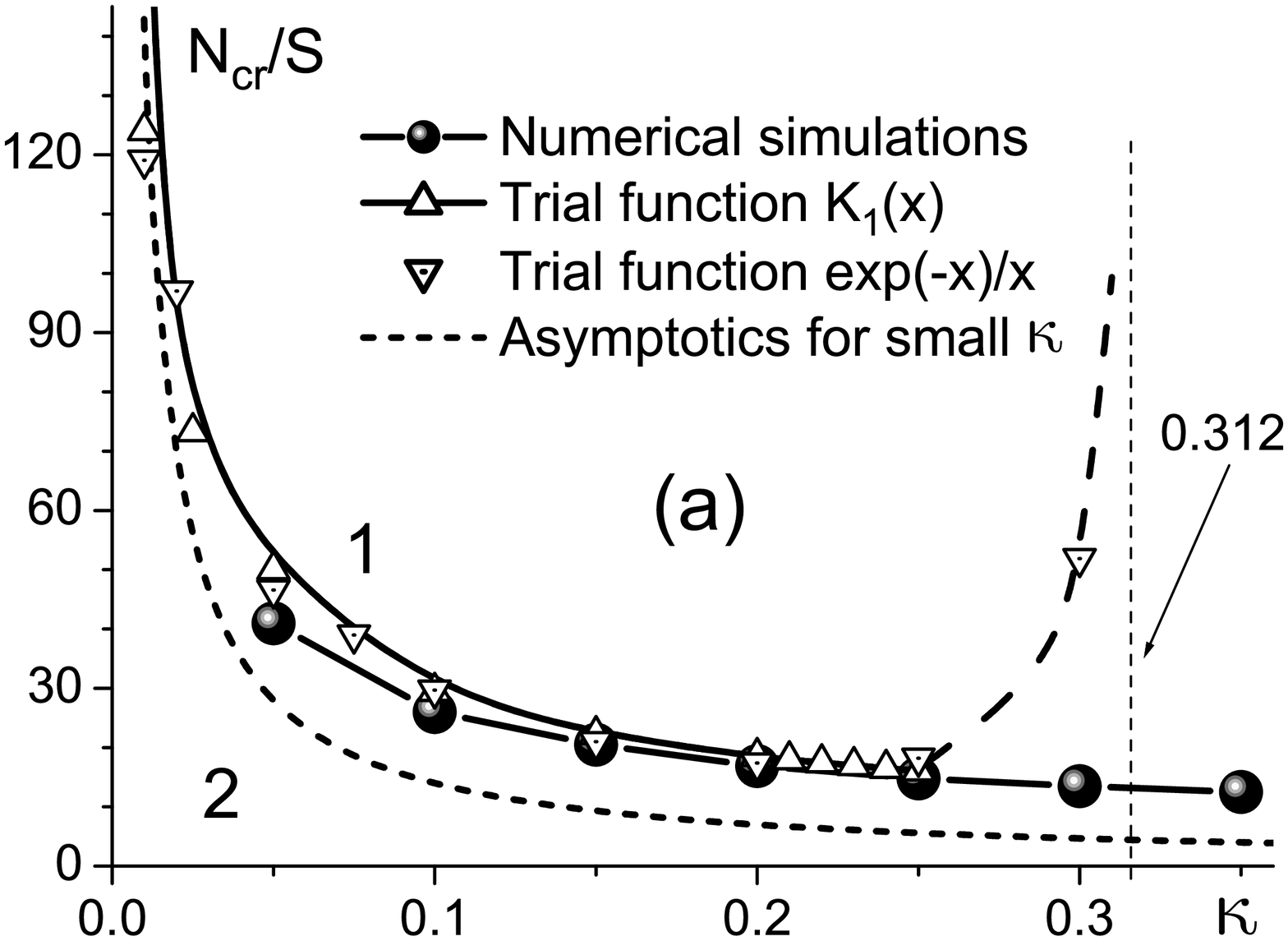}
\includegraphics[width=80mm]{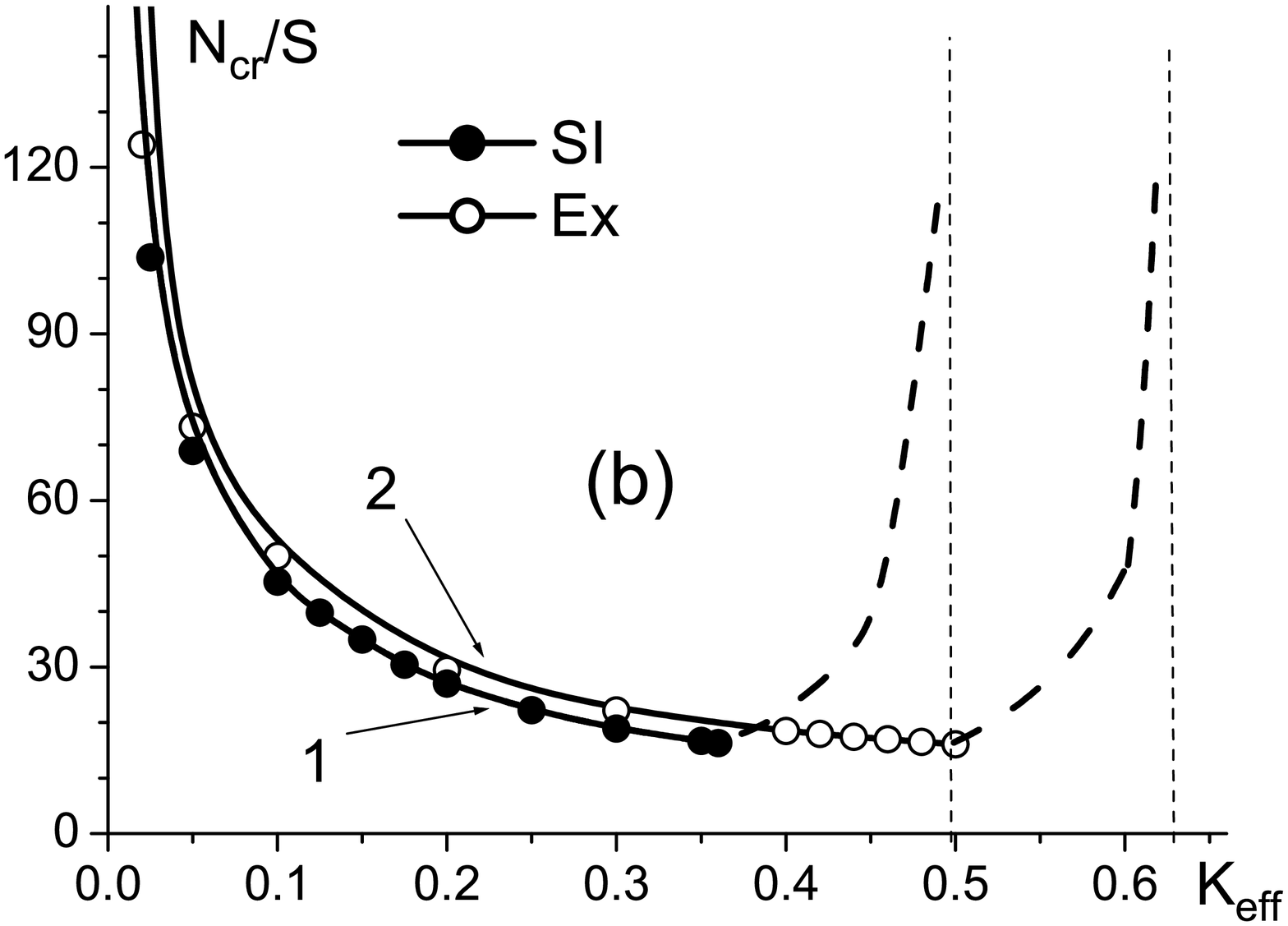}}
\caption{Soliton phase diagram. (a) - exchange anisotropy, solitons
exist in the region 1 and do not exist in the region 2. For small
$\kappa $ the value of $N_{\rm {cr}} \approx 1.4S/\kappa$. (b) -
comparison between exchange (curve 1) and single-ion (curve 2)
anisotropies. Limiting values of $N_{\rm {cr}}$ for both types of
anisotropies are shown by vertical dashed lines.} \label{fig:v2}
\end{figure}
Also, for comparison, we show in Fig. \ref{fig:v1} the result of
variational minimization of $W_2$ (i.e. energy, incorporating only
squares of magnetization gradients). It is seen, that in this case
there is no $N_{\rm {cr}}$, which coincides well with the previous
investigations of solitons in the continuum
models.\cite{Voronov83,Ivanov86,Kosevich90} This means, that mapping
of the initial discrete model on the simplest continuum model with
only squares of magnetization gradients can be wrong for some values
an $N$, and to get the correct description of solitons in a 2D
magnet one must take into account at least fourth powers of
gradients. On the other hand, for large enough $N$ and even moderate
value of anisotropy, the role of this higher derivative terms is
less important; it is in agreement with the recent numerical
simulations of soliton dynamics for easy-axial discrete models of
ferromagnets \cite{Sheka+ivanov+EuJP06}.

It is seen from Fig.\ref{fig:v1}, that the topological solitons in
such a model are not very "robust"\ - there is a quite large
parameter region where those solitons are unstable or do not exist
at all. That is why for comparison we show the dependence $E(N)$ for
non-topological solitons (open symbols on the figure) which have
been studied in detail analytically in
Ref.~\onlinecite{IvZaspYastr}. The main feature of non-topological
solitons is that while $\omega \to \omega _0$, the amplitude of such
solitons diminishes (so that at $\omega \geq \omega _0$ such soliton
decays to a number of noninteracting magnons) but both its energy
$E\equiv E_{\rm NT}\approx 11.7E_{\rm BP }$ and bound magnons number
$N\equiv N_c=E_\mathrm{NT}/\hbar \omega _0$ are left intact.
\cite{IvZaspYastr} This behavior is opposite to that of a
topological soliton, where as $\omega \to \omega _0$ the amplitude
still has its maximal value with decreasing soliton radius so that
soliton becomes "more localized". This tendency is seen in
Fig.\ref{fig:v1}, where a non-topological soliton exists up to
$N\equiv N_{NT}$ (corresponding to $\omega = \omega _0$, shown as
vertical dotted line on the figure) and then decays smoothly into a
noninteracting magnon cloud with the energy above  on the
corresponding curve (dotted line in the figure). This is because as
$\omega \to \omega _0$ the non-topological soliton has the same
energy as corresponding magnon cloud. But the soliton is "coherent"
(in a sense that it is a bound state of many magnons), while magnon
clouds are not coherent. Actually, the above remark reflects the
important point in a physics of solitons under consideration, namely
the difference in behavior of topological and non-topological
solitons.

As was mentioned above, there is a critical value of bound magnons
number $N_{\rm {cr}}$ such that the topological soliton exists with
$N\geq N_{\rm {cr}}$ only.  This threshold depends on $\kappa $.
This dependence, which we will call a phase diagram, is depicted in
Fig.\ref{fig:v2}a. The solitons exist in the region 1 above
corresponding curves and do not exist in the region 2 below them.
For comparison, on the same figure, we plot the points
$N_{\rm{cr}}$, corresponding to different trial functions. Very good
coincidence between these points shows that both trial functions are
well-suited for variational treatment of the solitons.

It is seen that the threshold number of bound magnons calculated
using the simple variational approach grows infinitely both for
$\kappa \to 0$ and for $\kappa \to \kappa _{\rm lim}\approx 0.312$.
The comparison with the numerical data shows that the divergence at
the large values of anisotropy is simply an artifact of continuous
description, and the corresponding ``numerical''  curve for discrete
model decreases monotonously with growing of anisotropy. For high
anisotropies, the soliton structure becomes strongly anisotropic,
and the description based on radially-symmetric trial functions
fails. We will discuss this in the next section.

On the other hand, the divergence of $N_{\rm{cr}}$ at small $\kappa
$ has a clear physical meaning and coincides  well with the
numerical data. The divergence of  $N_{\rm{cr}}$ at $\kappa \to 0$
is related to the fact that in BP soliton, which is the exact
soliton solution for purely isotropic ferromagnet, the integral
describing the value of bound magnons $N$ diverges logarithmically
as $r\to \infty$ due to slow decay of the function $\theta(r)
\propto 1/r$. Using the variational approach with the asymptotics of
the functions \eqref{asm7} it is possible to derive the asymptotic
formula for $N_{\rm{cr}}$ in this region, which reads
\begin{equation}\label{kas}
  N_{{\rm cr}}(\kappa)\approx \frac{1.4 S}{\kappa}.
\end{equation}
In contrast, the critical value of energy
$E_{\rm{cr}}=E(N_{\rm{cr}})$ is finite at the instability point,
\begin{equation}\label{Ecrit}
  E_{{\rm cr}}(\kappa)\approx 4\pi JS^2 (1 + 2.64\sqrt{\kappa / J }).
\end{equation}
This limiting energy contains non-analytical dependence on the
anisotropy constant $\kappa$. It appears due to \emph{simultaneous}
accounting of the anisotropy and fourth derivative terms in our
generalized continuum model. For typical values $\kappa / J =
0.1\div 0.2$, this energy is well above the Belavin-Polyakov
limiting energy, see Figs. \ref{fig:v6}, \ref{fig:v1}. The
correction to $E_{BP}$ becomes smaller then 1\% at extremely low
anisotropies like $\kappa \leq 10^{-5}J$ only. The situation here is
very similar to the analysis of heavy (less energetically favorable)
vortex decay in the cone state of an easy plane ferromagnet,
\cite{IvWysin02} which situation occurs in a magnetic field,
perpendicular to the easy plane.\cite{IvSheka,IvWysin02} There, in
the simple continuum model, the heavy vortices are stable in the
entire region of cone state existence ($0<H<H_a$, $H_a$ is an
anisotropy field)\cite{IvSheka}, but already for very small
anisotropies $\kappa \simeq 10^{-4}$, when $l_0\simeq 10a$, this
region has diminished substantially so that at $\kappa \simeq 0.1$
the heavy vortices have already become absent.\cite{IvWysin02}

In Fig.\ref{fig:v2}b, the phase diagrams for exchange (curve 1) and
single-ion anisotropies (curve 2) are shown. It is seen, that while
at small  $\kappa $ and $K$ the corresponding curves lie close to
each other, for larger anisotropies there is a drastic difference.
While the (unphysical) limiting value of $N_{\rm {cr}}$ for high
exchange anisotropy $\kappa$ equals 0.312, corresponding to $K_{\rm
{eff}}=0.624$, the same value for single-ion anisotropy constant $K$
corresponds to $K_{\rm {eff}} \approx$ 0.5.  This means that the
continuous description for exchange anisotropy is valid for larger
values of anisotropy constants, even in the region $K_{\rm {eff}}
\simeq 0.5$, where it already fails for single-ion anisotropy. This
is a consequence of  the fact, that the single-ion anisotropy
constant $K$ enters the problem only in the spatially homogeneous
term (via the combination $K_{\rm{eff}}$, see above), while the
constant $\kappa $ enters all terms of expansion $W_i$.

\section{Solitons in the discrete model with high anisotropy.
Magic numbers of bound magnons} \label{s:extreme}

For large anisotropies, the discreteness plays a decisive role so
that exact analytical treatment of the problem is impossible.
However, an account for the following two facts permits to
accomplish a comprehensive approximate study. First, as was
mentioned in the early publications,
\cite{IvKos76,KovKosMasl,IvDzyal} for the large number of bound
magnons $N\geq S(l_0/a)^2 $ , any solitons in two- and three
dimensional magnets, topological and non-topological, can be
presented as a finite region of flipped spins, separated by
180$^\circ$ domain wall (DW) from the rest of a magnet.

Second, for such solitons (magnon droplets) the difference between
topological and non-topological textures becomes vague. As the
anisotropy grows, $l_0 \sim a$ and the characteristic magnon number
becomes comparable with $2\pi S$. At the same time, the DW becomes
thinner so that at $\kappa ,K\sim J$ its width becomes comparable
with lattice constant $a$. It is clear that already at high
anisotropy the structure of bound states (solitons) with $N/S \geq
10 $ is above the described flipped area bordered by the DW (we
recollect here that we consider a soliton as a bound state of many
magnons) so that the soliton properties will be completely
determined by those of the DW. It is also clear that for this case
the difference between topological and non-topological solitons will
be negligible. We will see, that in this case the structure of bound
state is strongly dependent on the character of anisotropy. That is
why for a description of such bound states it is useful to study
first the structure and properties of the DW's in highly anisotropic
2D magnets with different types of anisotropy. First of all, of
importance is a notion of the DW pinning, i.e. the dependence of its
energy both on the DW center position in a lattice (positional
pinning) and on its orientation with respect to the lattice vectors
(orientational pinning).

The positional pinning of a DW can be discussed on the basis of a
simple 1D model of a spin chain. This case is obviously applicable
to the 2D square lattice with nearest-neighbor interaction for the
DW orientation along lattice diagonals (the directions of (1,1)
type). For a chain, it is natural to associate the DW coordinate $X$
with the total spin projection on the easy axis ${\cal S}_{tot}^{z \
(0)}$ and to define it as follows \cite{IvMik}. Let us choose some
lattice site and define the DW located on this site to have the
coordinate $X=0$. Let us then determine the coordinate of any DW via
its total spin $z$-projection ${\cal S}_{tot}^z$ from the expression
\begin{equation} \label{X-definition}
X = \frac{a}{2S} \sum_{n=-\infty}^{n=\infty} [S^z_n(X) -S_n^z(X=0)
],
\end{equation}
where $S_n^z(X=0)$ defines the spin distribution in a DW at a
reference point $X=0$.

For the solitons description, the different properties of DW placed
in different positions play a crucial role. First, consider a DW in
a middle between two arbitrary lattice sites, for which $X/a$ is
half-odd, $X = a(2n+1)/2$. Only domain walls of such type in the
highly anisotropic magnet can be purely collinear, i.e. it may have
$S_z=\pm S$, respectively, on the left and right sides of the DW
center. For the case of a spin chain with single-ion anisotropy this
is achieved for $K > 0.5$,\cite{Kovalev+} which was associated with
the destruction of non-collinear topological structure
\cite{Kamppeter01}.  According to the definition
(\ref{X-definition}), it is a vital necessity to have a non-integer
${\cal S}_{tot}^{z}$ and non-collinear component for the rest of the
DW positions. For example, for $X=(a/2)(2n)$ in the one of the sites
$S_z=0$, i.e. $\theta =\pi /2$.

The question regarding the character of the pinning potential $U(X)$
is also important. Gochev has shown, that there is no pinning, i.e.
$U=0$, in a spin chain with purely exchange anisotropy.\cite{Gochev}
On the other hand, for the anisotropy of a pure single-ion type, the
potential $U(X)$ has  minima in the points of type $a(2n+1)/2$,
which favors the appearance of collinear DW's\cite{IvMik}. For the
small perturbation of a problem with exchange anisotropy by
single-ion anisotropy with \emph{negative} sign $K<0$, it turns out
that $U(X)$ has the minimum at integer values of $X/a$, $X=0$, $\pm
a, \ ...$, i.e. the creation of a non-collinear structure becomes
favorable. The detailed investigation of the character of DW pinning
for general models with combined anisotropy shows, hoverer, that
this situation is mostly an exception and the intensity of such
pinning is never large. The pinning potential for the parameters
where it has minima at integer $X/a$  is much smaller (at least one
order of magnitude) than that for pure single-ion anisotropy.

Thus, for the two above mentioned cases one can expect substantially
different spin distributions in a soliton boundary. In particular,
DW pinning on  lattice sites can yield the conservation, at least
partial, of a soliton topological structure even with strong
discreteness effects. On the other hand, a DW pinned between lattice
cites becomes collinear and hence its topological structure can be
easily lost.
\begin{figure}
\vspace*{-5mm} \hspace*{-5mm} \centering{\
\includegraphics[width=80mm]{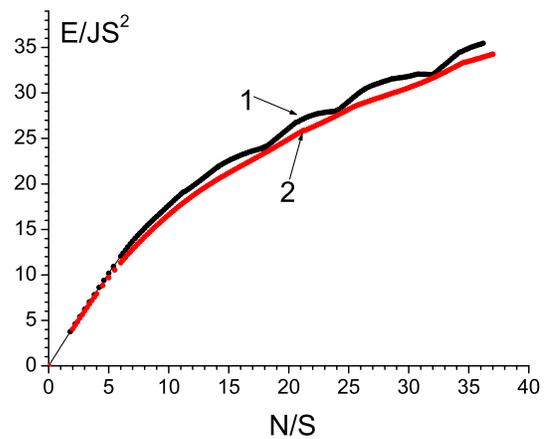}}
\caption{Dependence $E(N)$, obtained by numerical simulations for
$K_{\rm{eff}}=1$. Curve 1- single-ion, $K=1$, curve 2 - exchange,
$\kappa =0.5 $.} \label{fig:v3}
\end{figure}

It is clear, that for the question about the structure of a closed
soliton boundary, the important point is its angular pinning, i.e.
the dependence of the planar DW energy on its orientation in a
crystal. Gochev suggested, that the optimal DW direction is parallel
to the primitive vectors of lattice translation, (1,0) and
(0,1),\cite{Gochev} that differs from the conclusions of
Ref.~\onlinecite{Kamppeter01}. Our analysis has shown that almost
all numerical data about the properties of the solitons in highly
anisotropic FM's can be described under supposition that a DW tends
to be parallel to (0,1) direction. As we shall show, only $N/S$ is a
main parameter determining the soliton structure at high magnetic
anisotropy. It turns out that the variation of $N/S$ by around few
percent yields substantial variations of DW structure, which leads
to sharp dependence of energy $E$ (see Fig.\ref{fig:v3}) and
particularly the frequency $\omega $ on $N$, see Fig.~\ref{fig:v7}
below. This effect is different for single-ion and exchange
anisotropies, the most substantial manifestation is for the
single-ion case. These dependencies have a non-monotonic component.
Its analysis reveals certain specific numbers $N_{\rm {mag}}$,
which, analogously to nuclear physics, can be called the {\em magic
numbers}. To explain the origin of these magic numbers, we consider
the case when a DW tends to occupy a position between atomic planes
of (1,0) and/or (0,1) type so that both DW bend and non-collinear
structure formation are unfavorable. Then, the optimal (from the
point of view of DW energy) configuration, is that where the spins
with $S_z=-S$ occupy a rectangle $l_x\cdot l_y$, separated from the
rest of a magnet by a collinear DW. Clearly, the most favorable
configuration is that with spin square $l_x=l_y$, which yields
$N_{\rm {magic}}=2l^2S$, but configurations with $l_x\approx l_y$,
like those having $N/S = 2\cdot(l)(l+1)$, for example, $N/S =
2\cdot(3\cdot4)=24$ are also quite profitable, see
Fig.~\ref{fig:v3}. We will call such values of $N$
\emph{half-magic}. As we shall see, such a model describes the
solitons in magnets with single-ion anisotropy for $N/S=15\div50$
pretty well. If the energy of spatial and angular pinnings is not so
important, we may expect more or less cylindrical shape of a soliton
core and smooth dependencies $E(N)$ and $\omega (N)$. The numerical
analysis has demonstrated, that both above tendencies reveal
themselves in a FM with single-ion and exchange anisotropies
respectively, see Fig.~\ref{fig:v3} for $E(N)$ dependence.  Namely,
for SIA the dependence $E(N)$ has a non-monotonous component, while
for ExA this dependence seems to be more regular. The minima in the
non-monotonous dependence $E(N)$ for SIA occur for $N/S=18$ $=2\cdot
3^2$, $24=2(3\cdot 4)$ and $32=2\cdot 4^2$. Hence, the above magic
and half-magic numbers, related to the DW pinning between two atomic
planes, are clearly seen in the $E(N)$ dependence.
 \begin{figure}
   \subfigure[ ExA, 8]{\includegraphics*
  [bb = 180 320 520 660, width = 32mm]{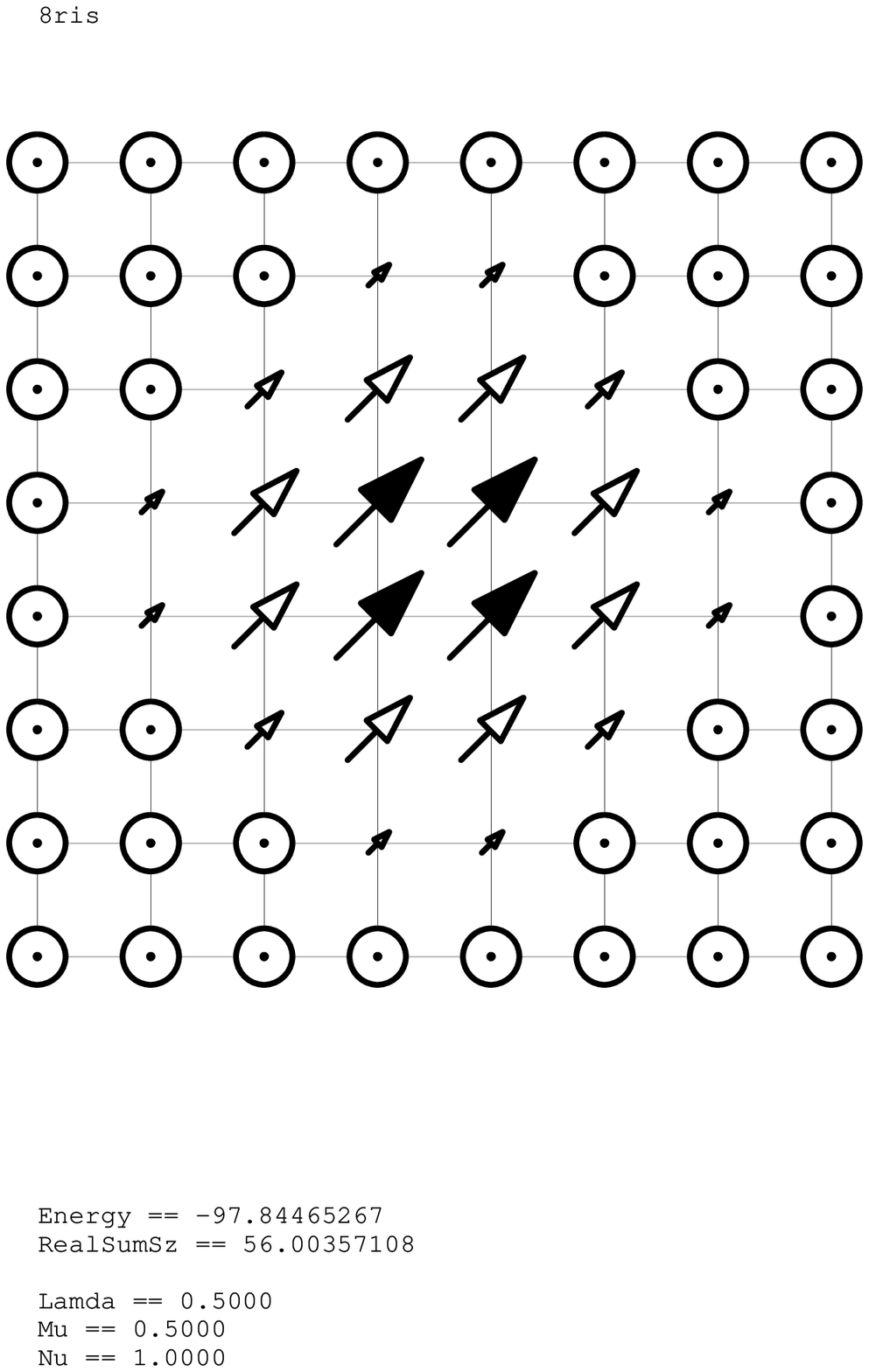}
    \label{ex1}}
    \subfigure[ SIA, 8]{\includegraphics*
   [bb = 180 320 520 660, width = 32mm]{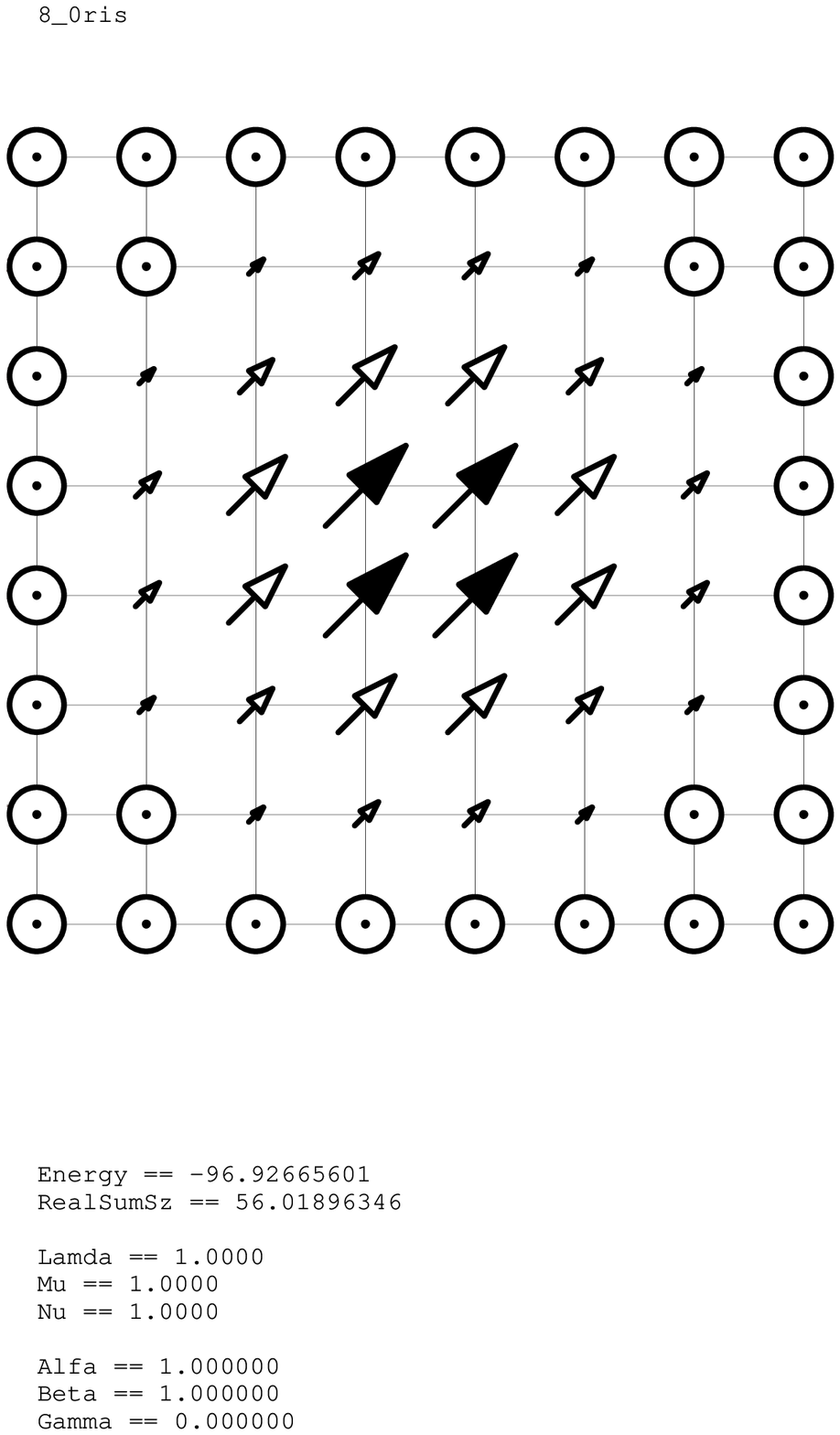}
    \label{si1}}
    \subfigure[ExA,  12.]{\includegraphics*
  [bb = 180 320 520 660, width = 32mm]{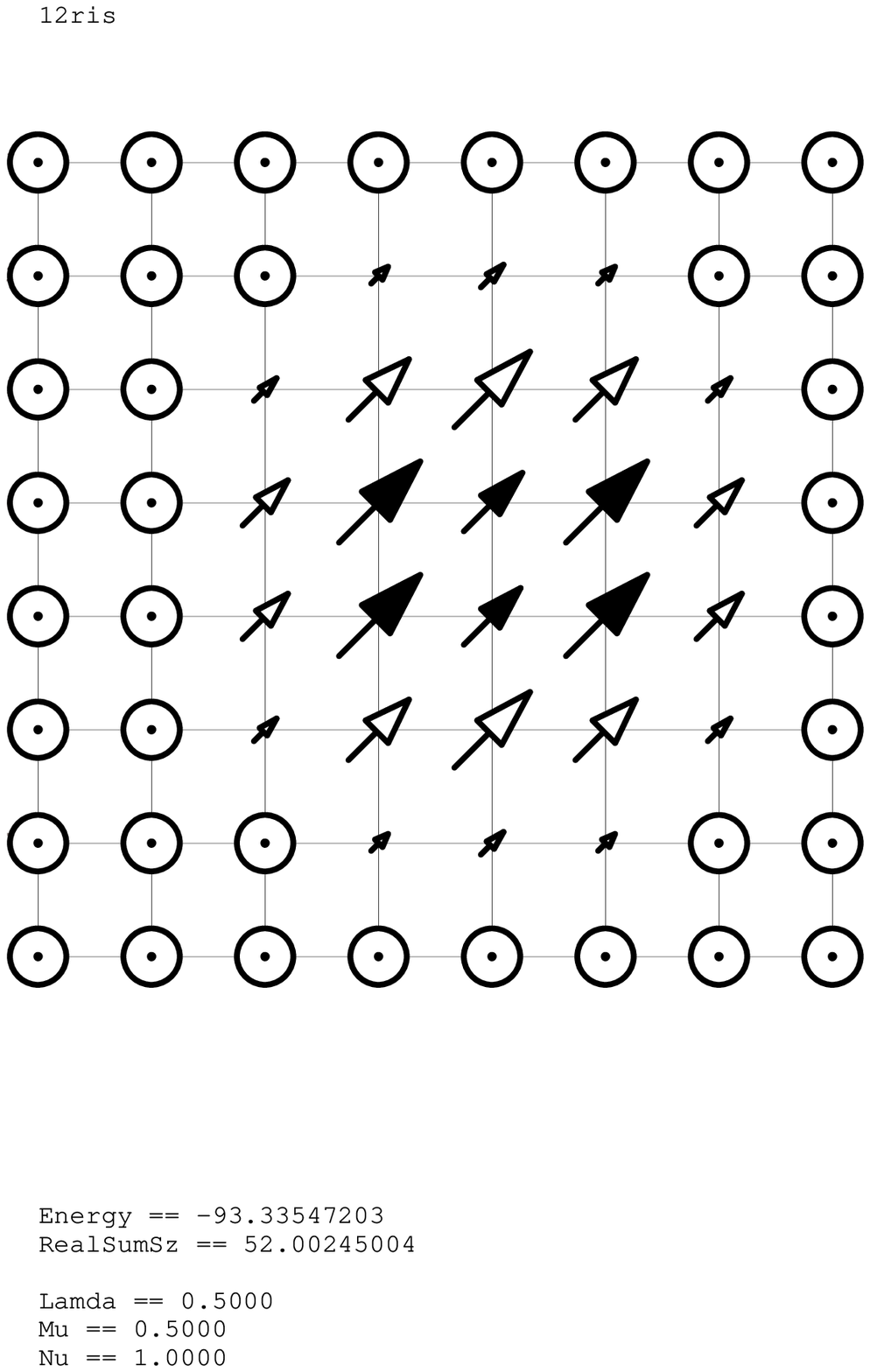}
    \label{ex2}}
    \subfigure[SIA, 12.]{\includegraphics*
   [bb = 180 320 520 660, width = 32mm]{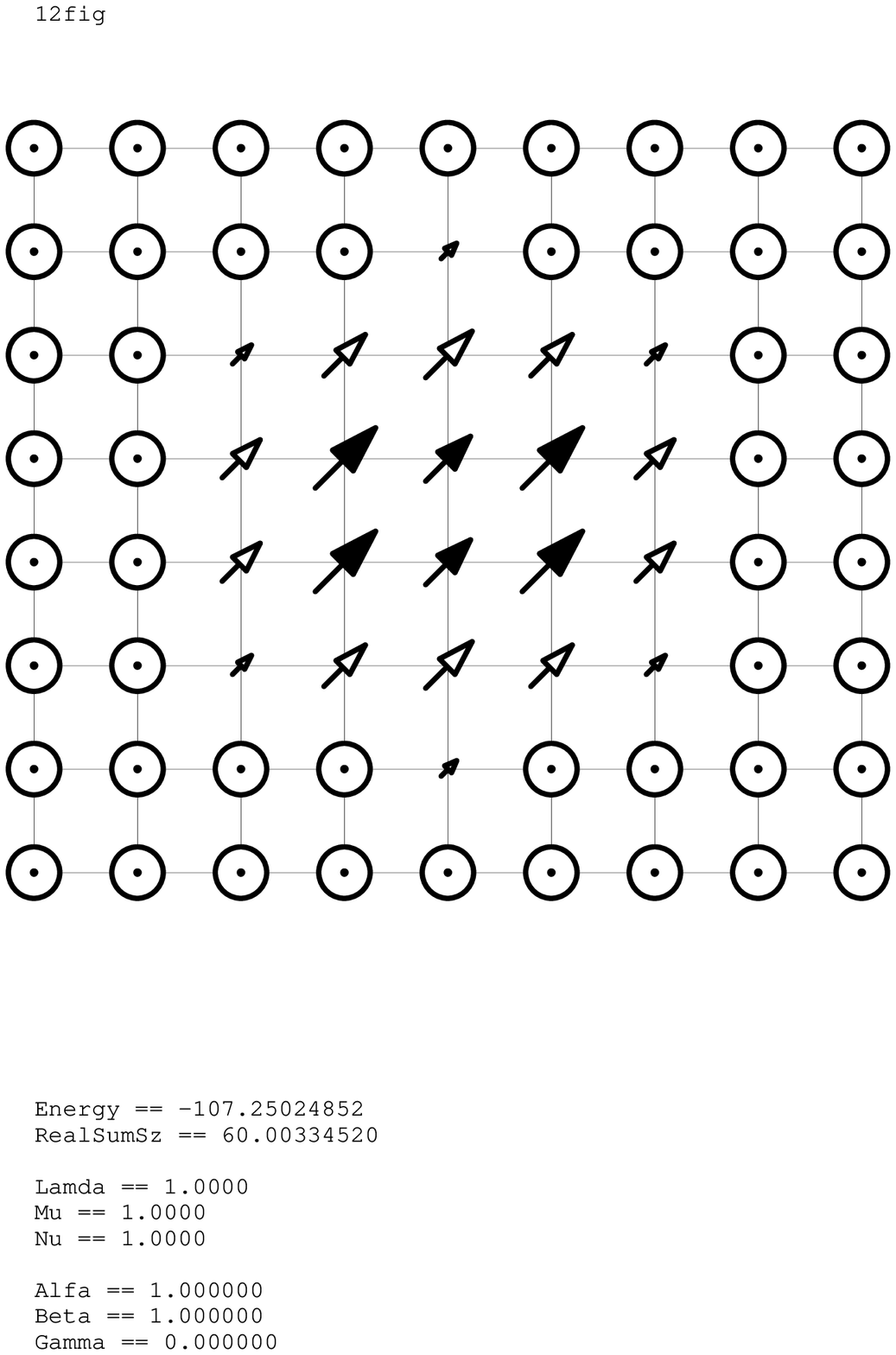}
    \label{si2}}
    \subfigure[ExA,  14.]{\includegraphics*
  [bb = 180 320 520 660, width = 32mm]{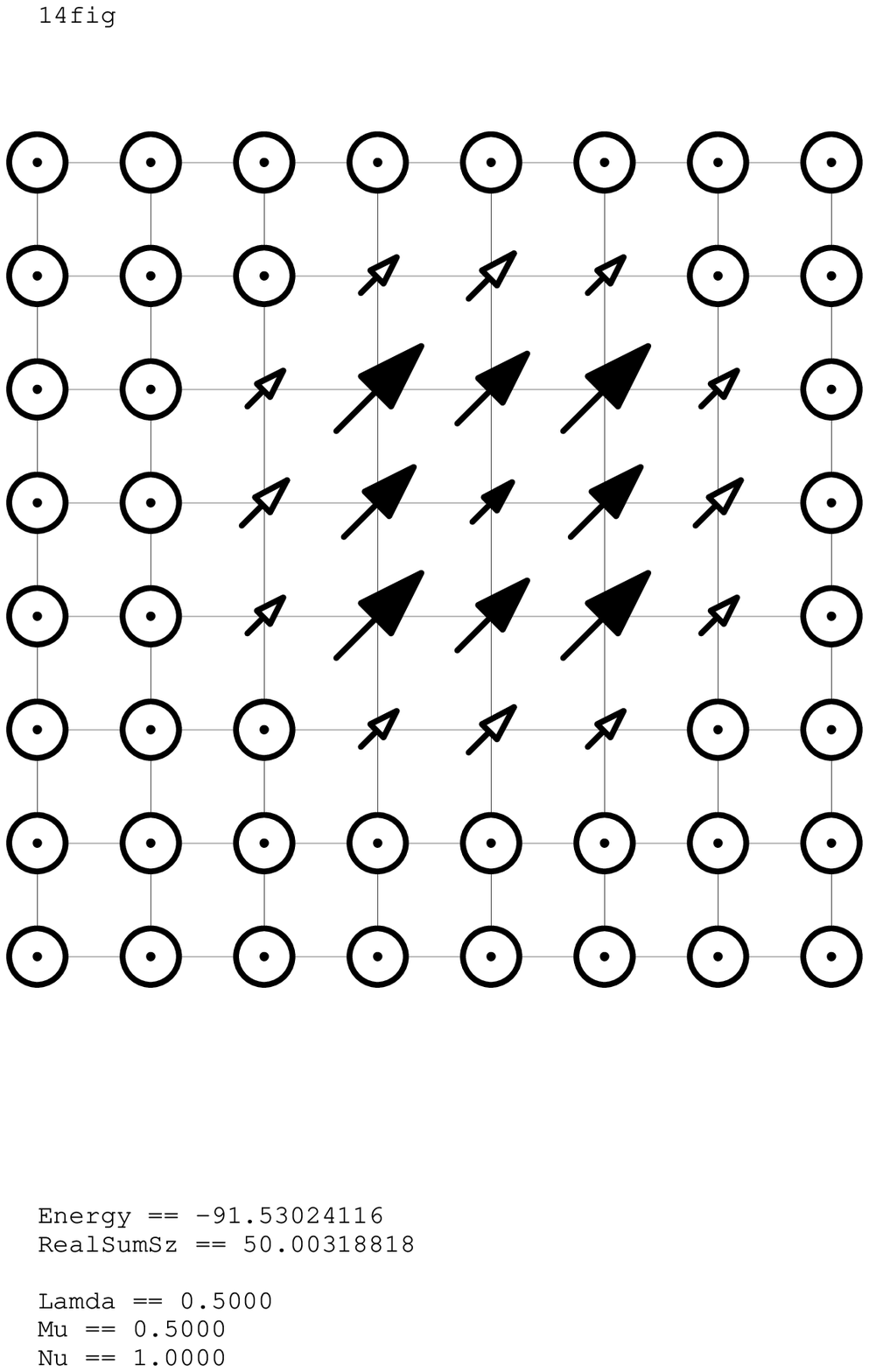}
    \label{ex4}}
    \subfigure[ SIA, 14]{\includegraphics*
   [bb = 180 320 520 660, width = 32mm]{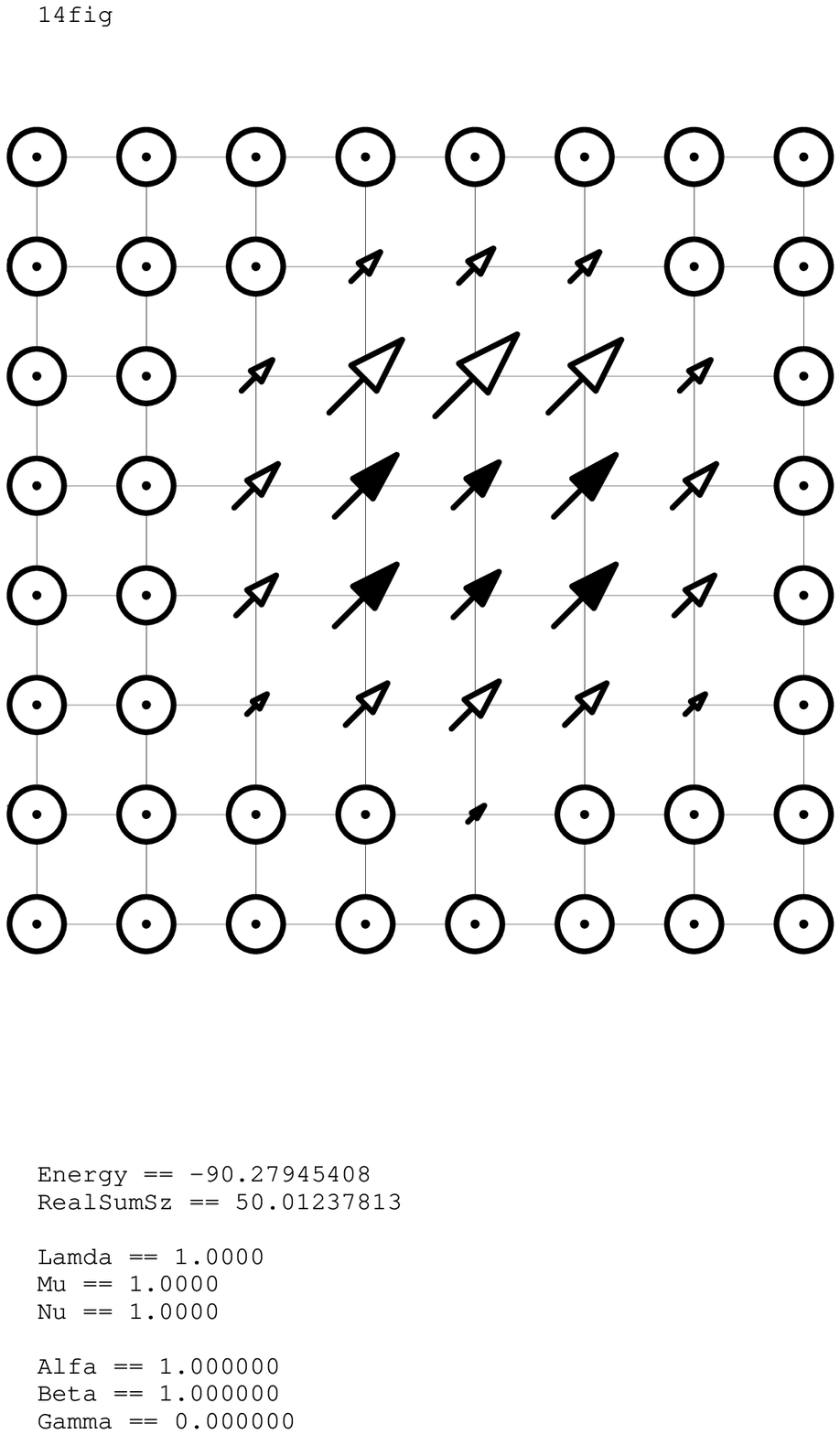}
    \label{si4}}
    \subfigure[ExA,  18.]{\includegraphics*
  [bb = 180 320 520 660, width = 32mm]{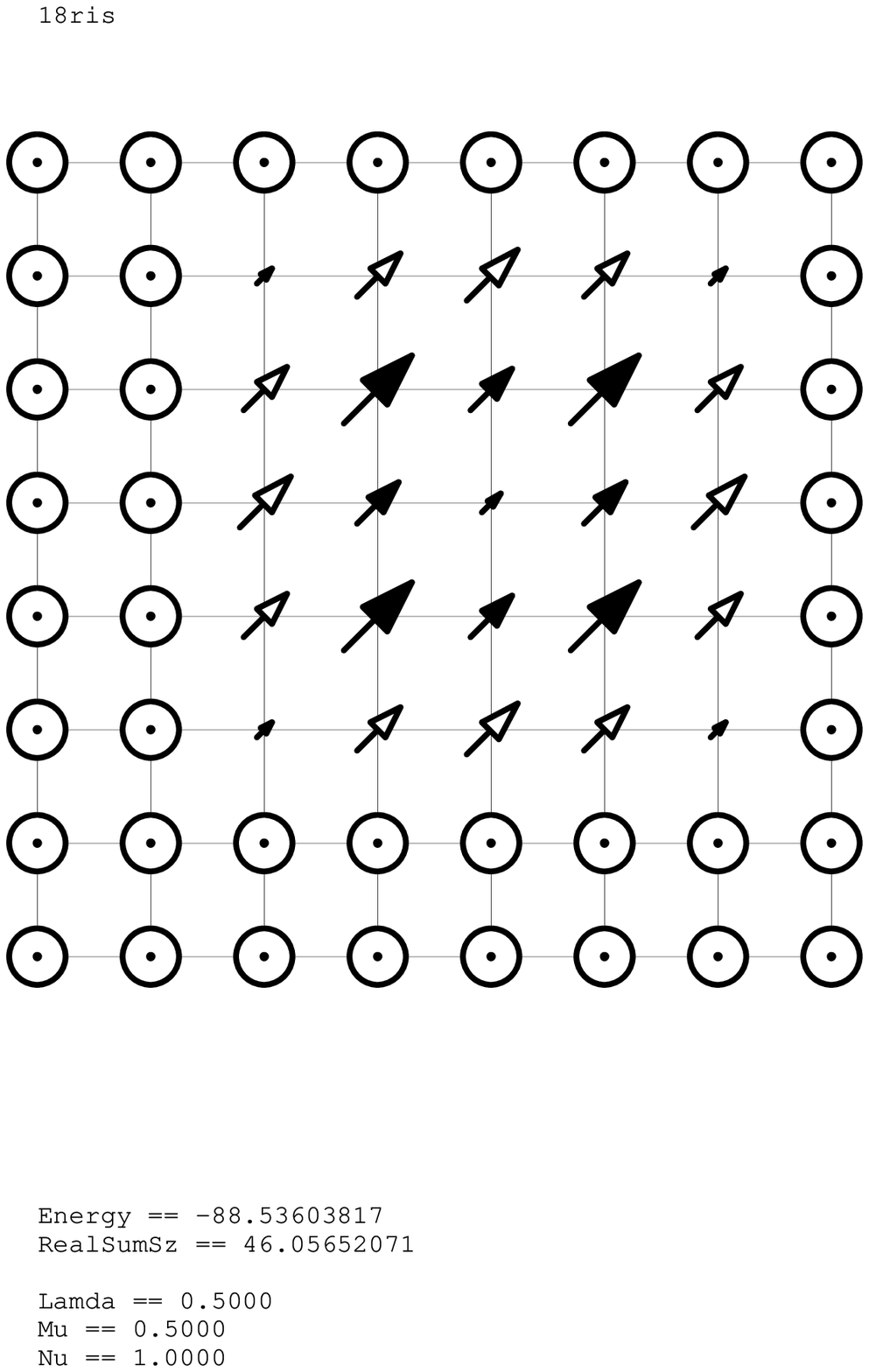}
    \label{ex3}}
    \subfigure[SIA, 18.]{\includegraphics*
   [bb = 180 320 520 660, width = 32mm]{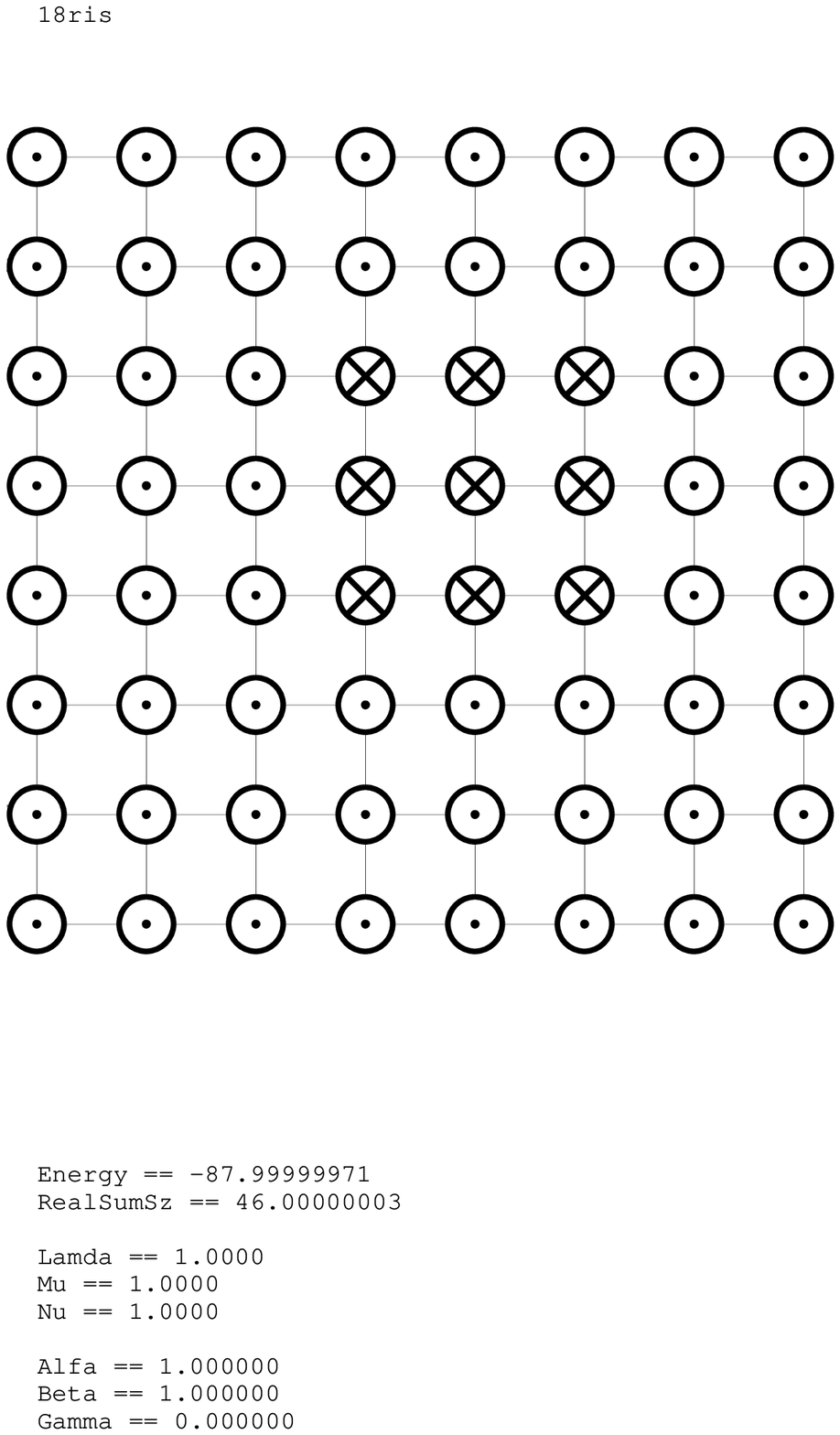}
    \label{si3}}
    \caption{The structure of the soliton textures for the magnets with exchange
   (left column) and single-ion anisotropy (right column) with the same value of
   effective anisotropy and different values of $n=N/S$ (shown
   as subfigure captions). The arrows present in-plane spin projections
   in 20 by 20 lattice. "Up" ($0 \leq \theta \leq 10^\circ$)\ and
   "down" ($170^\circ \leq \theta \leq 180^\circ$ )\ spins
   are presented by dotted and crossed circles, respectively.
   The in-plane projections of the spins with "up" \ and "down"
   \ $z-$ projections are depicted by arrows with open and full heads,
   respectively. \label{f:structure}}
\end{figure}

We now discuss non-topological spin configurations for different
numbers of bound magnons for ferromagnets with single-ion and
exchange anisotropy with large value of effective anisotropy
constant $K_{\rm{eff}}=J$, see Figs.\ref{f:structure} and
\ref{f:structure2}. Let us first consider the small values $n\equiv
N/S<20$. To economize the notations, hereafter we will use $n$
instead of $N$. For very small $n<10$ the soliton textures have the
same noncollinear structure for both types of anisotropies, SiA and
ExA, see Figs. \ref{ex1},\ref{si1}. The difference between spin
textures for SIA and ExA solitons becomes visible for $n\geq 10$. In
this case, the effects of lowering of a soliton symmetry with
respect to expected lattice symmetry of fourth order C$_4$, are
possible. For half-magic number $n=12$, corresponding to collinear
texture with flipped spins rectangle $2\times 3$, the symmetry C$_4$
is obviously absent both for SIA and for ExA, see Figs.
\ref{ex2},\ref{si2}. Also, there is no big difference between spin
textures in Figs. \ref{ex2},\ref{si2}. However, for $n$ far from
magic numbers, the difference between SIA and ExA is much more
pronounced. In particular, for the SIA case there is not even a
C$_2$ axis (Fig. \ref{si4}), while for ExA, the C$_4$ symmetry is
restored (Fig. \ref{ex4}). The explanation is pretty simple: the DW
pinning is weaker for ExA then that for SIA so that in former case
the symmetric closed DW is formed, while for SIA more favorable is
the formation of a piece of ``unfavorable'' DW, occupying only the
part of a soliton boundary, which makes possible optimization of DW
structure for the rest of the boundary. As $n$ increases further for
SIA the purely collinear structures of the above discussed type, can
appear, see Fig.\ref{si3},\ref{si5}. As for the ExA case, even at
sufficiently large $n=32$, the soliton texture does not contain a
purely collinear DW, see $n=18$ in Fig.\ref{ex3} and large $n=32$ in
Fig.\ref{ex5} below.

\begin{figure}
  \subfigure[ ExA, 21.]{\includegraphics*
  [bb = 180 320 520 660, width = 32mm]{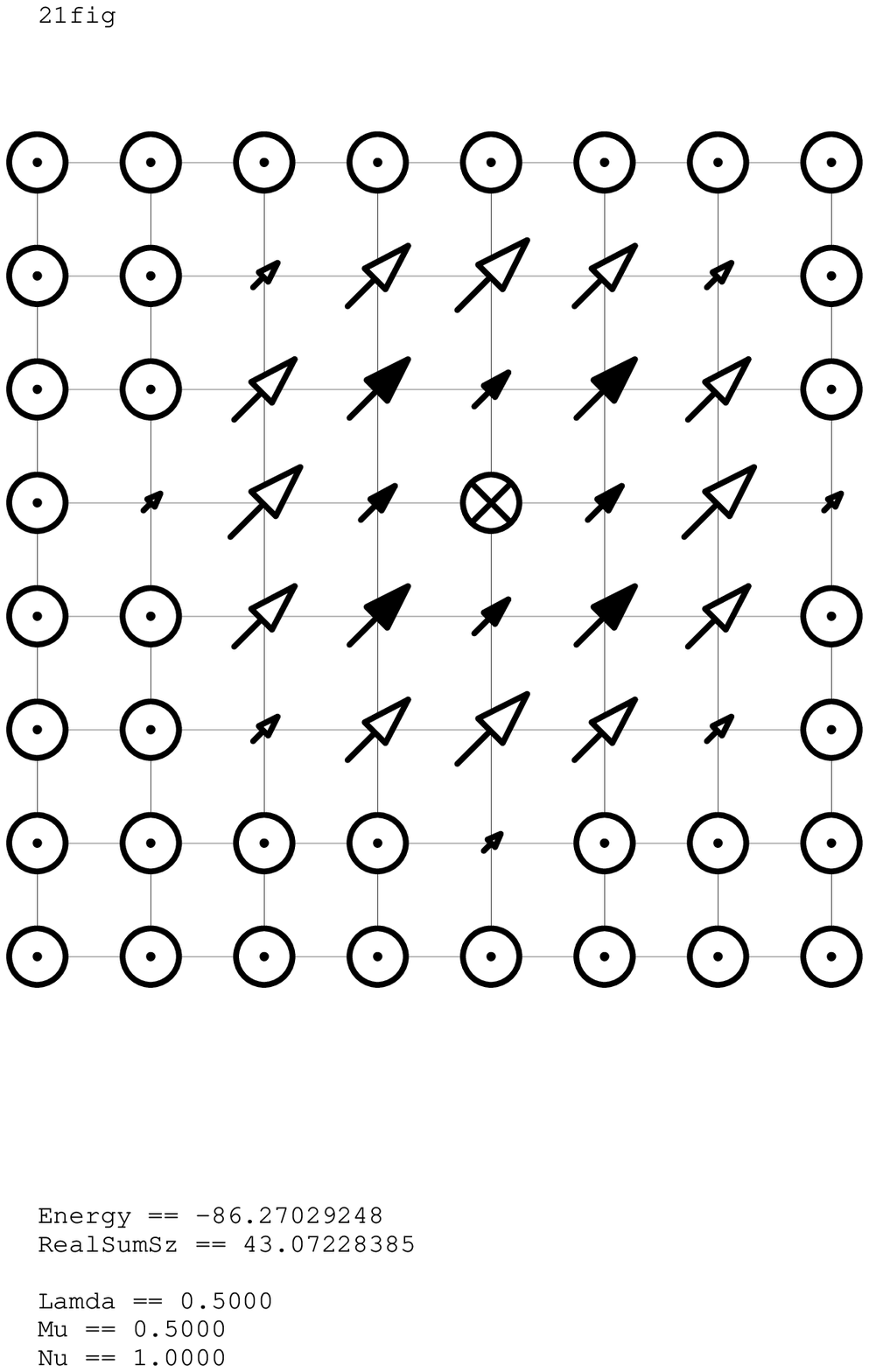}
    \label{exa1}}
    \subfigure[ SIA, 21.]{\includegraphics*
   [bb = 180 320 520 660, width = 32mm]{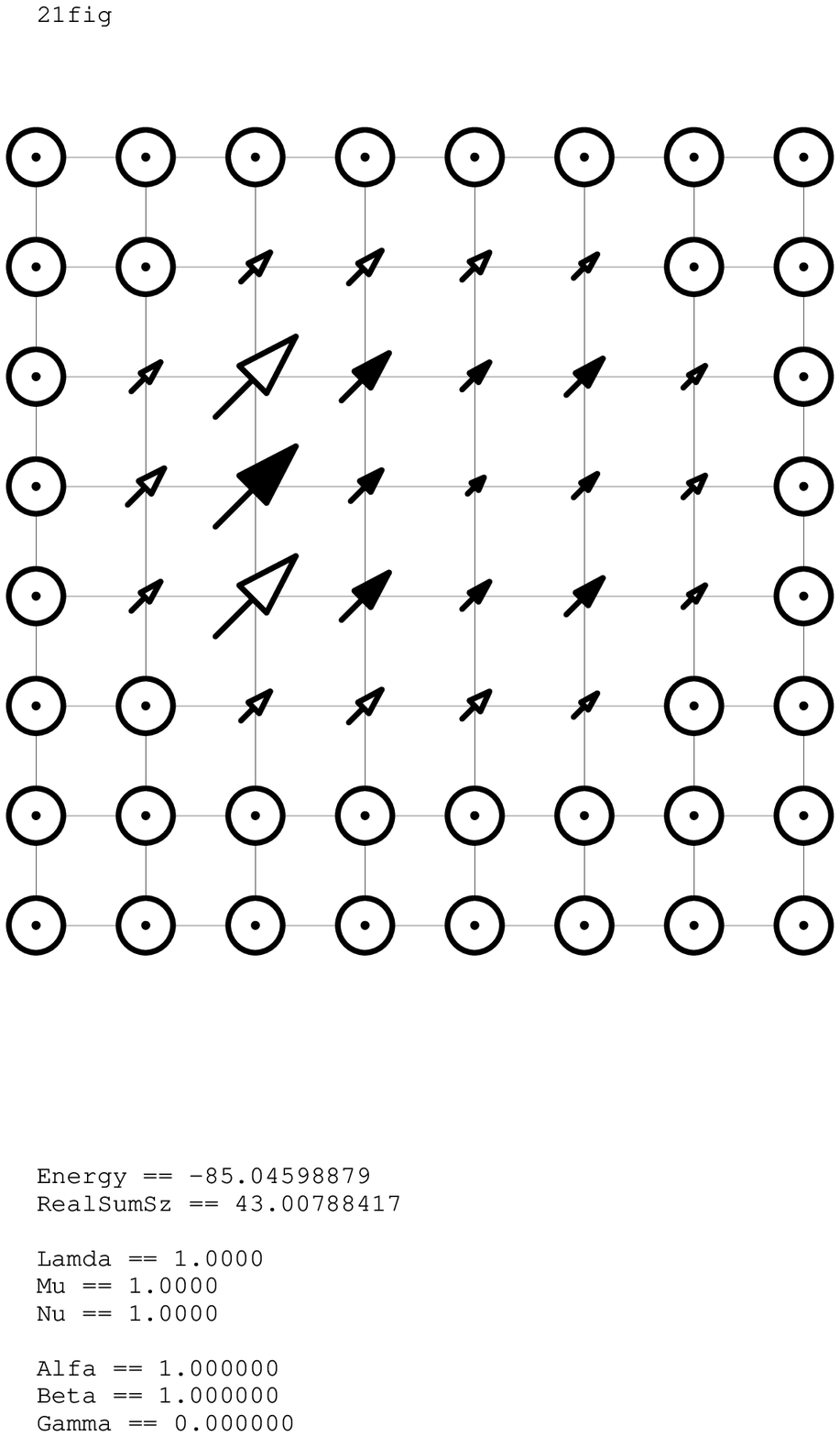}
    \label{sia1}}
    \subfigure[ExA, 24.]{\includegraphics*
  [bb = 180 320 520 660, width = 32mm]{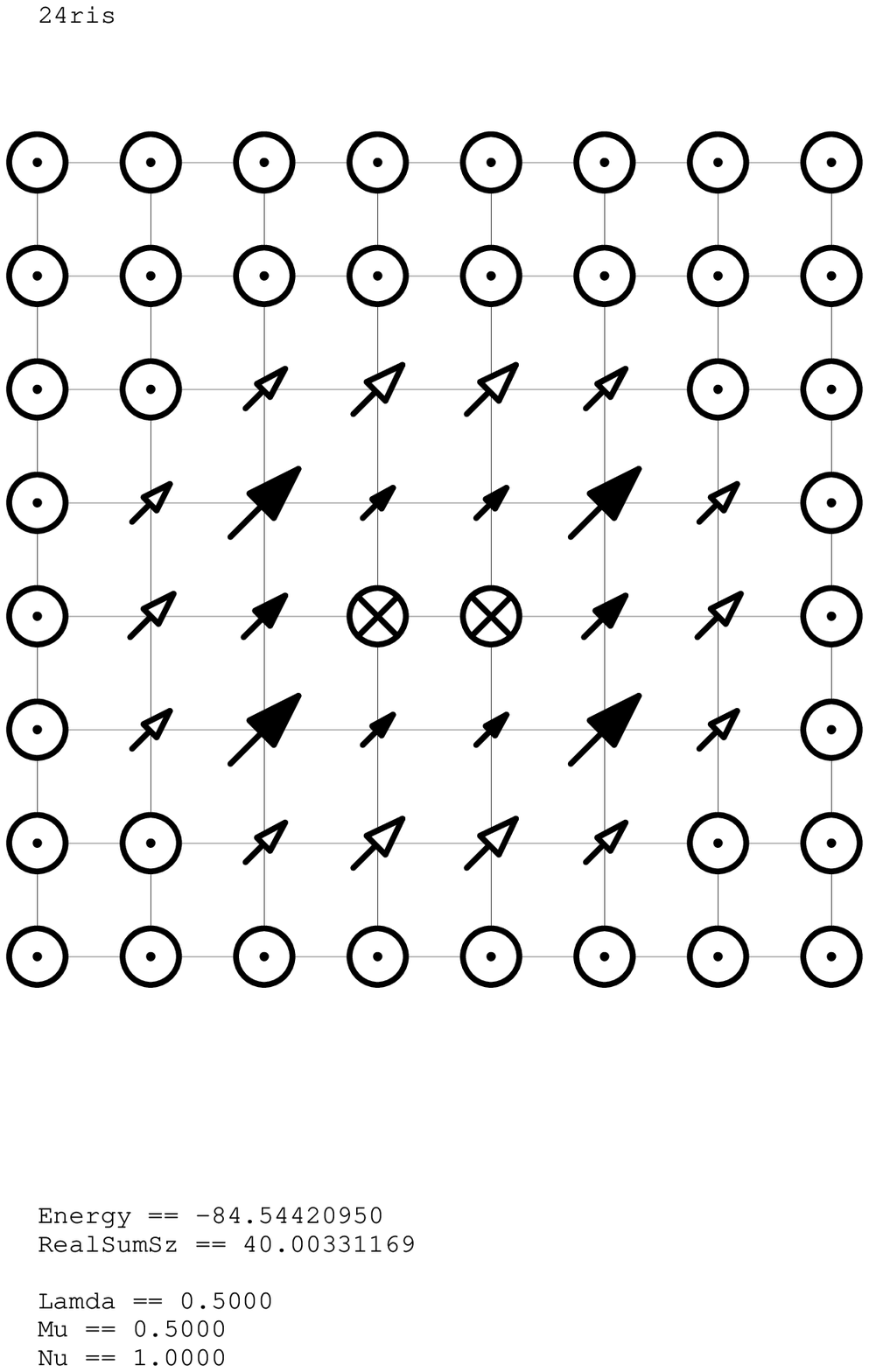}
    \label{exa4}}
    \subfigure[ SIA, 24.]{\includegraphics*
   [bb = 180 320 520 660, width = 32mm]{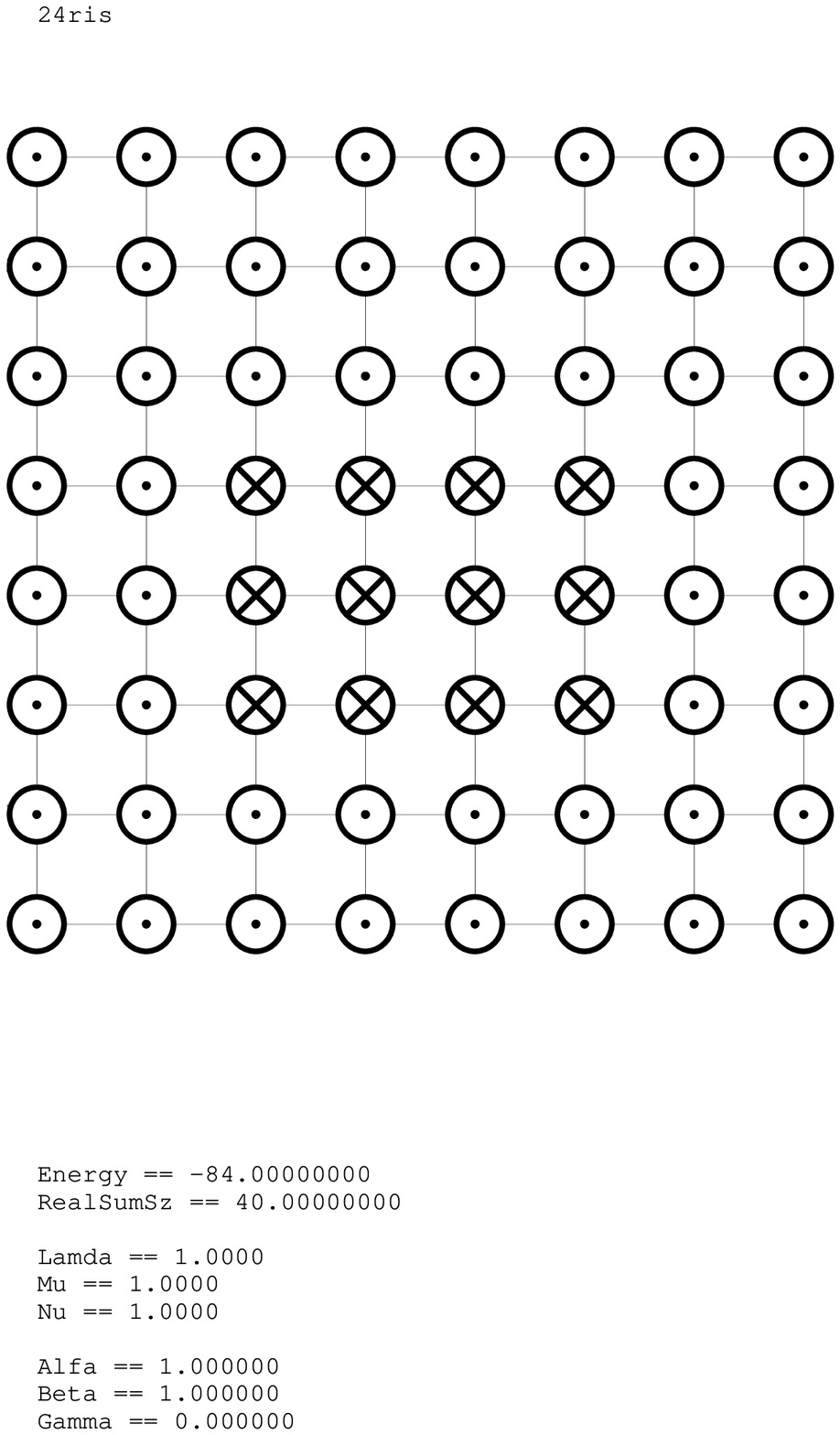}
    \label{sia4}}
    \subfigure[  ExA, 28.]{\includegraphics*
  [bb = 180 320 520 660, width = 32mm]{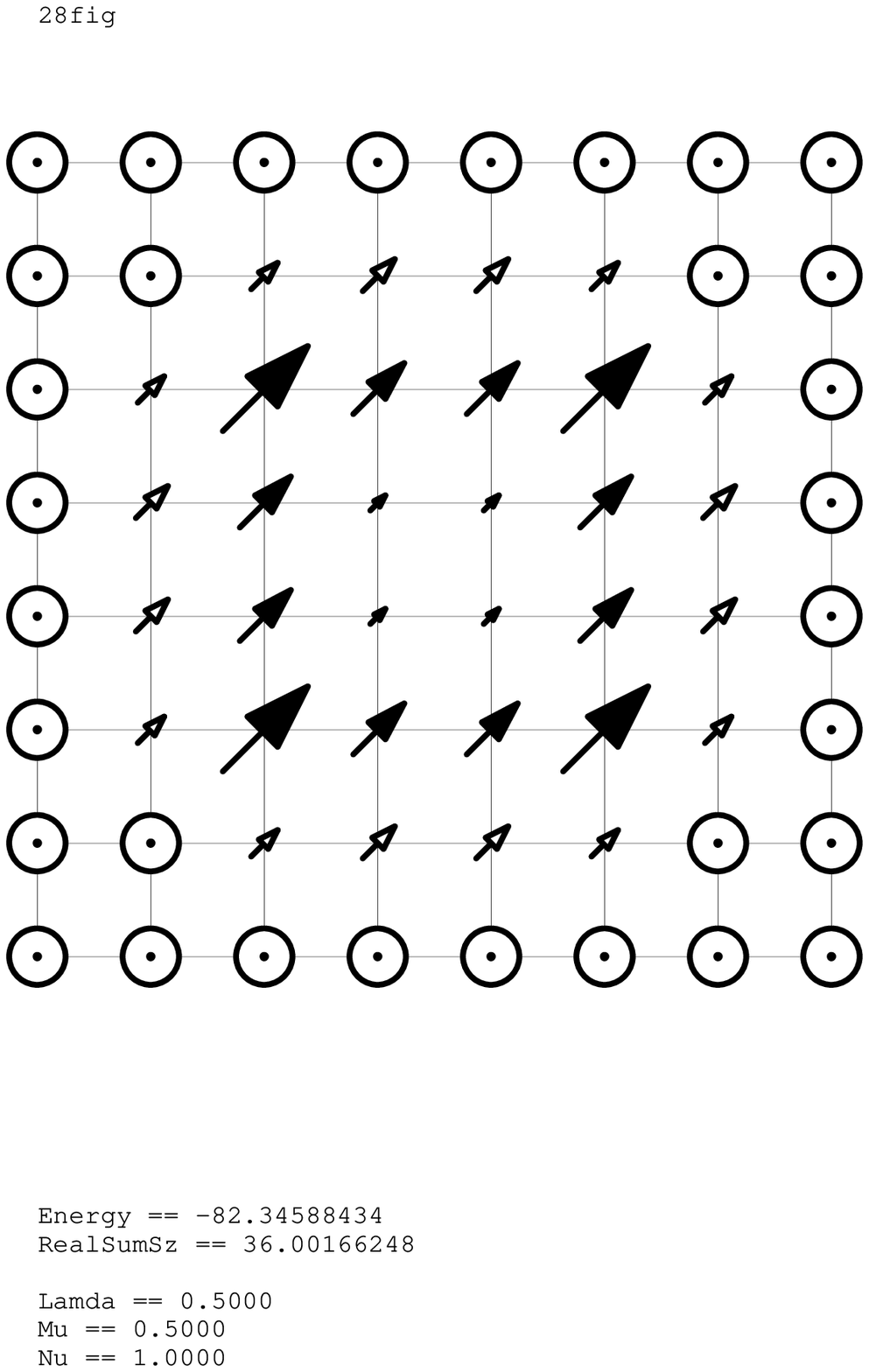}
    \label{exa3}}
    \subfigure[SIA, 28.]{\includegraphics*
   [bb = 180 320 520 660, width = 32mm]{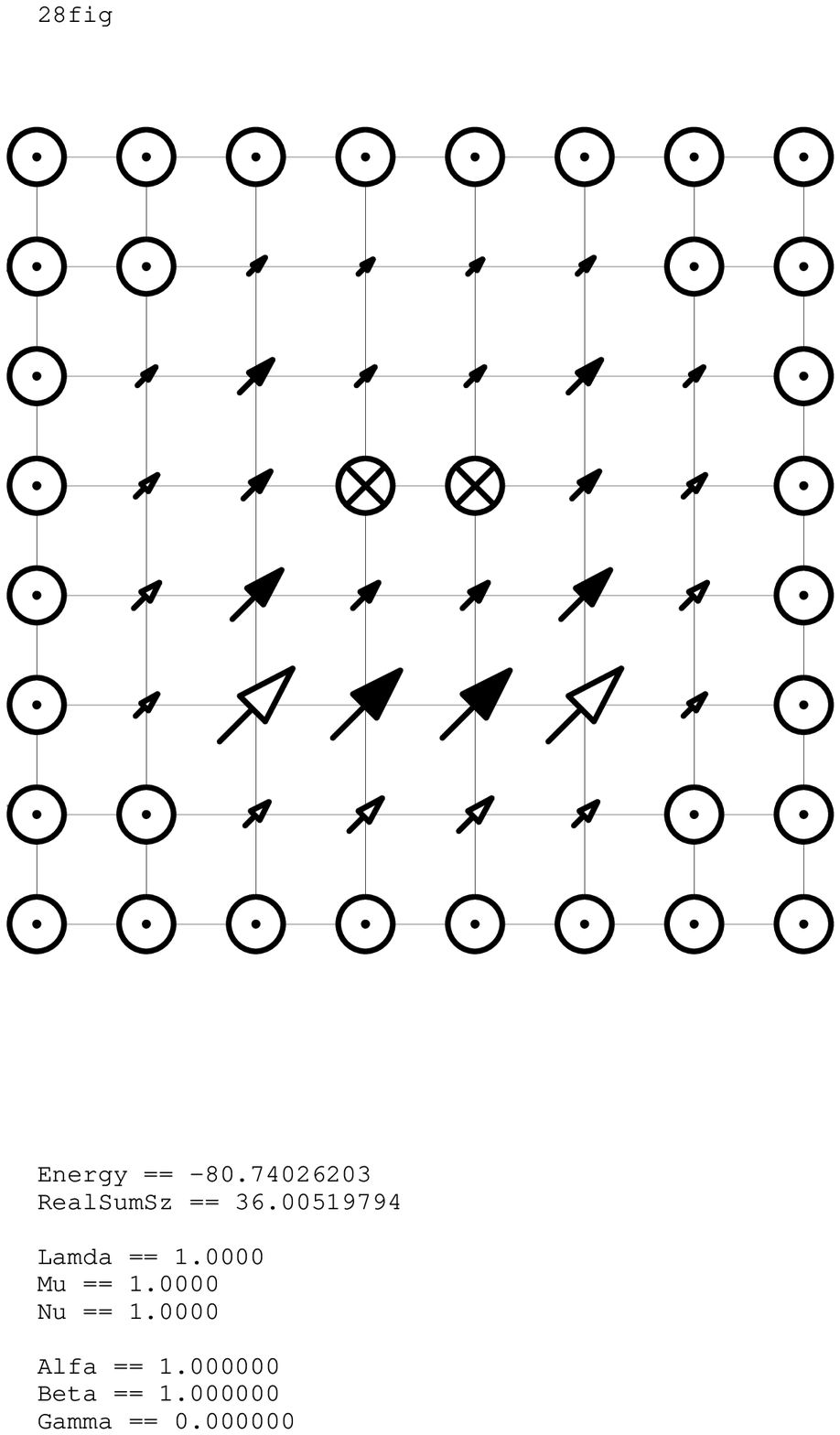}
    \label{sia3}}
    \subfigure[ExA, 31.]{\includegraphics*
  [bb = 180 320 520 660, width = 32mm]{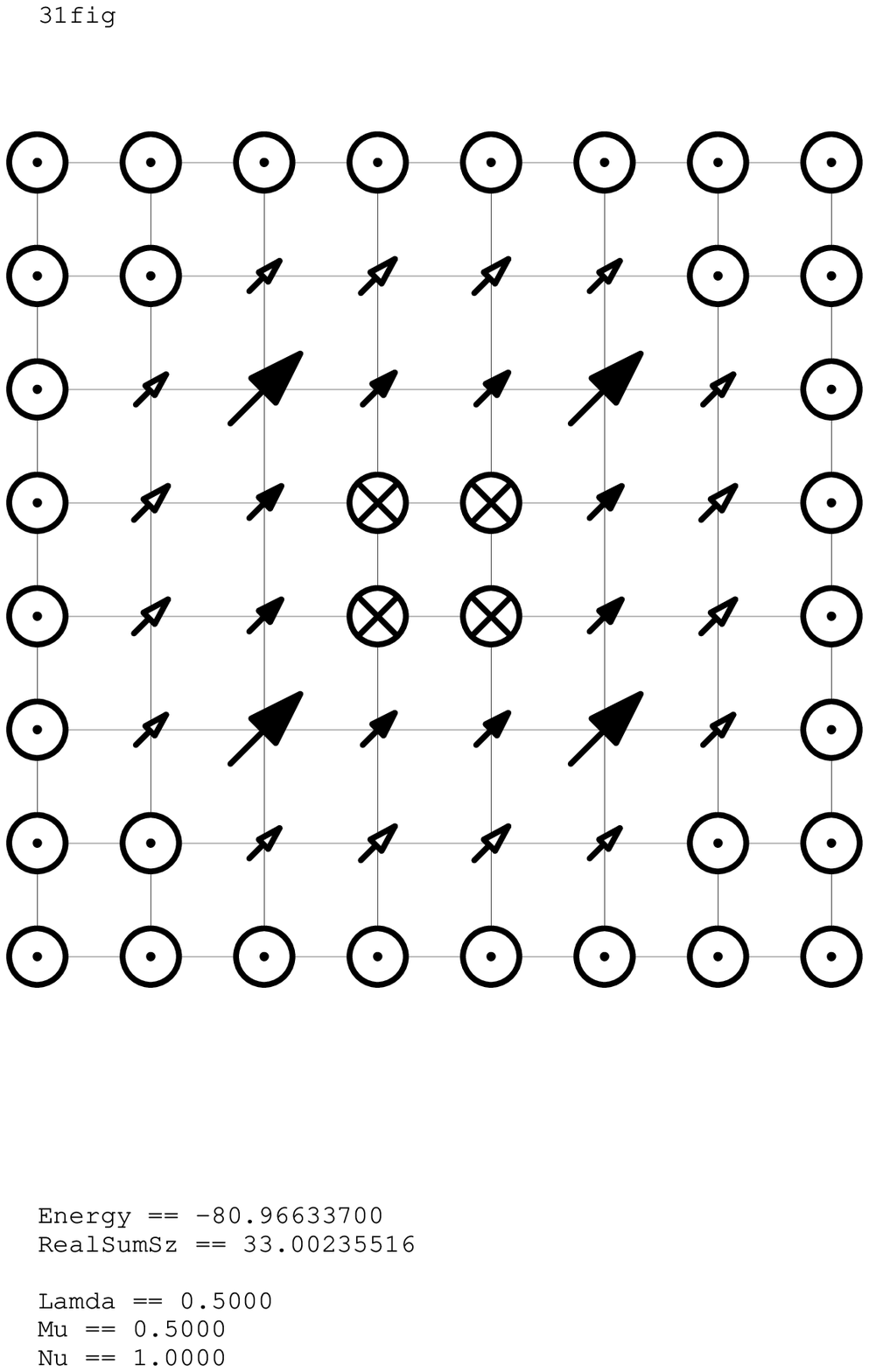}
    \label{exa5}}
    \subfigure[SIA, 31.]{\includegraphics*
   [bb = 180 320 520 660, width = 32mm]{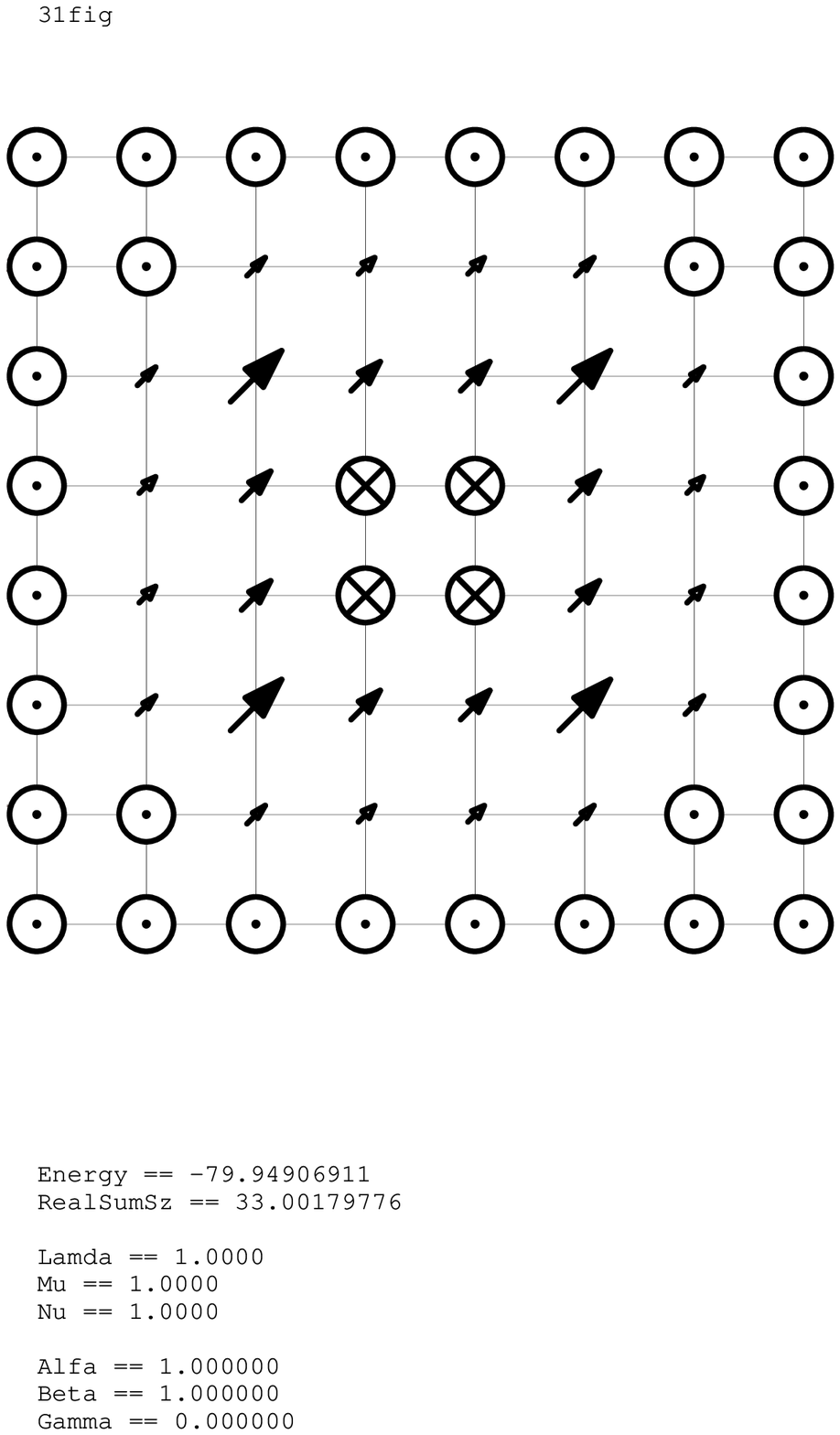}
    \label{sia5}}
    \subfigure[ExA, 32]{\includegraphics*
  [bb = 180 320 520 660, width = 32mm]{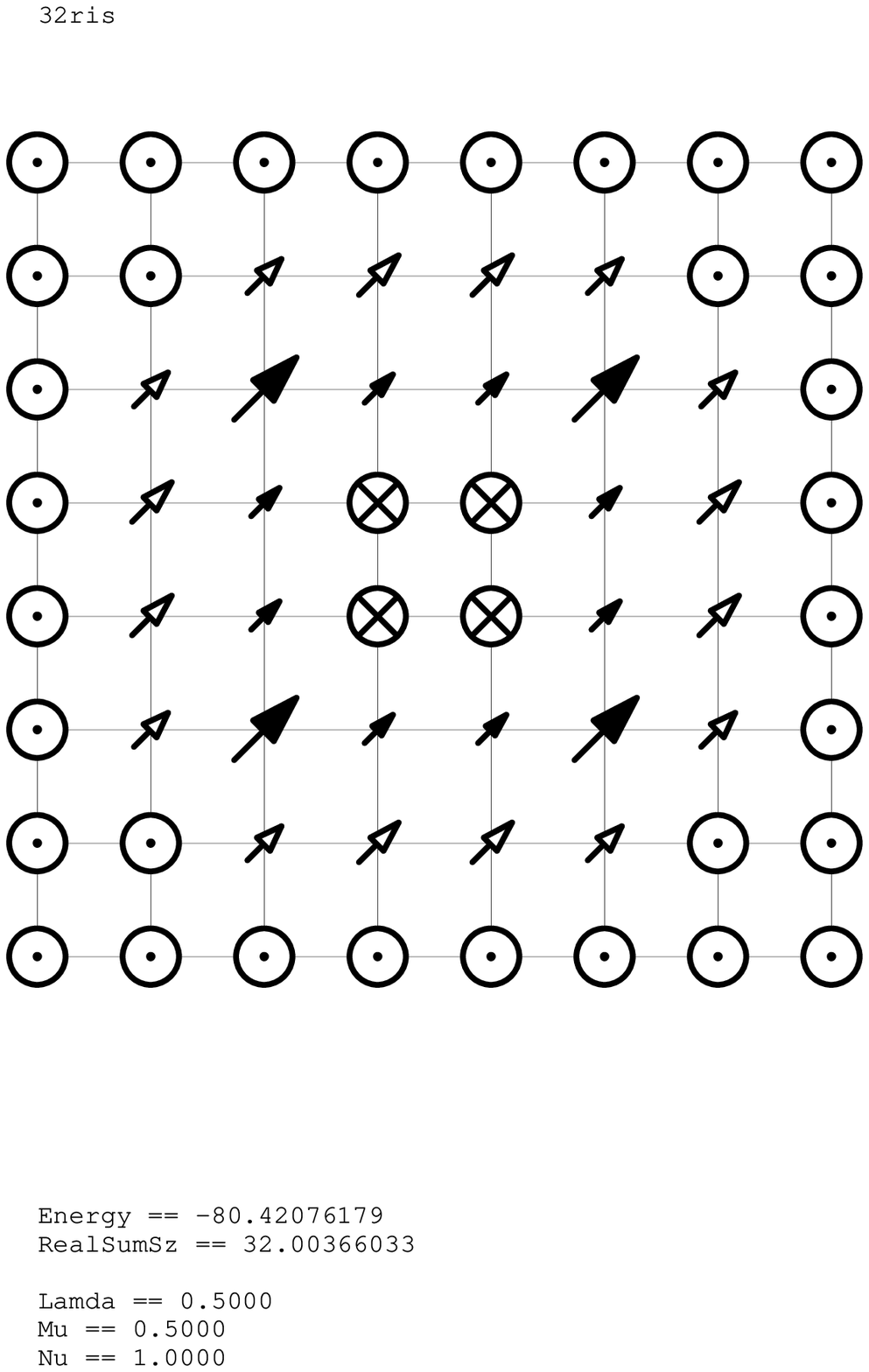}
    \label{ex5}}
    \subfigure[SIA, 32.]{\includegraphics*
   [bb = 180 320 520 660, width = 32mm]{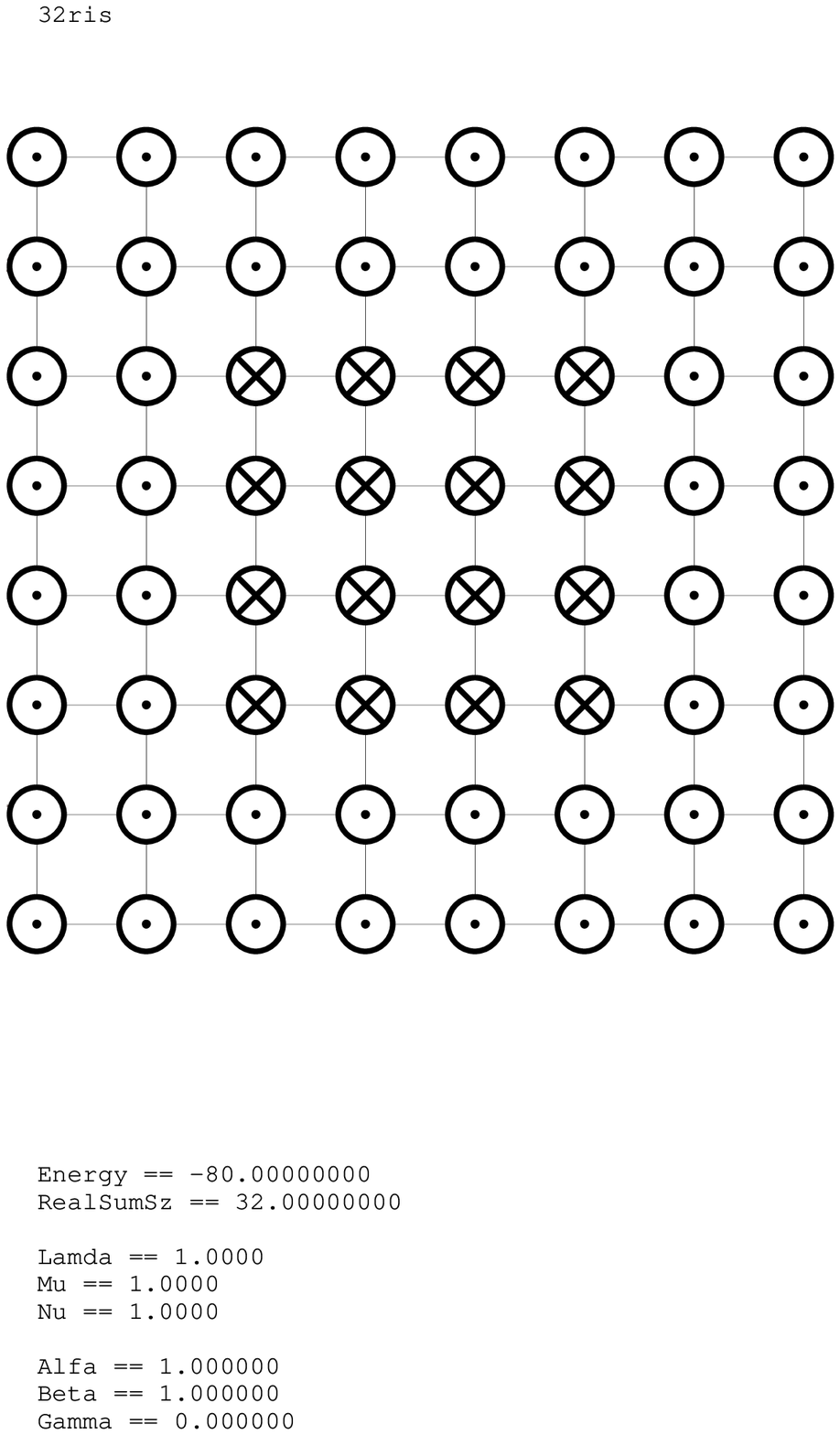}
    \label{si5}}
       \caption{The same as in Fig. \ref{f:structure} for larger values
       of $n$.  \label{f:structure2}}
\end{figure}

Thus, at small $n$, the certain tendency, which is confirmed at
large $n$ (see Fig. \ref{f:structure2}), is clearly seen. Namely,
the strong DW pinning for the SIA case yields almost always the
nonsymmetric configurations, where $n$ growth occurs due to increase
or decrease of ``DW pieces'' on a soliton boundary,
Figs.\ref{sia1},\ref{sia3}. The exceptions are magic numbers $n_{\rm
mag}=2l^2$ (Figs. \ref{si3}, \ \ref{si5}) or close to them
``half-magic'' numbers $n_{\rm hm}=2l(l+1)$ (Fig. \ref{sia4}), where
the collinear structure of  the flipped spins square or rectangle
type (with symmetries, respectively, C$_4$ and C$_2$) is present in
SIA case. For the ExA case the numbers $n_{\rm mag}$ and $n_{\rm
hm}$ are also revealed, but in a quite different manner. Namely, for
$n=n_{\rm mag}$ the soliton does not have a collinear structure but
resembles a square, see Fig. \ref{exa1}.  Here, however, there is a
fundamental difference with the SIA case. The same structures with
C$_4$ symmetry are formed also for $n$ non-magic, $n\neq n_{\rm
mag}$, $n_{\rm hm}$. This becomes clear if we recollect that in this
case the pinning is weak (see above) so that ``DW piece'' formation
is absolutely unfavorable. This means that C$_4$ symmetry occurs
both for $n$ magic and nonmagic, see Fig.\ref{ex3},\ref{ex5}.
However, for $n$ half-magic, the rectangular shape of a soliton core
occurs also for ExA both for small (Fig. \ref{ex2}) and large $n$
(Fig. \ref{exa4}). Thus, the soliton structure for the ExA case also
depends on $n$ nonmonotonously, but here the ``half magic'' numbers
are more important since close to these numbers the soliton symmetry
first lowers from C$_4$ to C$_2$ and then it restored back to C$_4$.
At the magic number $n=32$, see Figs. \ref{ex5} and \ref{si5} as
well as near this magic number, both SIA and ExA textures have quite
symmetric structure with almost collinear DW, see Figs. \ref{exa5}
and \ref{sia5} for $n=31$.

This complex and irregular picture of soliton behavior "maps"\ onto
the $E(N)$ dependence only as a local energy lowering at $n\approx
n_{\rm mag}$. The irregular behavior of soliton characteristics is
revealed much more vividly in the dependence $\omega (N)$, seen in
Figs. \ref{fig:v7}(a),(b). For both types of anisotropy, the
explicit traces of nonmonotonous behavior like jumps, regions with
$d\omega /dN>0$, and even those with $\omega <0$ (for SIA, see Fig
\ref{fig:v7}) occur. At this point it is useful to make two remarks.
First, we recollect, that for discrete systems, contrary to the
continuous ones, the condition $d\omega /dN<0$ is no more a soliton
stability criterion. Second, the condition $\omega <0$ just means
that in this region of parameter values, the soliton energy {\em
decreases} as $N$ increase, but says nothing about the soliton
stability. It is seen from Fig \ref{fig:v7}(a), that the
non-monotonous structure of the corresponding $\omega (N)$ curves
manifests itself most vividly at $n > 10\div 15$. For SIA, the
abrupt vertical up and downward jumps with growing of $n$ on the
curve $\omega(N)$ appear at certain values of $n$. The upward jumps
occur near magic and half-magic numbers introduced above. After
these upward jumps, the frequency has plateaus at $18<n<20$,
$24<n<26$, $32<n<34$, and then the deep minima. For large $n \sim 31
\div 32$, the frequency becomes negative, $\omega <0$. The fact,
that these negative values occur only for SIA for $n$, slightly less
then the ``magic'' value $n=32$, corroborates the above suggested
concept.
 \begin{figure}[!h]
 \vspace*{-5mm} \hspace*{-5mm} \centering{\
\includegraphics[width=1.1\columnwidth]{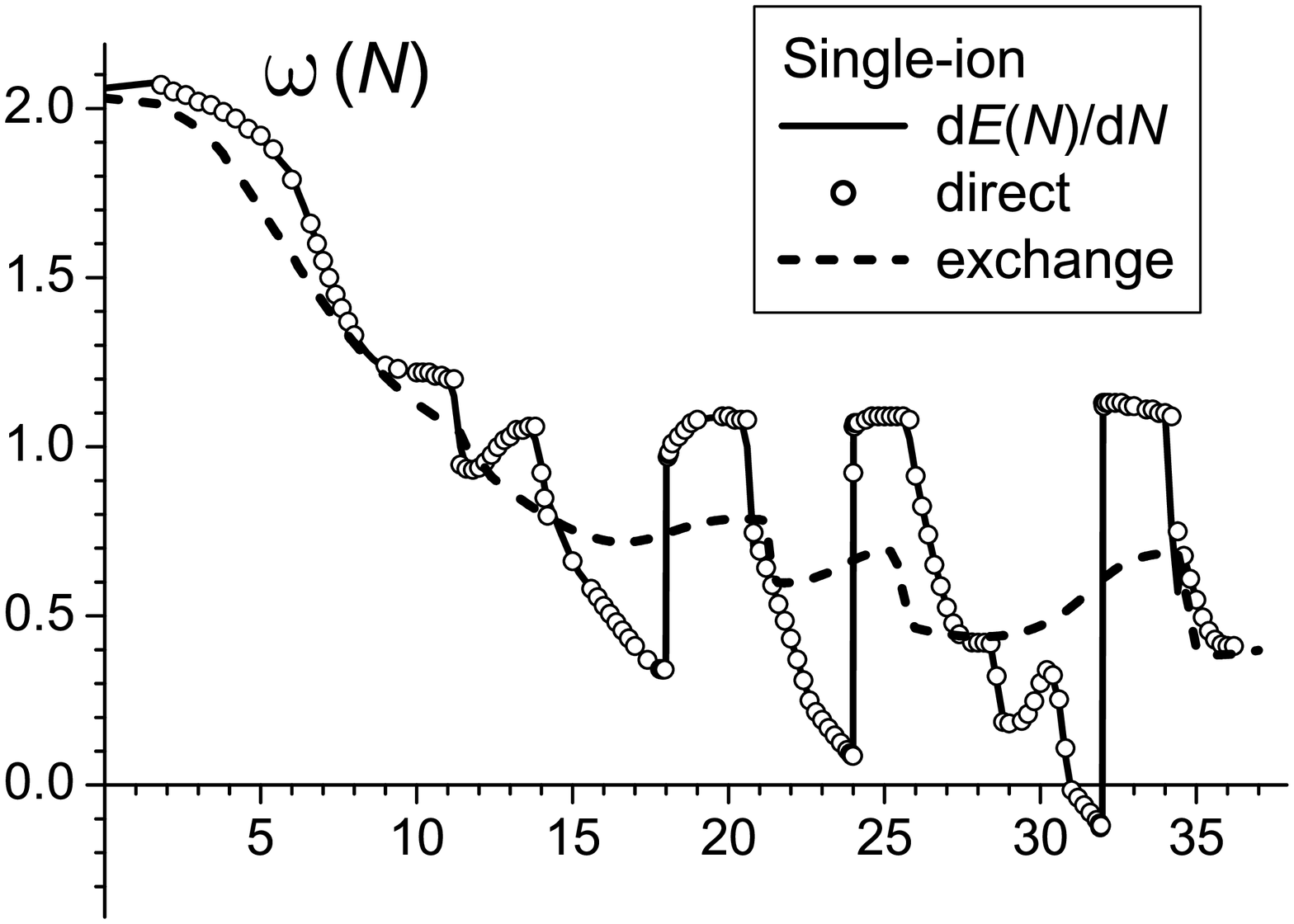}
\includegraphics[width=1.1\columnwidth]{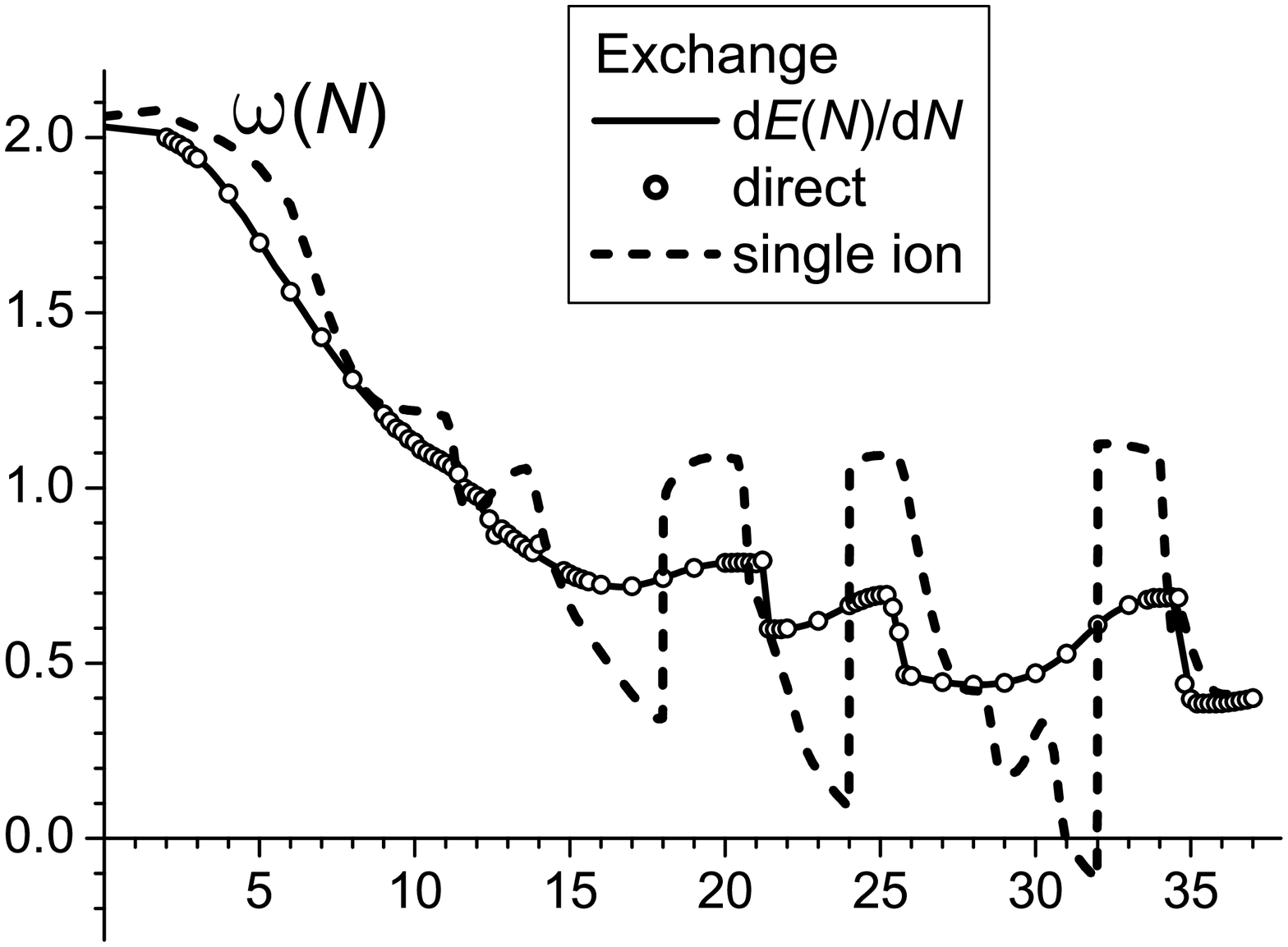}}
\caption{Dependence $\omega(N)$ for a soliton with $K_{\rm{eff}}=J$,
obtained from numerical simulations. (a)- Single ion anisotropy, (b)
- exchange anisotropy. Dashed lines are for better comparison and
correspond to exchange anisotropy (a) and single ion one (b).}
\label{fig:v7}
\end{figure}

For the case of exchange anisotropy (Fig.\ref{fig:v7}b), the upward
jumps on the curve $\omega (N)$ are almost absent, and the smooth
increase of $\omega$ occurs instead, i.e. the effect of "magic"\
numbers is absent. The downward jumps are clearly seen at the same
values of $n$ as those for SIA. These jumps sometimes are much
sharper that those for SIA, but for ExA  the amplitude of these
jumps are smaller, than  for SIA, and $\omega >0$ everywhere.

The physical explanation of the above quite complicated behavior can
be done based on the aforementioned picture.  First, the upward
jumps for $n=n_{\rm mag}$ with subsequent almost constant $\omega $
can be easily understood for SIA. In this case, at $n=n_{\rm mag}$
the formation of favorable collinear structure has already been
finished so that the plateau at larger $n$ is due to the creation
and growth of a "DW piece". It is clear that for ExA this scenario
does not occur, and this effect is completely absent. The downward
jump and general decrease of the soliton frequency is related to the
transition to more symmetric configurations, where the excessive
number, $n$  of magnons is easily spread along the soliton boundary.
Such transitions take place for both types of anisotropies, which
explains the behavior similarities.

Let us discuss now the possibility of the topological soliton
existence in high anisotropy ferromagnets. In this specific case, it
is useful to utilize a simplified obvious definition of the
topological invariant. The $\pi _2$ topological invariant
(\ref{eq:Pontryagin}) for the case of large anisotropy has simple
geometric meaning. Namely, only $\theta \neq 0,\pi$ make a
contribution into integral (\ref{eq:Pontryagin}) so that for its
evaluation it is sufficient to consider the DW region. Formally, the
integral (\ref{eq:Pontryagin}) can be represented as a contour
integral along the DW,
\[
Q \to \frac{1}{2\pi}\oint \frac{\partial \phi}{\partial\chi}d\chi.
\]
This value defines a mapping of the DW line (which is of necessity a
closed loop) onto the closed contour which is a domain of angle
$\phi $ variation. This representation makes it obvious that the
topological charge, $Q$ notion is meaningful only in the case when
the DW has a well defined noncollinear structure throughout its
length. The DW regions with collinear spins play the role of a "weak
link", where $\phi$ can change abruptly by $2\pi $ almost without
overcoming the potential barrier. Hence, even for nonsymmetric
soliton textures, when soliton has a quite large "piece"\ of
noncollinear DW, with "quasitopological" \ spin inhomogeneity,
literally topological structures are absent. Although the difference
between topological and nontopological solitons in the magnets with
$K_{\rm eff}\sim J$ is not that large and the question about
realization of topological solitons in such structures is rather
academic, this problem will be discussed  in more details.

\begin{figure}[!h]
\vspace*{-5mm} \hspace*{-5mm} \centering{\
\includegraphics[width=1.1\columnwidth]{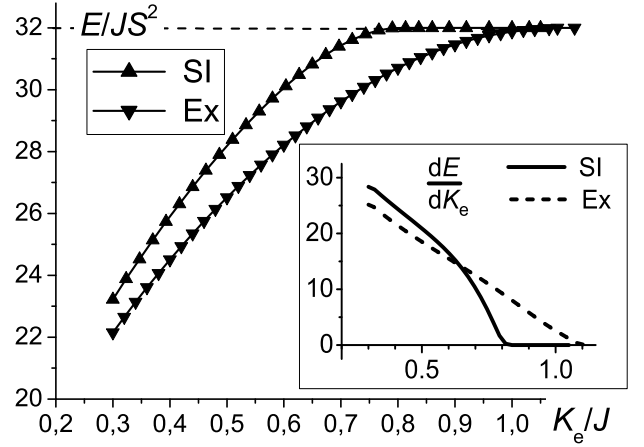}}
\caption{The energy of the soliton with $N/S=32$ (magic) as a
function of effective anisotropy constant. Inset shows the behavior
of a derivative $dE/dK_{\rm{eff}}$ near $K_{\rm{crit}}\approx 0.8$.}
\label{fig:v4}
\end{figure}

To answer the question about the presence or absence of a
topological texture, we have carried out the numerical minimization
both over a complete set of variables and over $\theta$'s only with
fixed in-plane spin directions. Contrary to the above considered
case of moderate anisotropy, the latter minimization (i.e. that over
$\theta$'s only) never gives the instability of a nontrivial
topological structure. Instead, the decrease of noncollinear
structure amplitude (either uniformly along the entire DW or on its
individual parts) occurs as $K_{\rm {eff}}$  increases. The behavior
is quite different for different  values of $n=N/S$ so that these
cases should be considered separately. For specific analysis, we
have chosen a few $n$ values, typical "magic" number $n=32$ and two
non-magic, $n=35$ and $n=36$. In spite of closeness of these
numbers, the soliton spin texture behavior in these cases differs
drastically as anisotropy increases.

\begin{figure}[!t]
\subfigure[ ExA, 0.6]{\includegraphics*
  [bb = 180 260 580 660, width = 35mm]{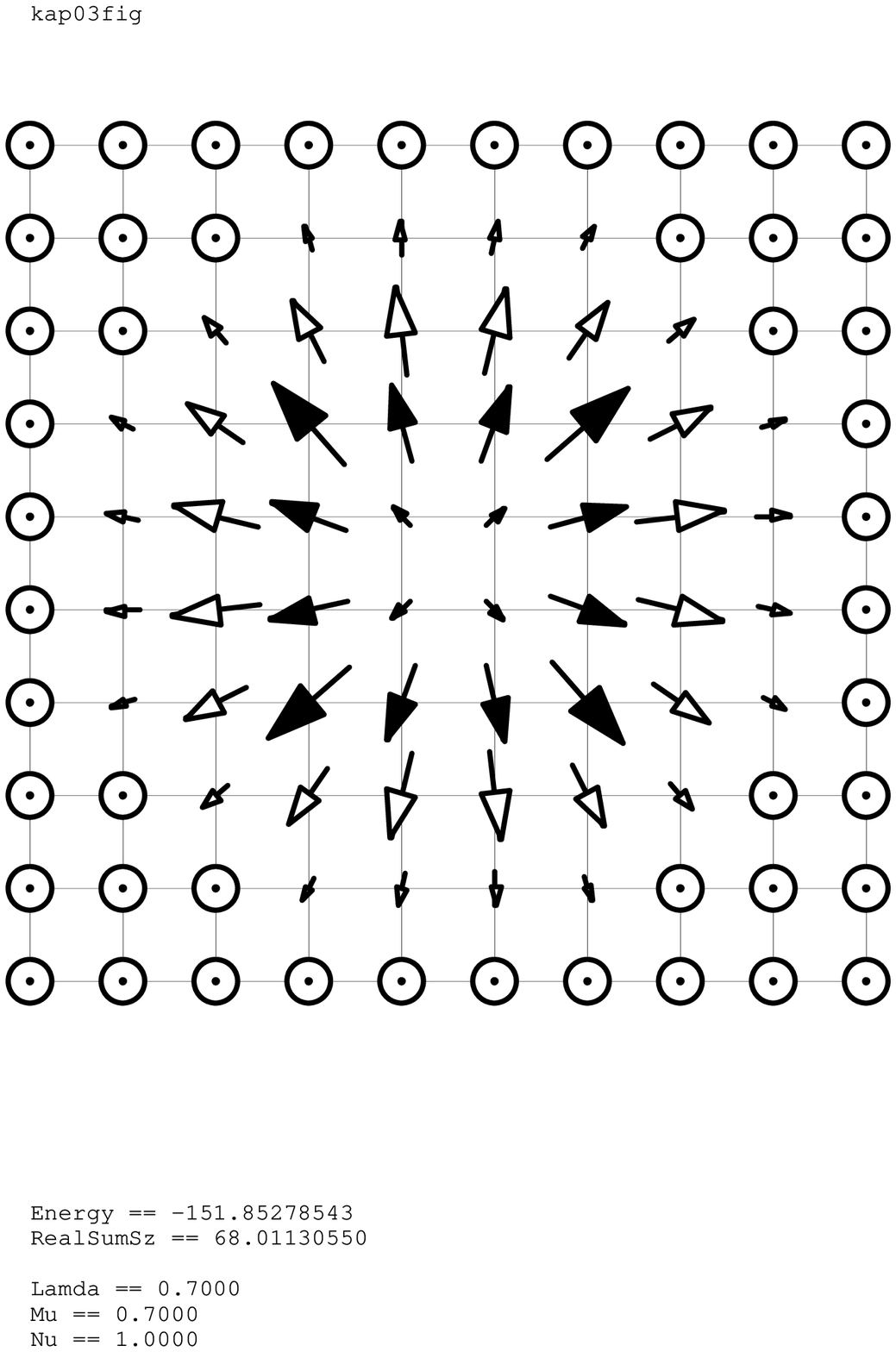}
    \label{32ex07}}
    \subfigure[ SIA, 0.6]{\includegraphics*
   [bb = 180 260 580 660, width = 35mm]{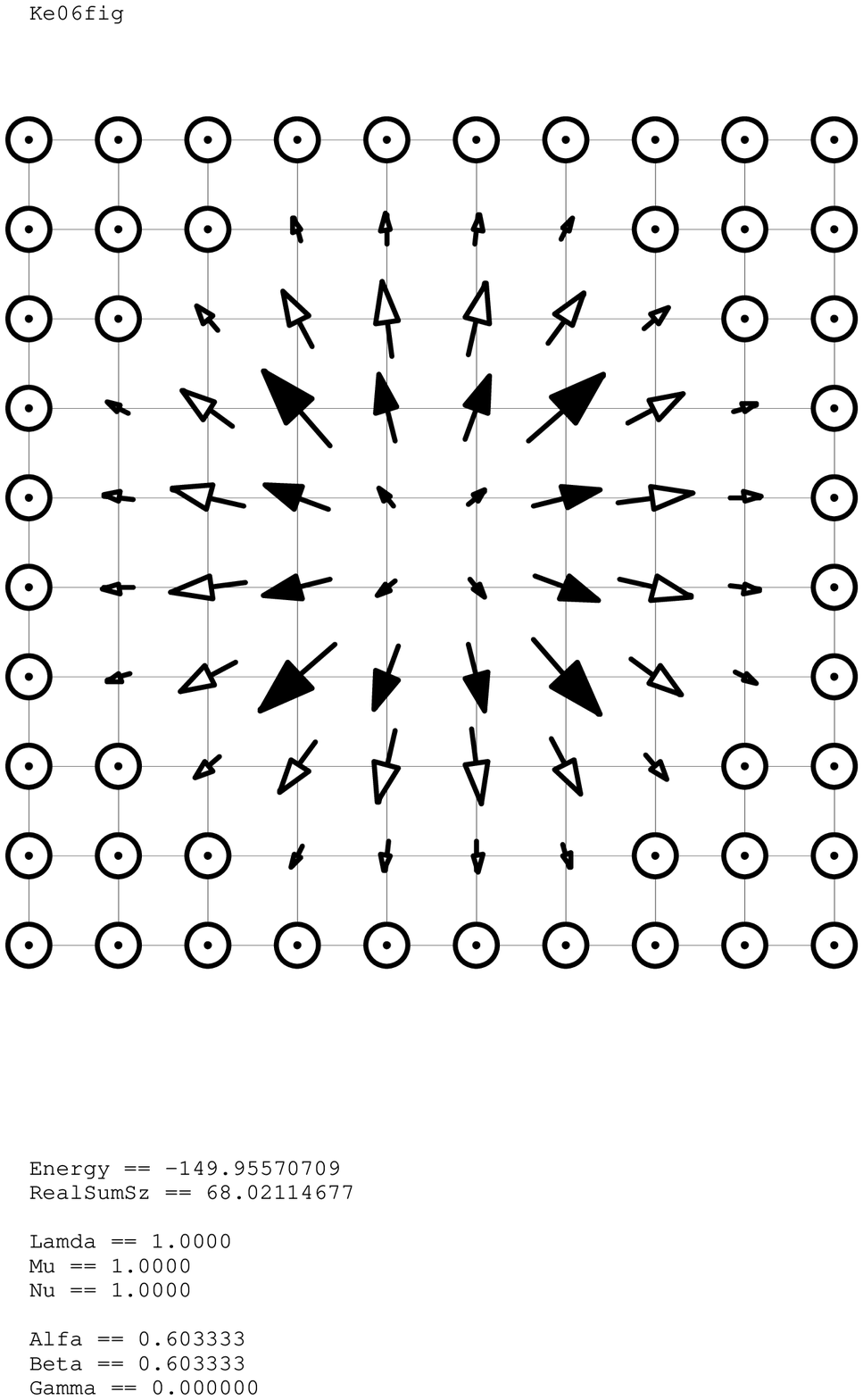}
    \label{32si07}}
    \subfigure[ ExA, 0.8]{\includegraphics*
  [bb = 180 260 580 660, width = 35mm]{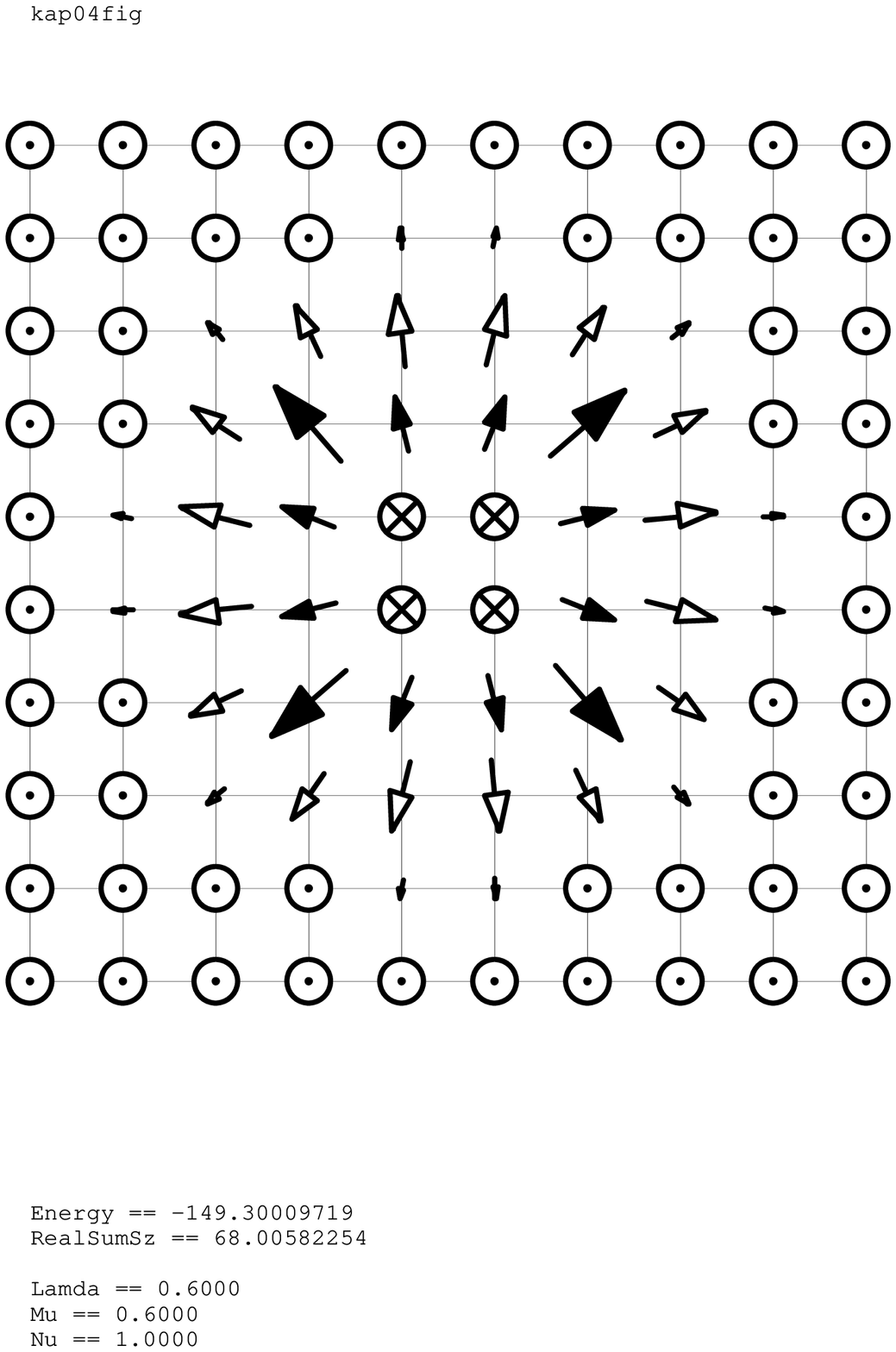}
    \label{32ex08}}
    \subfigure[ SIA, 0.8]{\includegraphics*
   [bb = 180 260 580 660, width = 35mm]{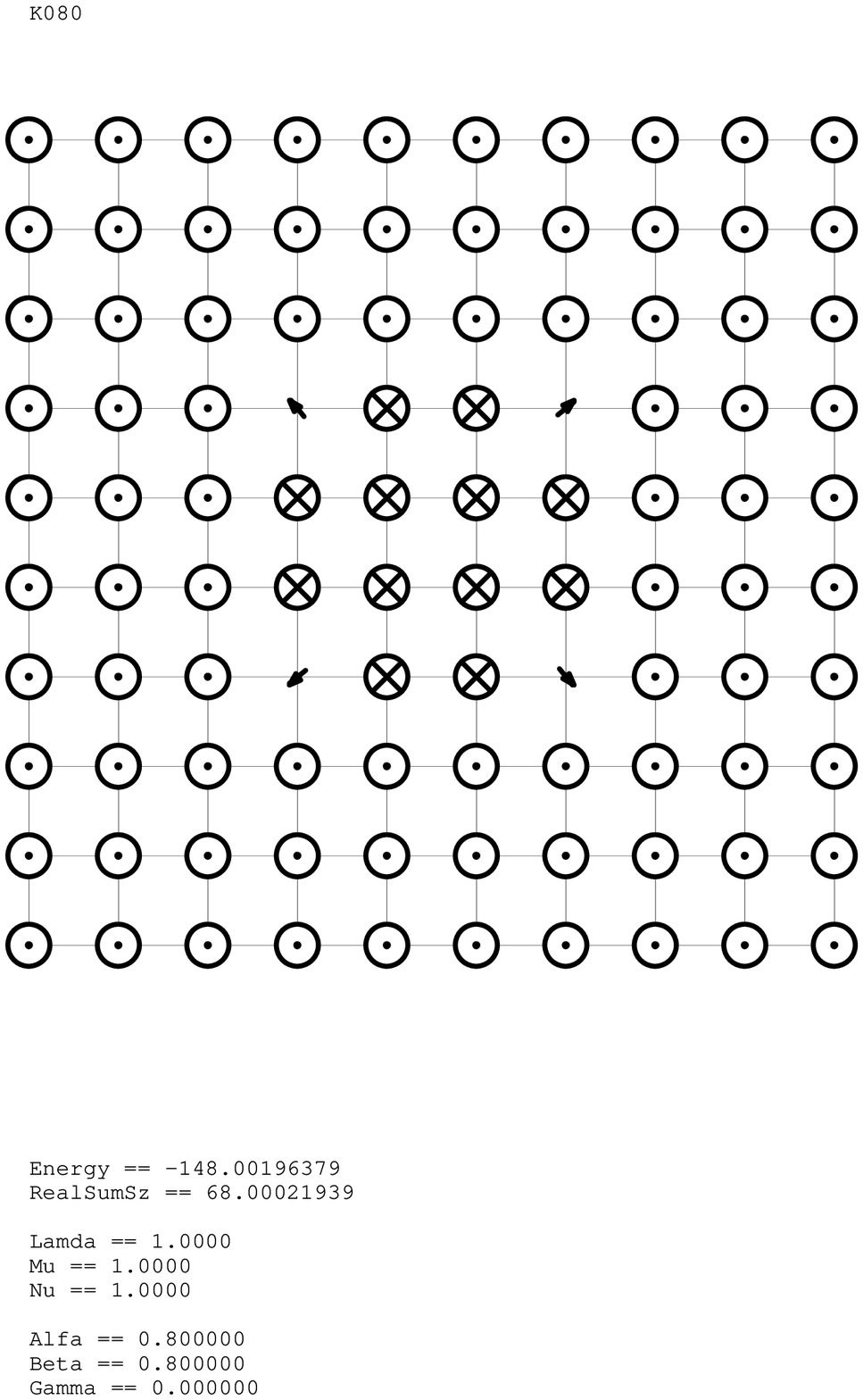}
    \label{32si08}}
    \subfigure[ ExA, 1.0]{\includegraphics*
   [bb = 180 260 580 660, width = 35mm]{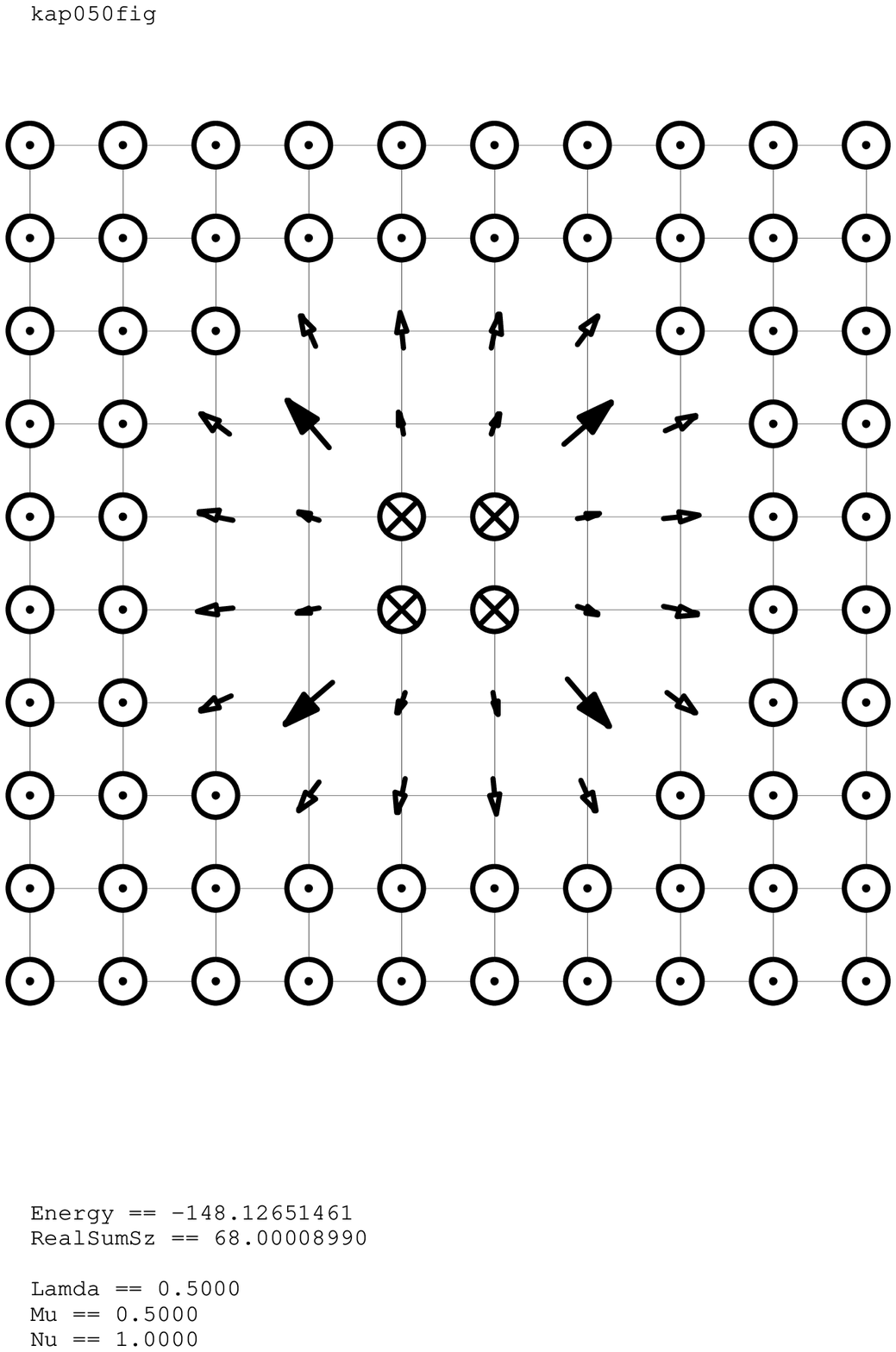}
    \label{32ex1.0}}
    \subfigure[ ExA, 1.08]{\includegraphics*
  [bb = 180 260 580 660, width = 35mm]{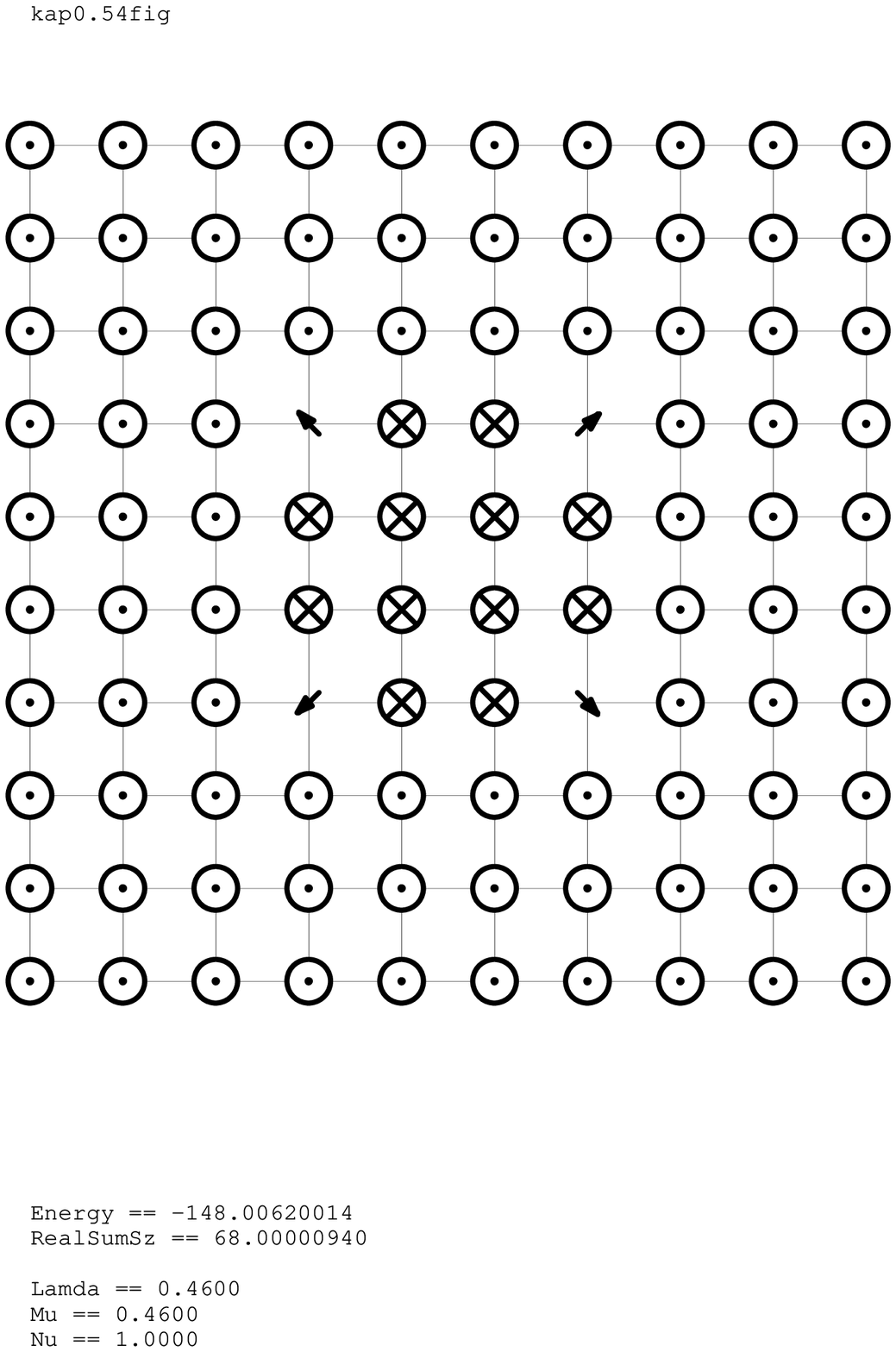}
    \label{32ex1.08}}
\caption{Soliton structure for  magic number of bound magnons
$N=32S$ and different values of effective anisotropy constant $K_e$,
the value of $K_e/J$, together  with the type of anisotropy, are
shown below pictures of spin distribution. \label{f:topol32}}
\end{figure}

The energy of the topological soliton as a function of
$K_{\rm{eff}}/J$ for magic number $n =32$ and two types of
anisotropy is shown in Fig.\ref{fig:v4}. These curves have been
obtained by numerical simulations on a lattice, for an anisotropy
increase from its small value, when there is a well-defined
topological soliton texture. It is seen that at large $K_{\rm{eff}}$
the soliton energy tends to some finite limiting value,
$E_0=32JS^2$, that is typical for a collinear structure with the
``magic number'' $n=32$. But the behavior of these functions near
this limiting value is different for single-ion and exchange
anisotropies, corresponding to the different scenarios of
annihilating of both non-collinear spin structure and topological
structure in a soliton. It is interesting to trace the disappearance
of a soliton topological structure at anisotropy increase, which is
shown in Figs.~\ref{f:topol32}. For small $K_{\rm{eff}}$ the
topological structure is well defined, and this structure is similar
for both SIA and ExA, see Figs.~\ref{32ex07} and \ref{32si07}. If
the value of $K_{\rm{eff}}$ increases to some critical value,
$K_{\rm{crit}}$, the in-plane spin amplitude decreases, but a
``vortex-like'' configuration with approximate symmetry C$_4$ is
still visible. For larger anisotropy $K\geq K_{\rm{crit}}$ the
soliton structure becomes purely collinear.  For SIA the critical
value is $K_{\rm{crit}}\approx 0.8$, but for ExA this value is much
larger, $K_{\rm{crit}}\approx 1.1$, compare Figs.~\ref{32si08} and
\ref{32ex08}.  For ExA, the topological structure is still visible
for such strong anisotropy as $2\kappa = K_e=J$, see
Fig.~\ref{32ex1.0}. However, for high enough $\kappa$ the structure
finally  becomes collinear and of the same structure as that for
SIA, as is shown in Fig.~\ref{32ex1.08}. In other words, for
``magic'' magnon numbers the topological structure of a soliton
decays smoothly as the anisotropy constant grows, which is seen for
both types of anisotropy, compare Figs.~\ref{32si08} and
~\ref{32ex1.08}.

For $K_{\rm{eff}}> K_{\rm{crit}}$ for both kinds of anisotropy there
is no more topological spin structure - all spins are directed
either up or down and energy does not depend on $K_{\rm{eff}}$ any
more (see Fig.\ref{fig:v4}). Such a picture resembles very much the
phase transition of a second kind with the collinear soliton texture
as more symmetric phase. This agrees with the behavior of the energy
$E(K_{\rm{eff}})$ near $K_{\rm{crit}}$, (see inset to
Fig.\ref{fig:v4})
\begin{equation}\label{ilm}
E-E_0\propto (K_{\rm{crit}}-K_{\rm{eff}})^2.
\end{equation}
To understand better the picture of the transition from a
topological (noncollinear) soliton texture to a collinear texture,
we discuss the analogy with the second order phase transition in
more detail. Note, that any collinear spin structure with {\em
arbitrary} positions of ``up'' and ``down'' spins has the same
symmetry element, namely, rotation  about the $z$ - axis in a
\emph{spin space}. This symmetry is due to the symmetry of the
Hamiltonian (\ref{eq:H-discrete}). On the other hand, the spin
textures, even noncollinear, with $n=32$ and C$_4$ symmetry, are
invariant with respect to a rotation by $(\pi/2)\cdot k$, $k\in
\mathbb{Z}$, \emph{simultaneously} in spin space and
\emph{coordinate space}. Thus, ``magic'' collinear textures with
spatial symmetry C$_4$ have higher symmetry, being invariant
relative to the {\em independent} rotation of the coordinate space
by $(\pi/2)\cdot k$ and of the spin space by arbitrary angle, while
noncollinear ``magic'' ones are invariant only relatively to {\em
simultaneous} rotation by $(\pi/2)\cdot k$ in spin and coordinate
spaces. This means that on a transition from a ``magic'' collinear
soliton to noncollinear one, spontaneous symmetry breaking occurs
and phase transition of  the second kind appears naturally.

For ``non-magic'' \ numbers $n=35$ and $n=36$, the symmetry of
soliton structures and their behavior is fundamentally different.
First, note that the collinear structure can be realized for
\emph{even} values of $n$ only. For any non-even $n$, odd-integer or
non-integer, some spins have to be inclined to the $z-$axis. Thus,
these two cases have to be considered separately.
\begin{figure}
\vspace*{-5mm} \hspace*{-5mm} \centering{\
\includegraphics[width=1.1\columnwidth]{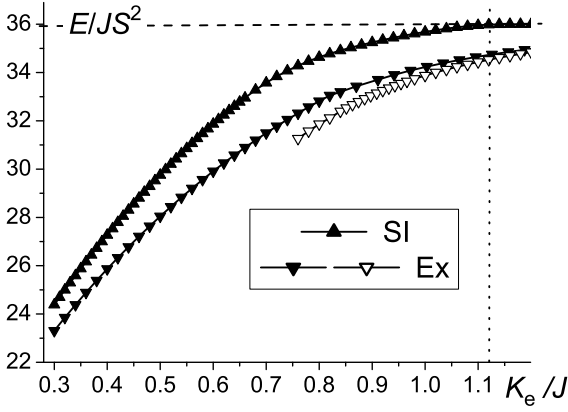}}
\caption{The energy of the soliton with $N/S=36$ (non-magic, even)
as a function of effective anisotropy constant $K_{\rm{eff}}$. For
exchange anisotropy, full symbols presents data for topological
solitons, open symbol non-topological solitons with $Q=0$. The
values of $K_{\rm{eff}} \approx 1.13$ for the transition from
non-collinear to collinear soliton for single ion anisotropy are
shown by vertical dotted lines.} \label{fig:v4+}
\end{figure}

The value $n=36$ is rather far both from "magic"\ number $n=32$ and
from the nearest "half-magic"\  $n=40$, corresponding to the
favorable configuration with a rectangular collinear DW. Here, for
small anisotropy $K_{\rm {eff}}\lesssim 0.6J$ we see the structure
with C$_4$ symmetry, similar to that for ``magic'' $n=32$ for both
types of anisotropy, see Fig.~\ref{32ex07} and \ref{32si07}.
However, with the increase of $K_{\rm {eff}}$, the evolution is
different, as seen in Fig. \ref{fig:v4+}. The difference in behavior
of $E(K_{\rm {eff}})$ is the largest for exchange anisotropy, where
the saturation, depicted in Fig.~\ref{fig:v4}, did not occur up to
quite large $\kappa \gtrsim 0.6J$ ($K_{\rm{eff}}\gtrsim 1.2J$).
However, for single ion anisotropy there is also a difference from
the above considered ``magic'' case $n=32$.

\begin{figure}[!t]
\subfigure[ExA, 0.7]{\includegraphics*
  [bb = 180 260 580 660, width = 35mm]{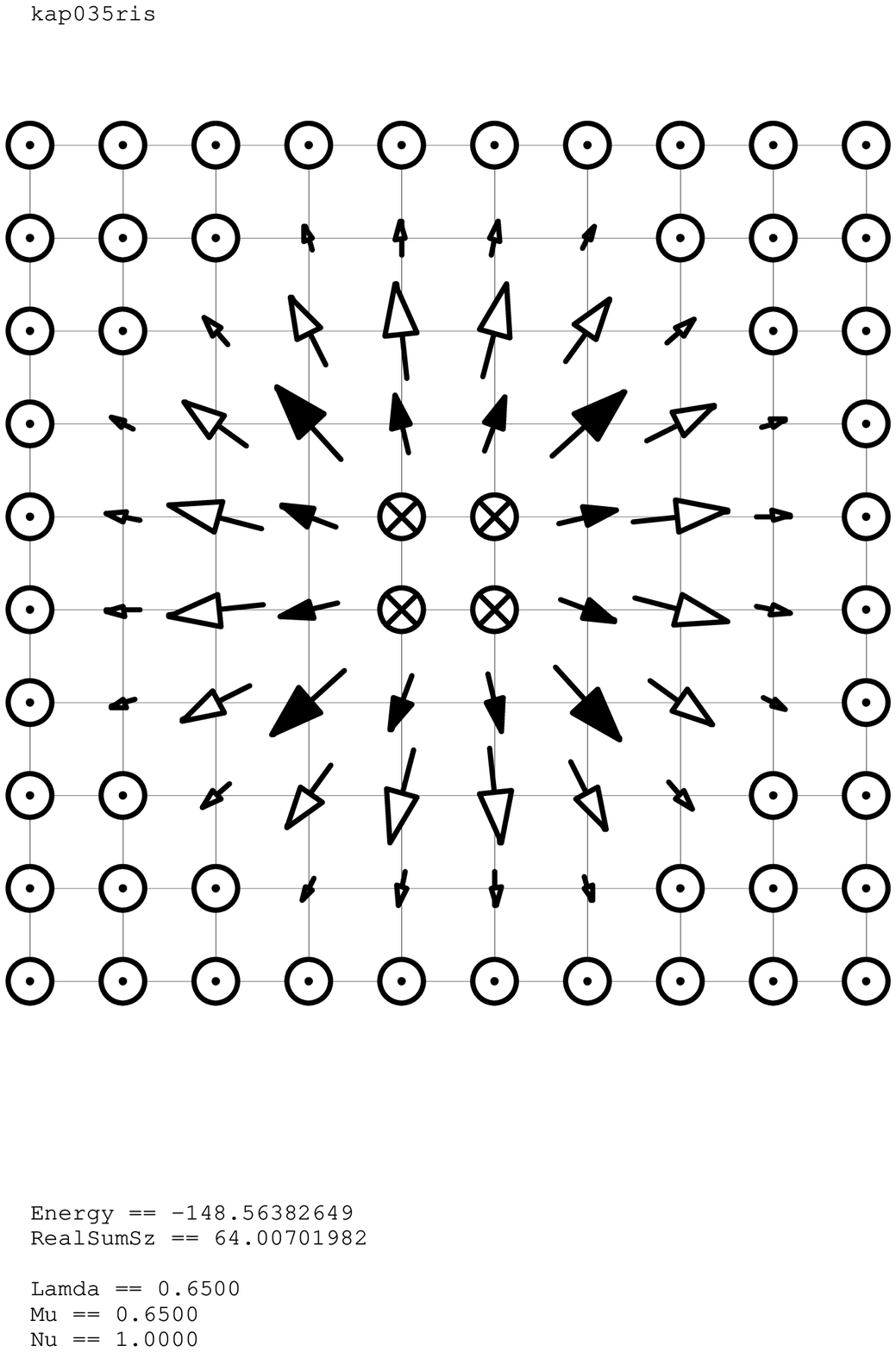}
    \label{ex362}}
    \subfigure[SIA, 0.7]{\includegraphics*
   [bb = 180 260 580 660, width = 35mm]{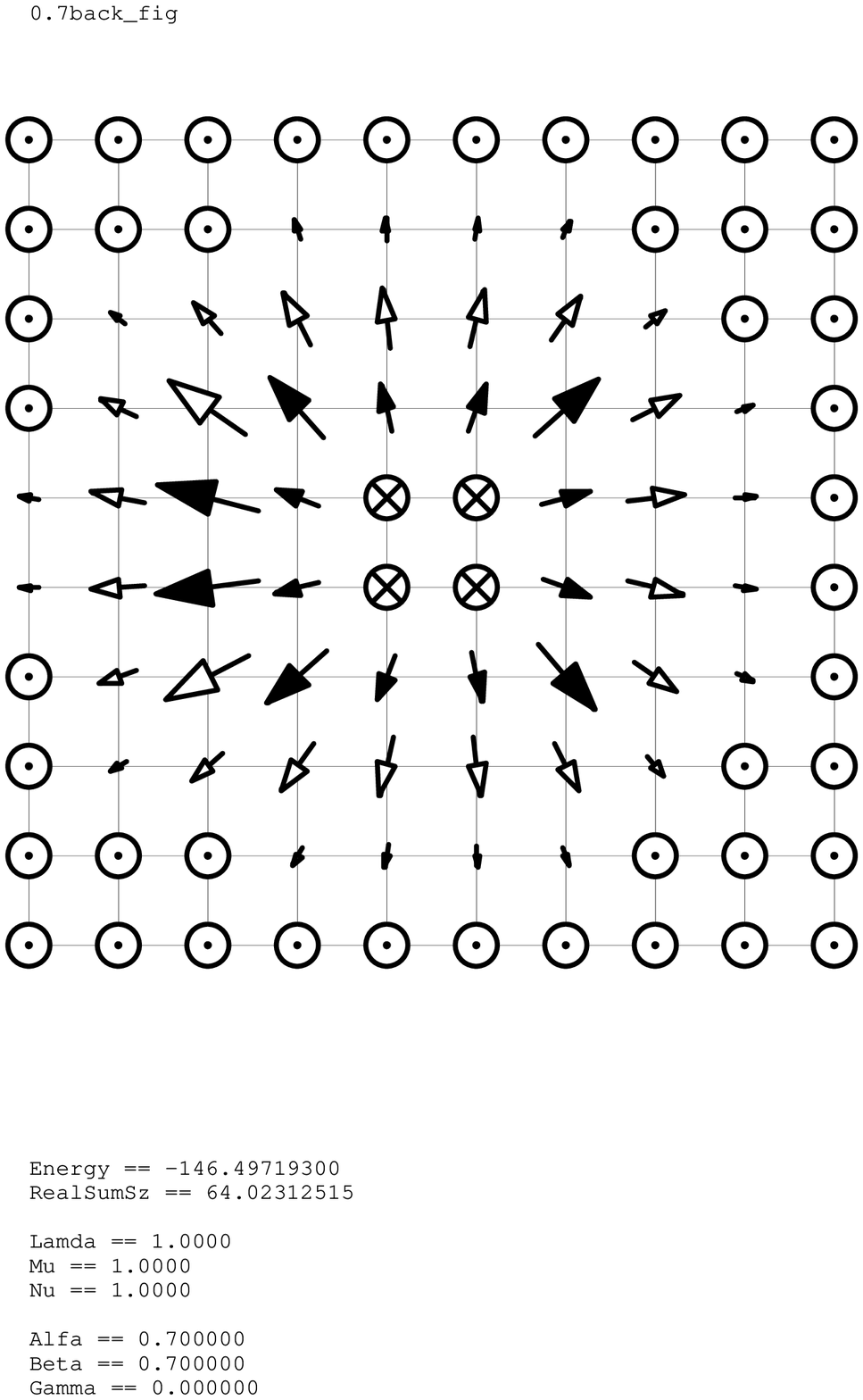}
    \label{si362}}
    \subfigure[ExA, 0.8]{\includegraphics*
  [bb = 180 260 580 660, width = 35mm]{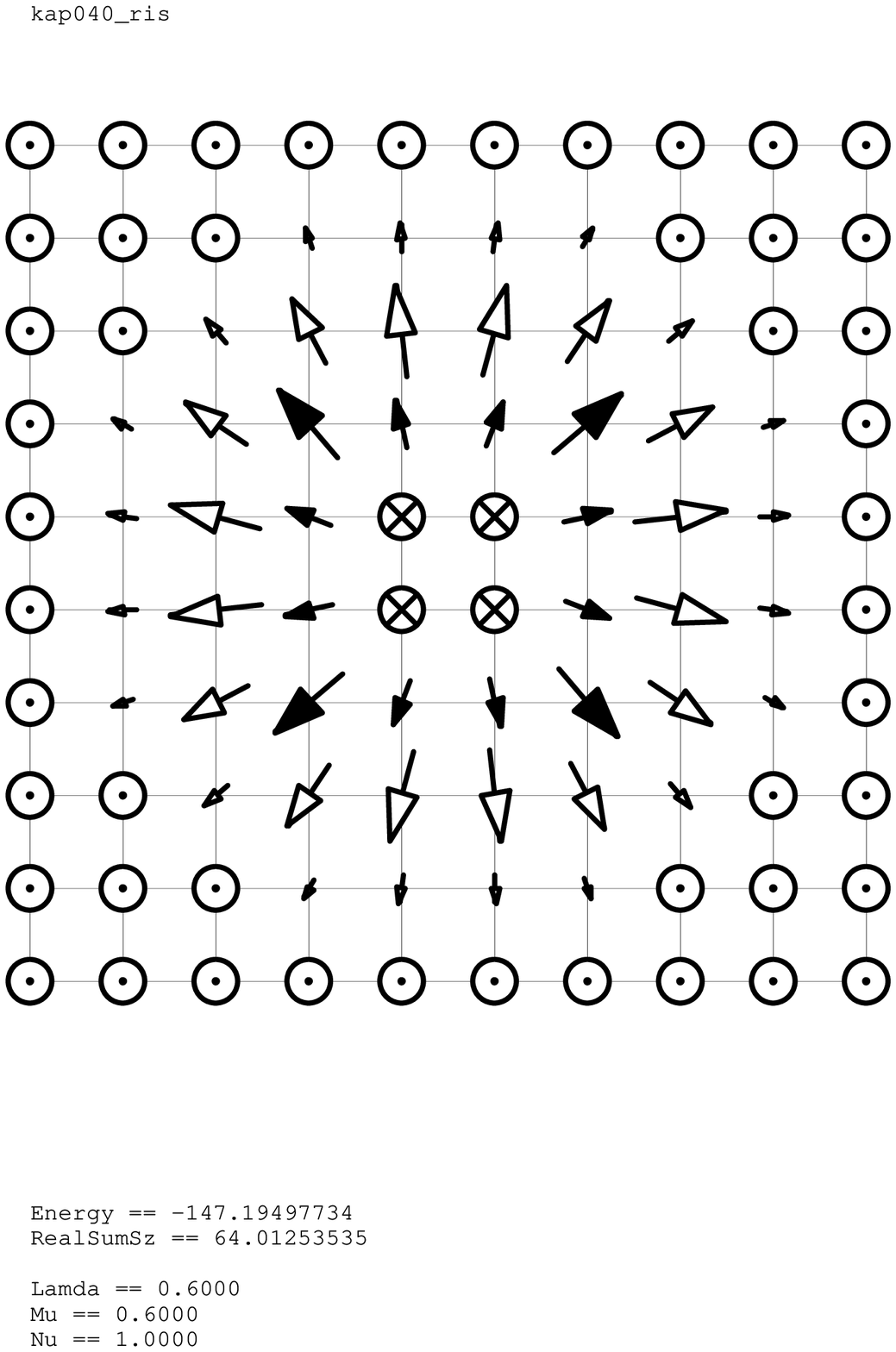}
    \label{ex363}}
    \subfigure[SIA, 0.8]{\includegraphics*
   [bb = 180 260 580 660, width = 35mm]{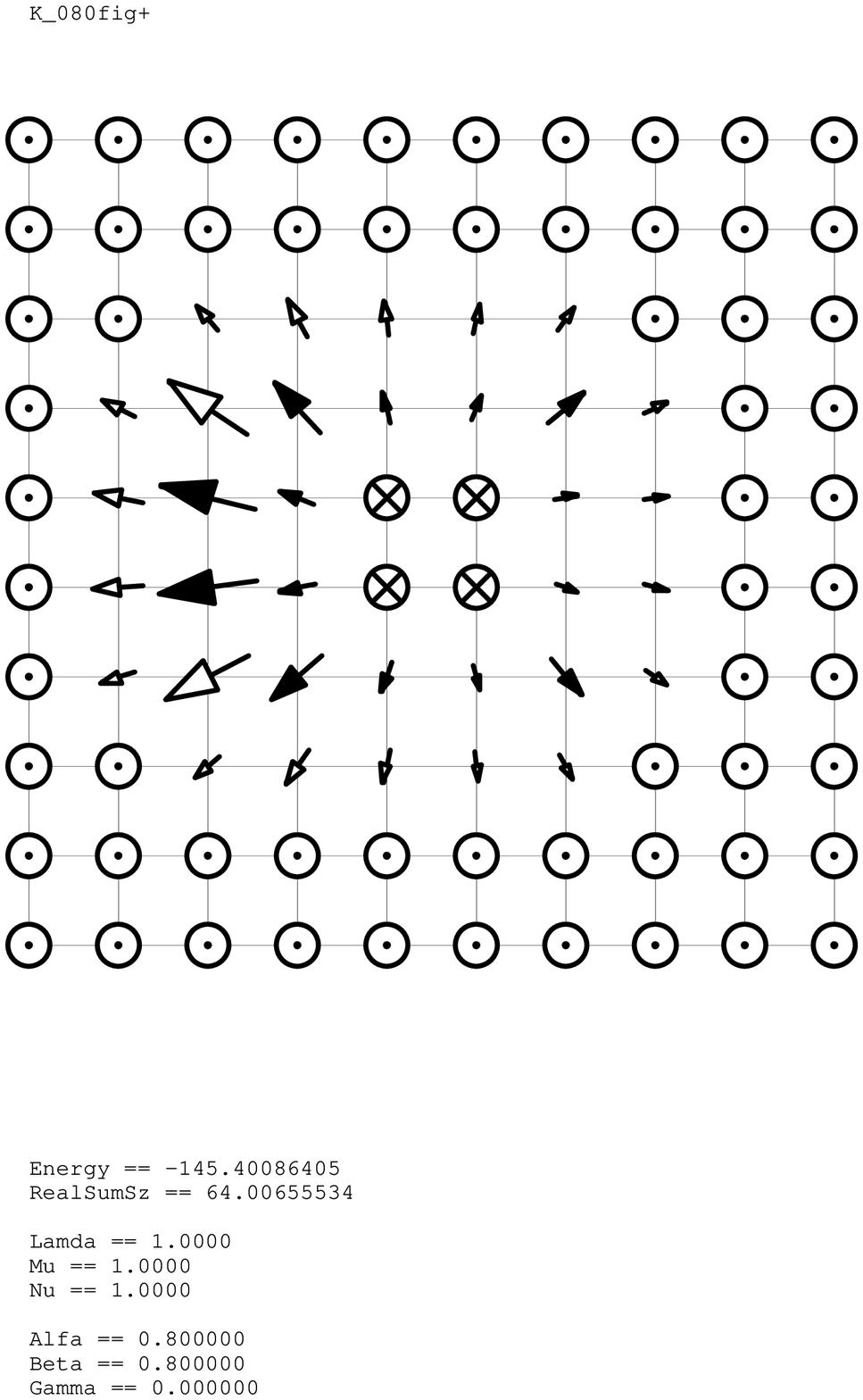}
    \label{si363}}
     \subfigure[ExA, 1.0]{\includegraphics*
  [bb = 180 260 580 660, width = 35mm]{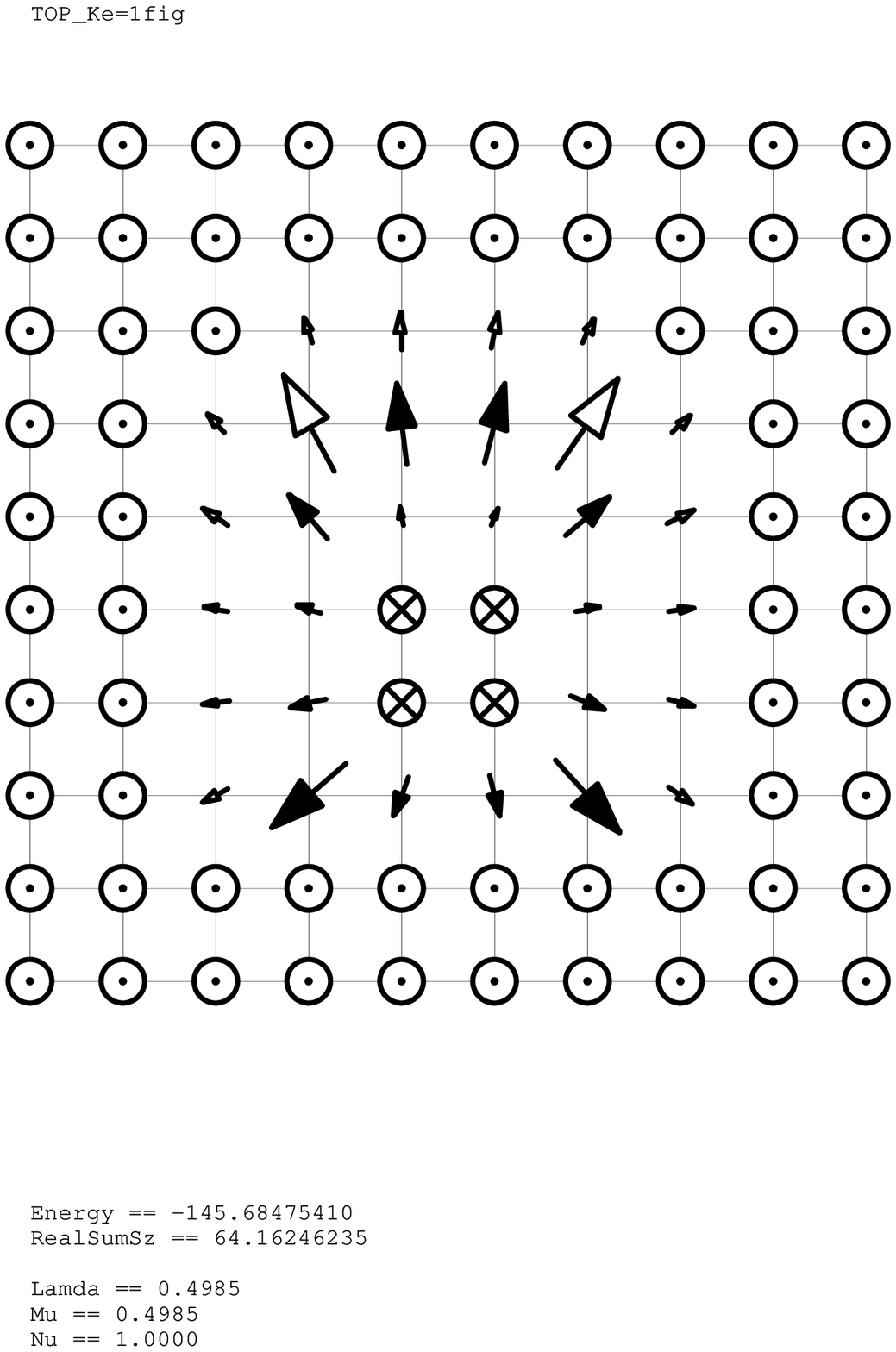}
    \label{ex364}}
        \subfigure[SIA, 1.0]{\includegraphics*
  [bb = 180 260 580 660, width = 35mm]{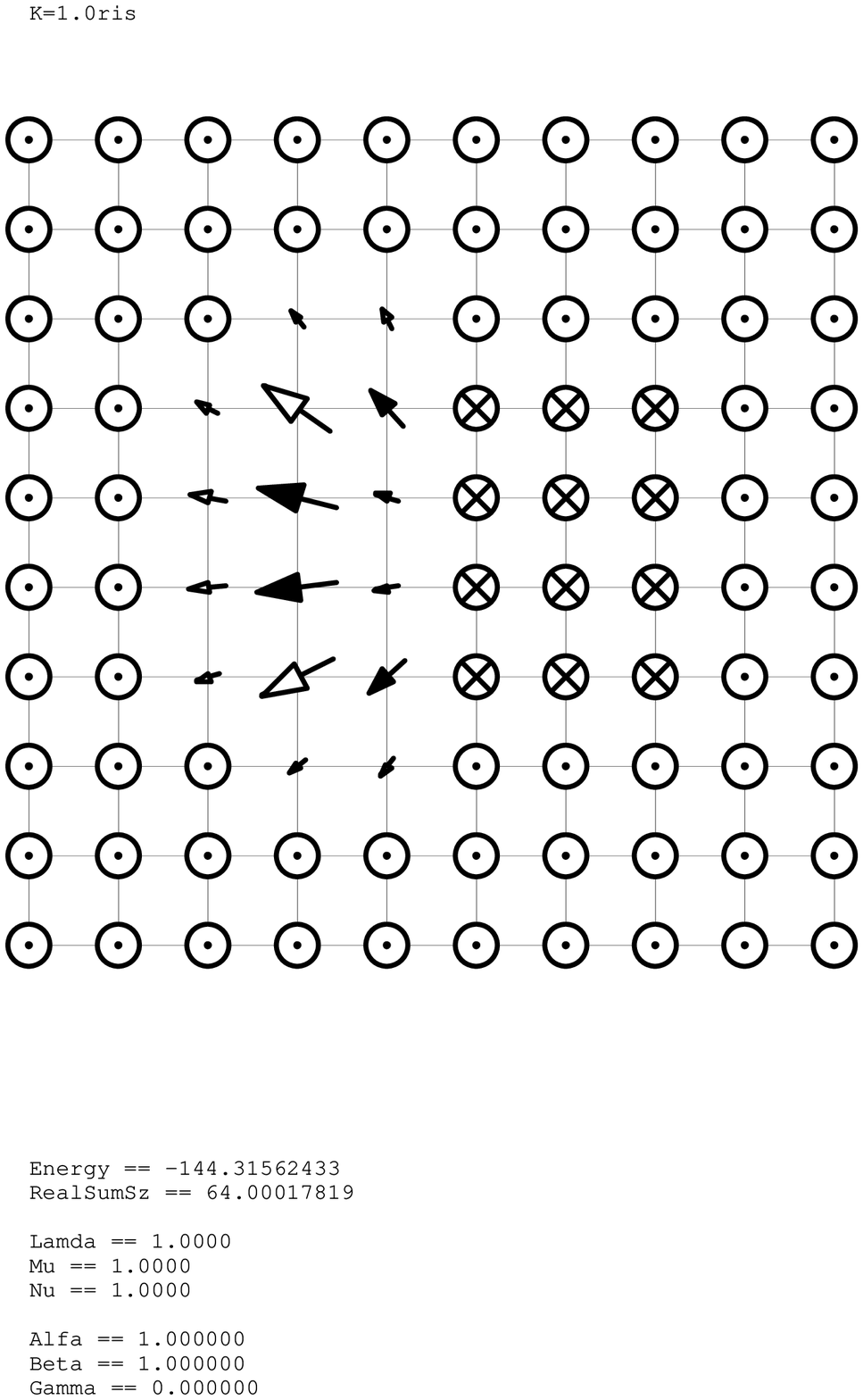}
    \label{si364}}
\subfigure[ExA, 1.2]{\includegraphics*
  [bb = 180 260 580 660, width = 35mm]{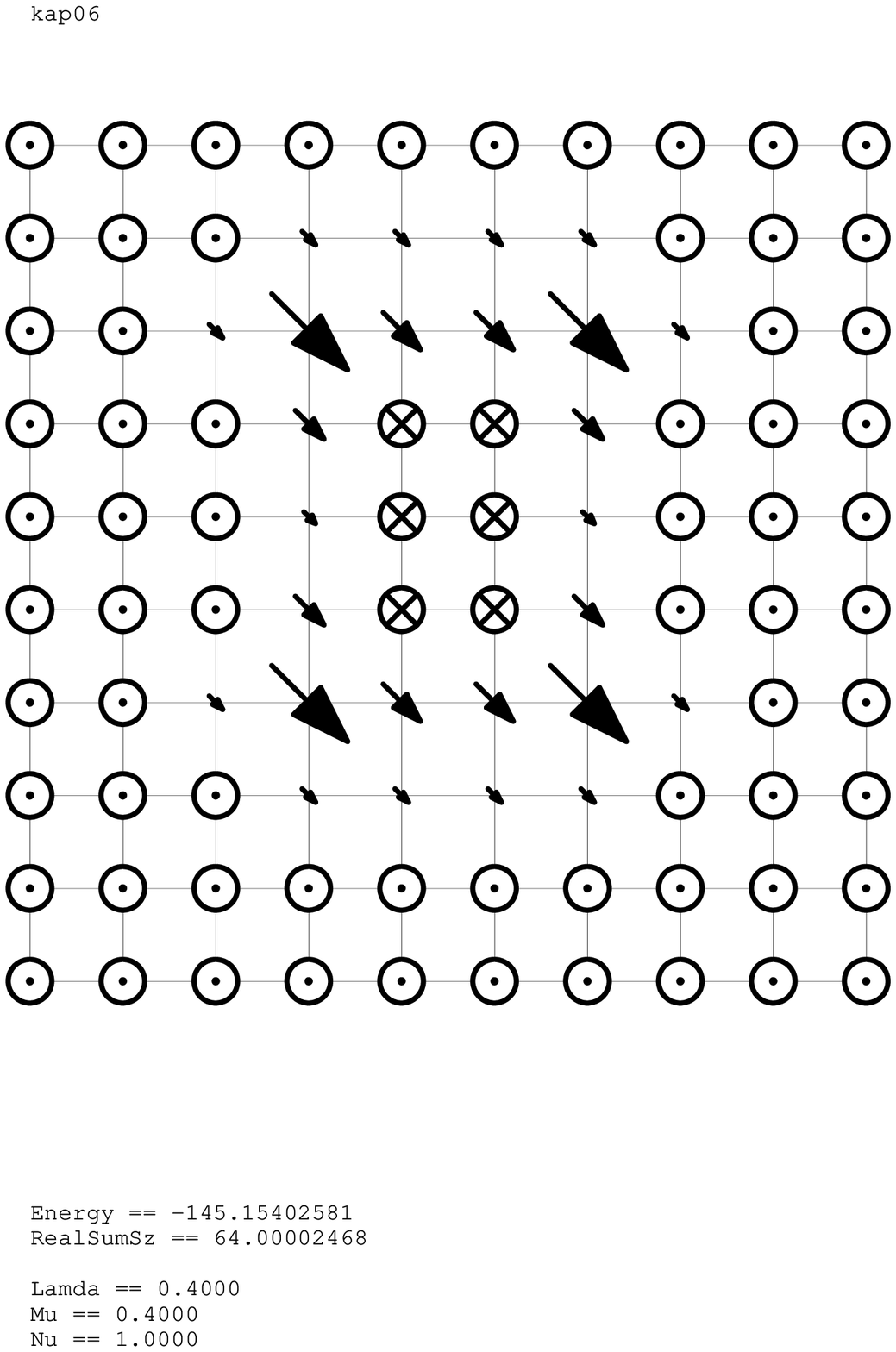}
    \label{ex365}}
    \subfigure[SIA, 1.2]{\includegraphics*
  [bb = 180 260 580 660, width = 35mm]{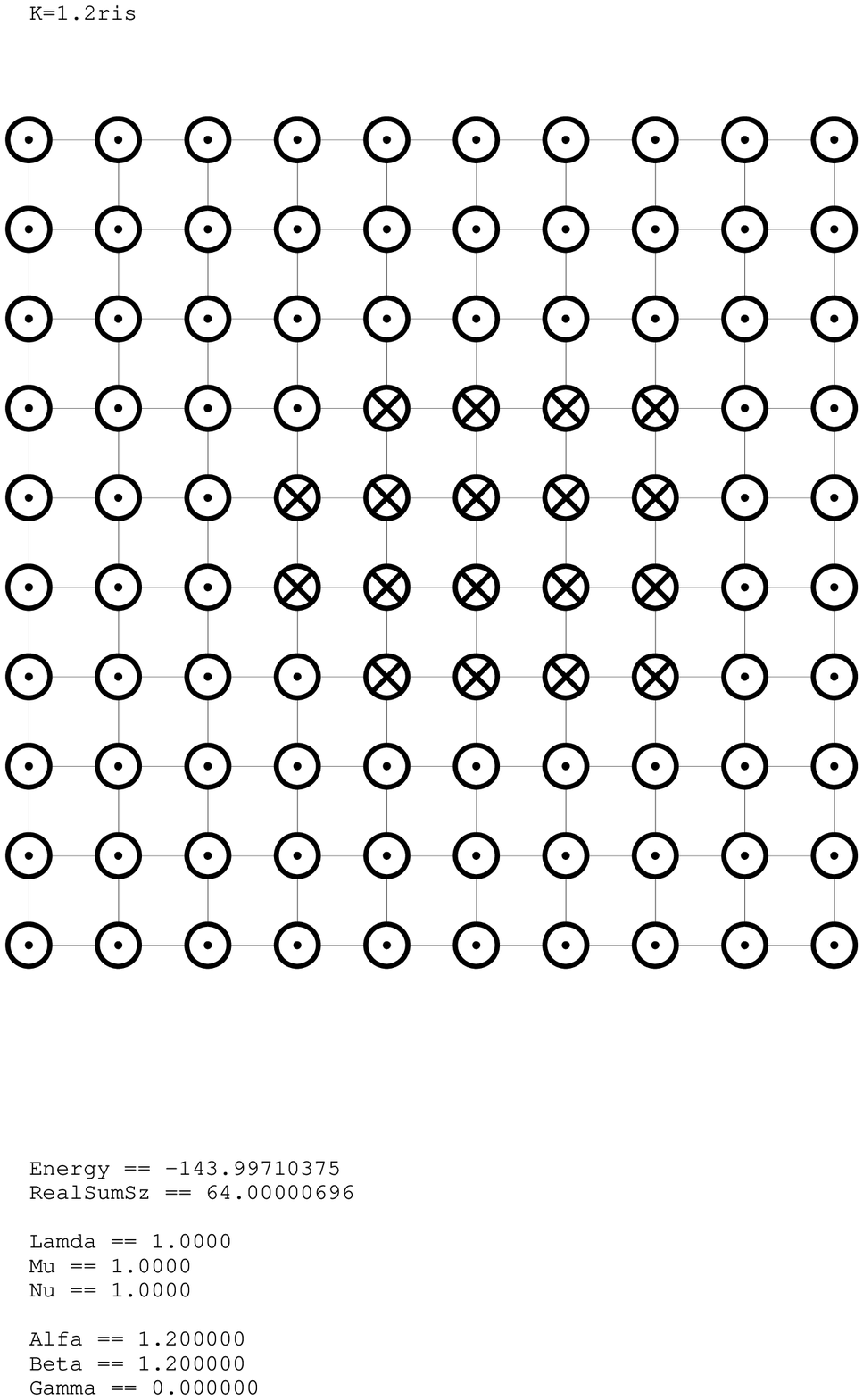}
    \label{si365}}
    \caption{Soliton structure for N=36 (even non-magic number) and different
    values of effective anisotropy constant $K_e/J$
    (shown in captions) \label{f:topol36}}
\end{figure}

Those differences could be explained in terms of previously
discussed fact that DW pinning is stronger for SIA compared to ExA.
Using symmetry arguments, we note that the purely collinear texture
for $n=36$ has a DW of complex form, lower symmetry
(Fig.\ref{si365}), and is much less energetically favorable, than
that for $n=32$. Hence, this state in its pure form occurs only for
sufficiently large SIA with $K>1.18J$ rather then for $K>0.8J$ as
for the "magic" \ case $n=32$. We have found that the textures with
sufficiently high symmetry (in any case they have a center of
symmetry or a couple of C$_2$ axes) are inherent to exchange
anisotropy. Such textures for non-magic $n$'s have DW's of strong
non-collinear structure with a great number of spins with $\theta
\simeq 90^\circ$. For the ExA case, the DW does not "tear apart" up
to quite large $\kappa >0.5J$ ($K_{\rm {eff}}\gtrsim J$). Here both
topological and non-topological soliton textures exist. The energy
of latter textures is shown in Fig. \ref{fig:v4+} by open symbols.
For $\kappa \lesssim 0.4$ both types of solitons have a rectangular
shape and rather high symmetry. Their difference is that for
non-topological soliton the inversion center is present even for
$K_{\rm {eff}}\simeq 1.2J$ while the symmetry of a topological
soliton lowers already for $K_{\rm {eff}}\simeq J$, see Fig.
\ref{ex364},\ref{ex365}. This difference can be understood as
follows. For a topological soliton, the DW containing spins with
$\theta \simeq 90^\circ $ (since the spin angle $\phi $ is
inhomogeneous) is less favorable than that for non-topological
solitons. Thus, for topological solitons the spins with $\theta \neq
0, \pi$ concentrate near one of the soliton edges. However, the
energy difference of these solitons at $K_{\rm {eff}}> 1.1J$ is
quite small. Thus we may assert that in the wide range of ExA
constants we have two soliton types with similar structure of spins
along the easy axis and approximately equal energies.

In the SIA case the DW pinning plays a much more important role so
that the low symmetry, which is characteristic of collinear texture
of the type shown in Fig.\ref{si365}, can be traced already for
sufficiently small $K_{\rm {eff}}\gtrsim 0.7J$, see
Figs.\ref{si363},\ref{si364}. Since the DW inhomogeneity, caused by
a topology, is localized in a small part of a boundary, its role is
not essential so that there is no big difference between topological
and non-topological solitons. At least this difference is much less
pronounced than that for ExA where the non-collinearity amplitude
decreases smoothly as $K$ increases. The new element of symmetry (a
rotation in a spin space) appears in a transition to the collinear
state at $K_{\rm {eff}} \geq K_{\rm {crit}}$. That is why this
transition resembles the phase transition of a second kind. This
behavior of $E(N)$, similar to that for ``magic'' magnon numbers,
can be seen in Fig.~\ref{fig:v4+} for SIA (but not for ExA!).

\begin{figure}[!h]
\vspace*{-5mm} \hspace*{-5mm} \centering{\
\includegraphics[width=1.1\columnwidth]{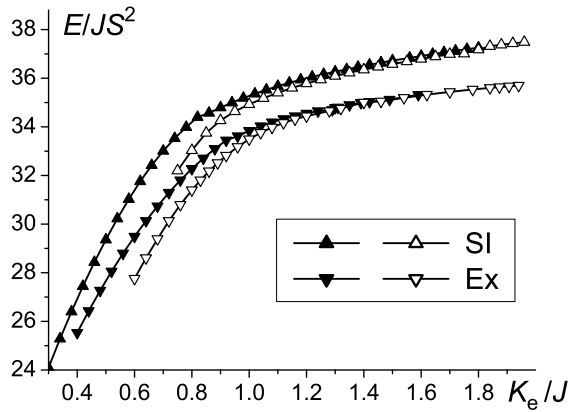}}
\caption{The energy of the soliton with $N/S=35$ (non-magic, odd) as
a function of effective anisotropy constant. Full symbols present
data for topological solitons with inhomogeneous DW, open symbols
correspond to non-topological solitons with $q=0$.} \label{fig:v12}
\end{figure}
As we have found out in the above example of $n=32$ and $n=36$, the
final stage of evolution of a soliton with even $n$ is a collinear
structure. Since the transition of a soliton into a collinear state
is related to the symmetry increasing, it formally resembles a
transition of the second kind. The collinear structure cannot be
realized, for obvious reasons, for any $n$, which is not even. In
this case, which has been considered for $n=35$, the smooth
variation of a soliton energy and structure occurs, see Fig.
\ref{fig:v12}. In particular, its energy is a monotonously
increasing function of $K_{\rm {eff}}$ for both types of
anisotropies. For high enough anisotropy, the difference between
topological and non-topological solitons diminish.

\section{Conclusions}
\label{s:conclusion}

In this paper we have studied the  topological and non-topological
solitons, stabilized by precessional spin dynamics in a classical 2D
ferromagnet with high easy--axis anisotropy on a square lattice. Our
analysis has been performed both analytically in the continuous
approximation and numerically. The main conclusion is,  that in the
2D case the solitons properties change drastically upon departure
from the "singular point" - isotropic continuous model with BP
solitons. Similar to previous studies, it turns out that the
presence of even weak anisotropy makes solitons dynamic, i.e. those
with nonzero precession frequency for any number $N$ of bound
magnons. It is very interesting and unexpected, that the role of
discreteness turns out to be of the same importance as that for
anisotropy, even for $\kappa$, $K\ll 1$, when $l_0\gg a$. The
minimal consideration of discreteness (via higher degrees of
gradients) yields the existence of some critical value (lower
threshold) of both soliton energy and the number of bound magnons.
Similar to the problem of cone state vortices, the instability is
related to the joint action of discreteness and anisotropy
"responsible" for $l_0$ formation (see Eq. (\ref{lzero})). As a
result, the nonanalytic, instead of linear, dependence of the
critical soliton energy (\ref{Ecrit}) on the anisotropy constant
appears.

On the other hand, for intermediate values of anisotropy, like
$K_{\mathrm{eff}} < (0.25\div 0.3 ) J$, and for the values of $N$
far enough from the  critical value of the bound magnons
$N_\mathrm{cr}$, we have checked and confirmed a number of results
regarding soliton structure from the continuous theory. In
particular, the relation between the number of bound magnons and
precession frequency of spins inside the soliton is common to these
two approaches. In agreement with Ref. \onlinecite{IvZaspYastr}
non-topological solitons have, as a rule, the lower energy than the
topological ones (see Fig.\ref{fig:v6}), and they are stable for any
$ N > N_{NT}$.

Some fundamentally new features in  soliton behavior appears as the
anisotropy grows. For "magic" numbers of bound magnons, $N=2l^2S$,
$l$ is an integer, the soliton texture corresponds to the most
favorable structure with a collinear DW of a square shape. In this
case the soliton topological structure disappears (as anisotropy
grows) continuously (to some extend analogously to the second order
phase transition) giving purely collinear state. Such behavior is
inherent to both types of anisotropy,  but for SIA the critical
value $K_{\rm{crit}}$ is lower and the transition is sharper. For
non-magic numbers, when the collinear structure is certainly less
favorable, it still appears for even numbers $N/S$ and the SIA case
by a smooth transition, but with essentially larger $K_{\rm{crit}}$.
Note that in all cases this value of $K_{\rm{crit}}$ is much larger
then that for 1D spin chain, $K_{\rm{crit}}^{(1D)}=0.5J$
\cite{Kamppeter01,Kovalev+}.  Finally, if $n$ is not an even number,
then for $K\gtrsim J$ the soliton always has a non-collinear
structure in the form of DW "piece"  or even a single spin with
$S_z\neq \pm 1$. The energy of such a soliton is almost independent
of inhomogeneity of planar spin components in the DW.

Let us note one more interesting property of the solitons in magnets
with strong anisotropy, namely the existence of the {\em stable
static} solitons in this case. It is widely accepted, that such
solitons are forbidden both in the model with $W_2$ (Hobart-Derrick
theorem) and in the generalized model with negative contribution of
fourth powers of magnetization gradients ($W_2+W_4$ in this
article). However, the region of $N$ values, where the soliton
frequency changes its sign, is clearly seen on the Fig. \ref{fig:v7}
for SIA. This means the existence of certain characteristic values
of $N$, where $\omega =0$ and the soliton is actually static. Also,
for given number of $N$ in some region one can find the
corresponding value of anisotropy constant (high enough) to fulfill
the condition $\omega = 0$ and to realize this static soliton with
non-collinear state. In addition, all purely collinear
configurations, existing for sufficiently strong anisotropy of both
types, are indeed static since the precession with the frequency
$\omega $ around the easy axis does not define any actual
magnetization dynamics. Hence, in the magnets with sufficiently
strong anisotropy the stable static soliton textures can exist. But,
contrary to Ref. \onlinecite{Ivanov86}, they are due to discreteness
effects, namely due to DW pinning.

To finish the discussion of the problem considered in this work, let
us consider the effects taking place at the transition from
classical vectors $\vec S_{\vec n}$ to quantum spins. Usually the
simple condition of integrity of the total $z$-projection of spin
$N$, see \eqref{eq:N}, is considered as a condition of
semi-classical quantization of classical soliton solutions of
Landau-Lifshitz equation for uniaxial ferromagnets
\cite{Kosevich90}. Moreover, the dependence $E=E(N)$ is used to be
interpreted as a semiclassical approximation to the quantum result.
Note, that for some exactly integrable models, for example, $XYZ-$
spin chain with spin $S=1/2$, the exact quantum result coincides
with that derived from the semi-classical approach, see Ref.
\onlinecite{Kosevich90}). This simple picture was elaborated for
solitons having the radial symmetry. Now we consider how this
picture may be modified for solitons with lower spatial symmetry
discussed in this article. Solitons with low enough symmetry, say
"rectangular" solitons for "half-magic" numbers, or even lower
symmetry, which occurs for "non-magic" numbers and especially odd
ones, apparently occur for high-anisotropy magnets. It is clear,
that they can be oriented in a different way in a lattice, that may
be construed as $k$-fold degeneracy of a corresponding state in the
purely classical case, with $k=2$ for a "rectangular" soliton or $k>
2$ for less symmetric states like those reported in the Fig.
\ref{f:topol36}. With the account for effects of coherent quantum
tunneling transitions between these states, this degeneration should
be lifted.  According to the semiclassical approach, valid for bound
states of a large number of spin deviations in high anisotropy
magnets, the transition probability is low and can be calculated
using instantons concept with Gaussian integration over all possible
instanton trajectories \cite{MQT}. As a result one can expect the
splitting of states degenerated in the purely classical case, with
creation of k-multiplet and lifting of the symmetry of the soliton.
A detailed discussion of these effects is beyond the scope of the
present work.

This work was partly supported by the grant INTAS Ref.05-8112. We
are grateful to the Institute of Mathematics and Informatics of
Opole University for generous computing support.


\end{document}